\documentclass{jfm}
\usepackage{epsf,color,supertabular}
\usepackage{amssymb,gensymb}
\usepackage{amsbsy}
\usepackage{verbatim}
\usepackage{enumerate,siunitx}
\usepackage{amsmath,bm,bbding}
\usepackage{graphicx}
\usepackage{natbib}
\usepackage{datetime}
\usepackage{multirow}

\usepackage[colorlinks=true, linkcolor=blue, citecolor=blue, filecolor=blue, runcolor=blue, urlcolor = blue]{hyperref}

\newcommand{\vindex}[2]{#1_{#2}}

\newcommand{\mb}[1]{\bm{#1}}
\newcommand{\tensor}[1]{\underline{\mathbf{#1}}}
\newcommand{\stress}[1]{\underline{\bm #1}}

\newcommand{\vacf}{C_{{\bm U}_{\alpha}}(t)}
\newcommand{\vacff}{$C_{{\bm U}_{\alpha}}(t)$}
\newcommand{\avacf}{C_{{\bm \varOmega}_{\alpha}}(t)}
\newcommand{\avacff}{$C_{{\bm \varOmega}_{\alpha}}(t)$}
\newcommand{\sis}{{\color{blue} Supplementary Information}}

\def \rhof {\rho^{(f)}}
\def \rhop {\rho^{(p)}}

\def \Rep {{\rm Re}^{(p)}}
\def \kbt {k_{\rm B}T}
\def \mum {\,\mu{\rm m}}
\def \asp {${\bm \varepsilon}$}
\def \aspp {{\bm \varepsilon}}

\begin{document}

\shorttitle{Brownian and flow driven motions of nanoellipsoids.}
\shortauthor{Ramakrishnan, Wang et al.}
\title{Motion of  a nano-ellipsoid in a cylindrical vessel flow: Brownian and hydrodynamic interactions.}
\author{
N. Ramakrishnan\aff{1,*}
Y. Wang\aff{2,*}\footnote{Presently at Center for Applied Mathematics, Tianjin University, Tianjin, China, 300072.}
D. M. Eckmann\aff{1,3}
P. S. Ayyaswamy\aff{2}
R. Radhakrishnan\aff{1,4,5}
\corresp{\email{rradhak@seas.upenn.edu}}
}

\affiliation
{
\aff{*} These authors contributed equally
\aff{1}{Department of Bioengineering, University of Pennsylvania, Philadelphia, PA-19104}
\aff{2}{Department of Mechanical Engineering and Applied Mechanics, University of Pennsylvania, Philadelphia, PA-19104}
\aff{3}Department of Anesthesiology and Critical Care, University of Pennsylvania, Philadelphia, PA-19104
\aff{4}Department of Chemical and Biomolecular Engineering, University of Pennsylvania, Philadelphia, PA-19104
\aff{5}Department of Biochemistry and Biophysics, University of Pennsylvania, Philadelphia, PA-19104
}

\maketitle

\begin{abstract}
We present comprehensive numerical studies of the motion of a buoyant or a nearly neutrally buoyant nano-sized ellipsoidal particle in a fluid filled cylindrical tube without or with the presence of imposed pressure gradient (weak Poiseuille flow). The Fluctuating hydrodynamics approach and the Deterministic method are both employed. We ensure that the fluctuation-dissipation relation and the principle of thermal equipartition of energy are both satisfied. The major focus is on the effect of the confining boundary. Results for the velocity and angular velocity autocorrelations (VACF and AVACF), diffusivities, and drag and lift forces as functions of shape, aspect ratio, inclination angle, and proximity to the wall are presented. For the parameters considered, the boundary modifies the VACF and AVACF such that three distinct regimes are discernible --- an initial exponential decay, followed by an algebraic decay culminating in a second exponential decay. The first is due to thermal noise, the algebraic regime is due to both thermal noise and hydrodynamic correlations, while the second exponential decay shows the effect of momentum reflection from the confining wall. Our predictions display excellent comparison with published results for the algebraic regime (the only regime for which earlier results exist). We also discuss the role of off-diagonal elements of the mobility and diffusivity tensor that enables the quantification of the degree of lift and margination of the NC in the vessel. Our study covers a range of parameters that are of wide applicability in nanotechnology and in targeted drug delivery related to the health sciences.
\end{abstract}



\section{Nomenclature}
\sloppy{
\begin{enumerate}[]
	\item $x$, $y$, $z$ : Cartesian Coordinates
	\item $1$, $2$, $3$ : Principal or body fitted coordinates
	\item $a$, $b$, $c$ : dimensions of the ellipsoid along the $x$, $y$, and $z$ directions
	\item ${\bm \varepsilon}$ : aspect ratio of the ellipsoidal particle, defined by $a/b$
	\item $d_{\rm eq}$ : equivalent spherical diameter computed as $\sqrt[3]{abc}$
	\item $m$ : mass of the ellipsoid
	\item $m^{*}_{\alpha}$ : effective mass of the ellipsoid along principal direction $\alpha$, with $\alpha=1,2,3$
	\item $I_{\alpha\alpha}$ : moment of inertia of the ellipsoid for rotation about $\alpha$, with $\alpha=1,2,3$
	\item $I_{\alpha\alpha}^{*}$ : effective moment of inertia of the ellipsoid for rotation about $\alpha$, with $\alpha=1,2,3$
	\item $D$ : diameter of the cylindrical tube
	\item $L$ : length of the  cylindrical tube
	\item $r$ : radial position of the particle with respect to the central axis of the tube
	\item $h$ : shortest distance between the curved wall and the centroid of the particle
	\item $\zeta_0$ : shortest distance between the curved wall and any point on the surface of the particle
	\item $\widetilde{h}$ : $(h-\zeta_0)/\zeta_0$, the non-dimensional particle separation from the curved wall
	\item $l_P$ : discretization length on the particle
	\item $l_W$ : discretization length on the bounding wall
	\item ${\bm u}$ : velocity of the fluid
	\item ${\bm U}_{\alpha}$ : translational velocity of the particle along the $\alpha$ direction, $\alpha=x,y,z$
	\item ${\bm \omega}_{\alpha}$ : rotational velocity of the particle in Cartesian coordinates, $\alpha=x,y,z$
	\item ${\bm \varOmega}_{\alpha}$ : rotational velocity of the particle in the principal coordinates, $\alpha=1,2,3$
\end{enumerate}
}
\section{Introduction}

Nanoparticles of various sizes and shapes are employed in many technologies. In certain applications, it is important to predict the diffusivity and the trajectory of the particle in a fluid medium close to confining boundaries where hydrodynamic interactions with the wall gain prominence. The fluid medium itself may be stationary or flowing.

In targeted drug delivery, for example, ligand functionalized nano-sized particles, or nanocarriers (NCs) are commonly used to deliver drugs to specific locations inside the vasculature. The dynamics of these particles in a confined environment, such as in a narrow blood vessel, is governed by a complex interplay between the hydrodynamic forces,  Brownian interactions, wall effects, and adhesive interactions of the ligands with specific receptors expressed on the vessel wall. The magnitude of each of these effects is governed by a number of factors including the size and shape of the NC, size of the vessel, flow rate, hematocrit density, and expression levels of receptor molecules on the vascular surface~\citep{Ayyaswamy:2013ki}.

In this study, we will be concerned with the effect of shape (which is taken to be an ellipsoid) and confinement on the dynamics of the NC. Compared to a spherical NC, a non-trivial shape such as an ellipsoidal NC has been shown to have a higher efficacy of binding to the cell~\citep{Champion:2006ic,Dasgupta:2013iz,Shah:2011hg}. It is essential to quantify the hydrodynamic forces acting on an ellipsoidal NC under confinement in order to evaluate the role of hydrodynamic interactions in mediating such highly efficacious binding~\citep{Liu2012}. For simplicity, the bulk medium is considered to be a Newtonian incompressible fluid in a cylindrical vessel. When pressure gradients are present, the flow is taken to correspond to a weak Poiseuille (parabolic) profile.

At present, very limited numerical studies exist that accurately evaluate the shape effect of an NC on the momentum transport under confinement with or without the presence of bulk fluid flow. As a consequence, the data on how the diffusivity of such particles  are renormalized by the hydrodynamic interactions due to confinement (i.e. wall effect for anisotropic particles) are largely unavailable. A  major objective of this study is to fill this void.

Numerical simulations of a finite-sized ellipsoid immersed in a fluid medium have been carried out by employing the Stokesian dynamics method \citep{Wakiya57}, the finite element method \citep{Sugihara96,XuMichaelides96,Glowinski01,Swaminathan2006}, boundary integral method \citep{HsuGanatos89}, Lagrange-multiplier-based fictitious domain schemes \citep{Glowinski2001363}, and the lattice Boltzmann method (LBM) \citep{HuangYangLu14,Aidun00,Xia09}, among others.

For nano-sized particles, thermal effects should be considered. To account for the effects of thermal fluctuations on a mechanical system, one can add the thermal force terms to the governing equations of the system based on the formulations of non-equilibrium statistical mechanics \citep{Kubo:1966dq}. In order to achieve thermal equilibrium, the spatial and temporal correlations in these systems should satisfy a balance between the thermal random force and the dissipation of system which is required by the fluctuating-dissipation theorem \citep{Kubo:1966dq}. To add the thermal force describing the Brownian motion of a particle immersed in a fluid,  we adopt the fluctuating hydrodynamic approach \citep{Landau1980}, which essentially adds a stochastic stress to the stress tensor in the fluid momentum equation. Numerical simulations of fluctuating hydrodynamics governing a spherical NC have been carried out by employing the finite volume method \citep{Sharma2004466,Donev10}, LBM \citep{Ladd93,Ladd94a,Ladd94b,Patankar02,Adhikari05,Dunweg08,Nie09}, stochastic Eulerian-Lagrangian method \citep{Atzberger11}, and stochastic Arbitrary Lagrangian-Eulerian (ALE) \citep{Uma2011}.

For an ellipsoidal particle, there are special requirements based on the shape that have to satisfied. We employ quaternions  to account for the orientations of the ellipsoid~\citep{Kuipers99,Chou92,Swaminathan2006}. Both translational and rotational motions of the ellipsoidal NC in: (i) a quiescent fluid medium and (ii) a weak Poiseuille flow  are investigated. The Delaunay-Voronoi method \citep{George91} is employed to generate an unstructured finite element mesh. Thermal fluctuations are represented by adding a stress tensor as the white noise in space and time \citep{Landau1980,Espanol09} to the stress term in the Navier-Stokes equations. The fluctuation-dissipation theorem is satisfied by discretizing the fluctuating hydrodynamic equations in terms of finite element shape functions based on the Delaunay triangulation \citep{Espanol09}. Although the particle Reynolds number is very small, the presence of thermal effects requires a treatment of the  full Navier-Stokes equations together with the random stress tensor in the problem formulation. As discussed in Uma et al. \citep{Uma2011}, for a spherical particle, the added mass of the displaced fluid should be considered along with the mass of the particle. It must be emphasized that for ellipsoidal particles, the accounting for the added masses and added moments of inertia is non-trivial due to their dependence on the shape and orientation of the particle. The appropriate expressions for these quantities are  provided in detail in this manuscript.

For the fluctuating hydrodynamics approach, a large number of realizations are required to develop adequate statistics of the dynamics. Under certain conditions, the relaxation behavior of the velocity autocorrelations may also be obtained in the absence of the imposed random stress tensor. This procedure, the deterministic method, is based on the Onsager regression hypothesis, which states that the regression of microscopic thermal fluctuations at equilibrium follows the macroscopic law of relaxation of small non-equilibrium disturbances~\citep{Onsager:1931uu,Onsager:1931vt}. The deterministic method affords computational ease to develop relevant results with a stationary medium. The details are described in a subsequent subsection. As stated earlier, the approaches described in this paper may be extended to biological applications which additionally require the treatment of a non-Newtonian fluid such as blood.

The paper is organized as follows. Section \ref{sec:formulation} describes the mathematical formulation of the problem, the Galerkin finite element method for solving the fluid momentum equations, and the generation of the random stress tensor for a tetrahedron mesh. Section \ref{sec:resuls} presents the validations, numerical results and discussion. We conclude in Section \ref{sec:conclusions} with a detailed discussion on the applications of our methods.

\section{Formulation of the problem and solution methodology}\label{sec:formulation}
\subsection{Governing equations and boundary conditions for the fluctuating hydrodynamics study}
\par We consider an ellipsoidal NC immersed in an incompressible,  quiescent or flowing Newtonian fluid contained in a cylindrical tube $\Sigma$, as shown in Fig.~\ref{fig_surfacemesh}. The inlet and outlet boundaries are denoted by $\Sigma_i$ and $\Sigma_o$, respectively, $\Sigma_w$ is the wall boundary, and the particle surface is denoted by $\Gamma_p$. The dimensions of the particle  are denoted by $a$, $b$, and $c$, and the length and diameter of the tube are $L$ and $D$, respectively, as shown in Figs.~\ref{fig_surfacemesh}(a) and (b). The position of the particle (i.e., its center of mass) is expressed either in terms of $r$, the radial distance from the tube axis, or $h$, the radial distance as measured from the wall boundary. The angular orientation of the particle is measured in terms of the inclination angle $\theta$ which denotes an in-plane tilt (in the $x-z$ plane). In view of the asymmetric shape and the orientation of the ellipsoid, yet one more measure of length becomes relevant in our problem. With reference to Fig.~\ref{fig_surfacemesh}(a) it may be noted that $\zeta_{0}$ is the maximum value from among the projections of $a$, $b$, and $c$ on a plane perpendicular to the cylinder axis (see appendix ~\ref{app:a0derivation}). For example, $\zeta_{0}=b/2$ when $\theta=0$\degree, and $\zeta_{0}=a/2$ when $\theta=90\degree$. For notational simplicity, we define the non-dimensional separation between the NC and the wall, in terms of $h$ and $\zeta_0$, as $\widetilde{h}=(h-\zeta_0)/\zeta_0$. \\

\begin{figure}
\centering
\includegraphics[width=\textwidth]{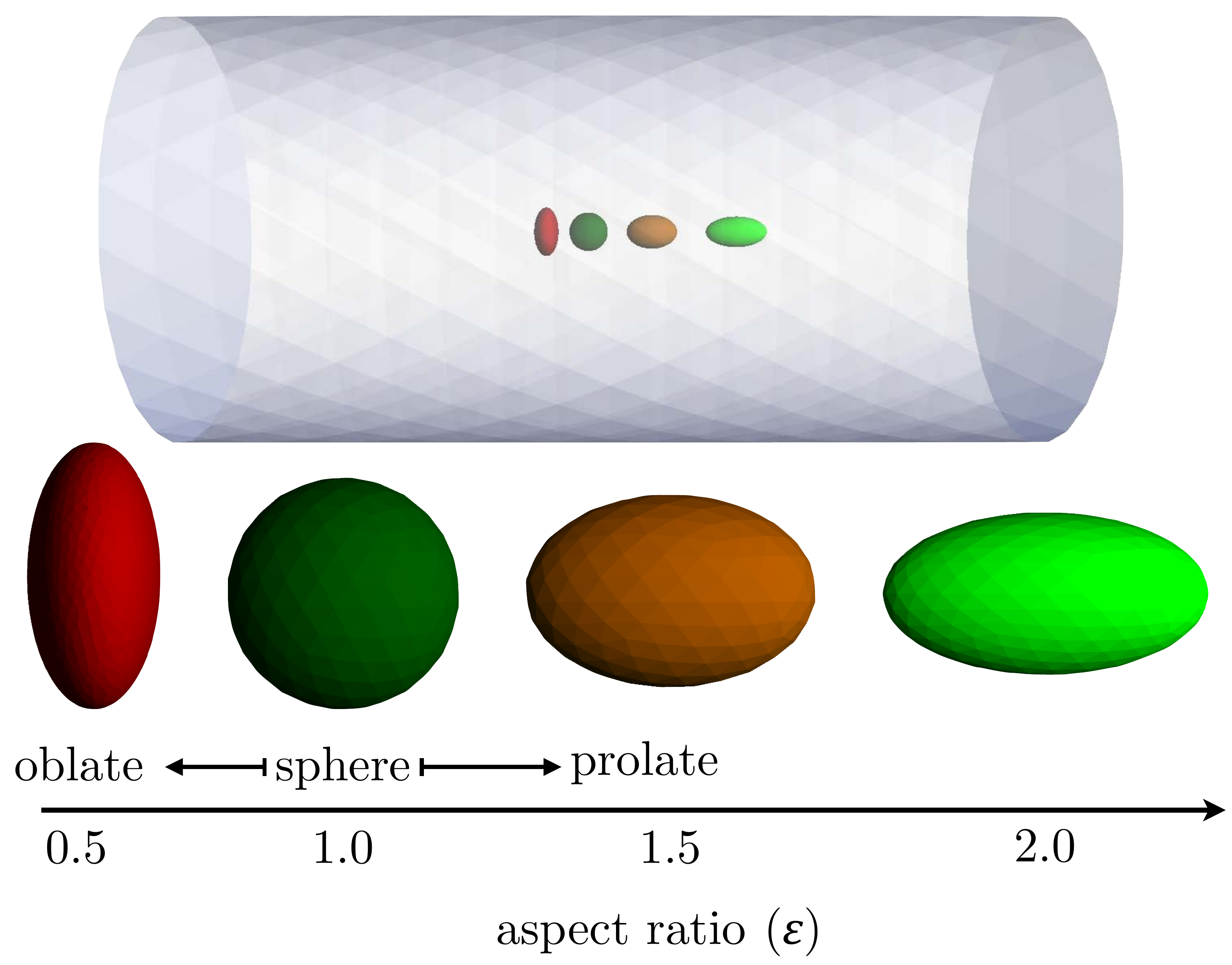}
\caption{Ellipsoidal NCs with aspect ratio $\aspp=0.5$, $1.0$, $1.5$, and $2.0$  at the center of a cylindrical tube of diameter $D=5\mum$, and oriented along the axis of the tube.}
\end{figure}

 \begin{figure} 
	\centering
	\includegraphics[width=1.0\textwidth]{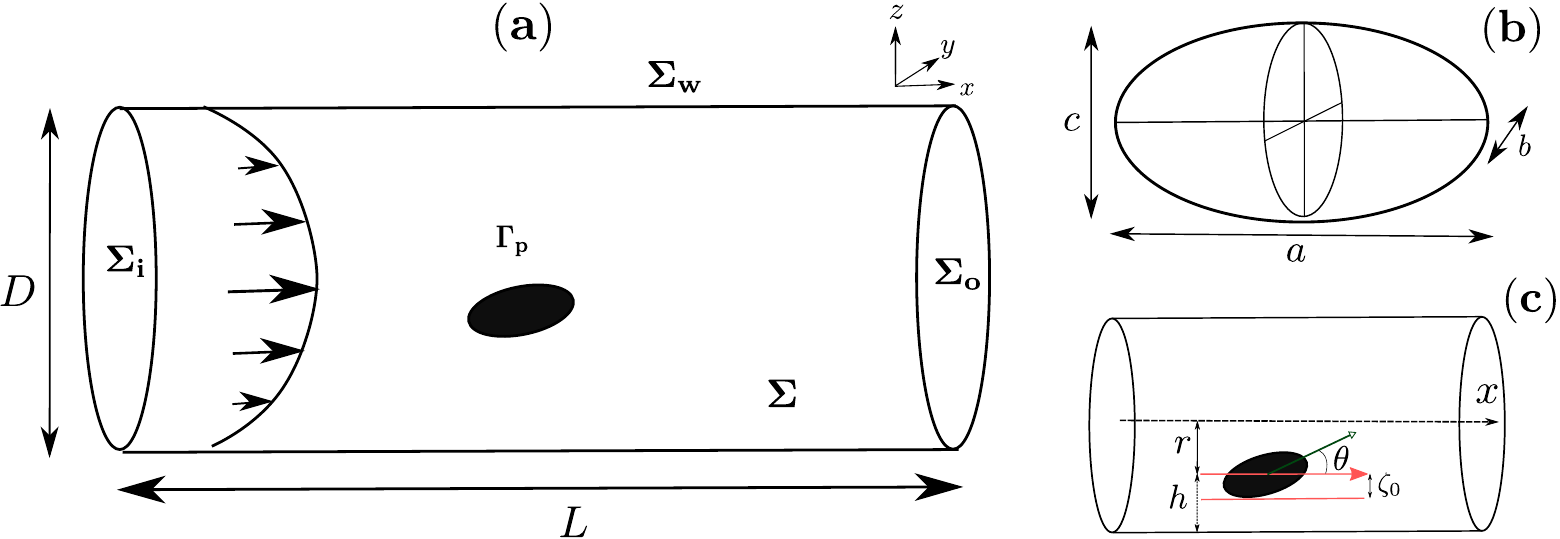}
	\caption{Schematic representation of  (a) an ellipsoid bounded by a circular tube of length $L$ and diameter $D$ with a Poiseuille flow along the $x$ direction, (b) the dimensions of an ellipsoid denoted by $a$, $b$, and $c$. Panel (c) shows the various length scales in the system: (i) the proximity of the particle from the wall boundary described either in terms of its radial distance $r$ or its separation from the wall $h=D/2-r$, and (ii) $\zeta_0$ which denotes the maximum radial size of the particle -- the value of $\zeta_0$ is a function of $\theta$.}
	\label{fig_surfacemesh}
\end{figure}

\noindent The fluid domain satisfies:
\begin{eqnarray}
\label{equ_fluid_incomp}
\nabla\cdot\mb{u} &= &0,\\
\label{equ_fluid_mome}
\rho\frac{D\mb{u}}{Dt} &= &\nabla\cdot \stress{\sigma},
\end{eqnarray}
where $\mb{u}$ and $\rho$ are the velocity and density of the fluid respectively. $\stress{\sigma}$ is the  stress tensor given by:
\begin{eqnarray}
\label{eqn_stress}
 \stress{\sigma}=-p\tensor{J} + \mu[\nabla\mb{u}+(\nabla\mb{u})^T] + \tensor{S}.
 \label{eqn:stress-relation}
\end{eqnarray}
Here, $p$ is the pressure, $\tensor{J}$ is the identity tensor, and $\mu$ is the dynamic viscosity. The random stress tensor $\tensor{S}$ is assumed to be a Gaussian white noise that satisfies:
\begin{eqnarray}
 \langle S_{ij}(\mb{x},t) \rangle & = & 0, \\
 \langle S_{ik}(\mb{x},t)S_{lm}(\mb{x}',t') \rangle & = & 2k_BT\mu(\delta_{il}\delta_{km}+ \delta_{im}\delta_{kl}) \delta(\mb{x}-\mb{x}')\delta(t-t'),
 \label{eqn:fluct-dissip}
\end{eqnarray}
where $\langle\cdot \rangle$ denotes an ensemble average, $k_B$ is the Boltzmann constant, $T$ is the absolute temperature, and $\delta_{ij}$ is the Kronecker delta. The Dirac delta functions $\delta(\bm{x}-\bm{x}')$ and $\delta(t-t')$ denote that the components 
of the random stress tensor are spatially and temporally uncorrelated. The mean and variance of the random stress tensor are chosen to be consistent with the fluctuation-dissipation theorem for an incompressible fluid~\citep{HaugeMartin73}. \\

The translational and rotational motions of a rigid particle suspended in the fluid satisfy,
\begin{eqnarray}
\label{eqn_part_trans}
 m\frac{d\mb{U}}{dt} &=& \mb G-\int_{\Gamma_p}\stress\sigma\cdot\hat{\bm{n}}\, ds, \\
 \label{eqn_part_rot}
 \tensor{I}\frac{d\mb\Omega}{dt} +\mb\Omega\times(\tensor{I}\mb\Omega)&=& -\tensor{R}^T \int_{\Gamma_p} (\mb{x} - \mb{X})
 \times (\stress{\sigma} \cdot \hat{\mb{n}})\, ds,
\end{eqnarray}
where $\mb{U}=\left(\mb{U}_x,\mb{U}_y,\mb{U}_z\right)^T$ and $\mb{X}=\left(\mb{X}_x,\mb{X}_y,\mb{X}_z\right)^T$ are the translational velocities and the position of the center of mass of the NC, respectively, in the Cartesian frame $(x,y,z)$.
$\mb\varOmega=\left(\mb\varOmega_1,\mb\varOmega_2,\mb\varOmega_3\right)^T$ is rotational velocity of the particle in the body fitted frame of reference given by $(1,2,3)$. The mass and the moments of inertia of the particle are given by $m$ and $\tensor{I}$, respectively, and $\mb{G}$ represents a body force such as gravity. $\hat{\mb{n}}$ is the outward drawn unit normal  to the particle surface. Here, the moment of inertia $\tensor{I}$ is also defined with respect to the body frame attached to the particle.  $\tensor{R}$ is the rotational matrix that transforms  the body frame quantities to the inertial frame $(x,y,z)$. In this study, the rotational matrix is defined in terms of the quaternions 
$\mb{q}=(q_0,q_1,q_2,q_3)^T$ with $\| q \|^2=q_0^2+q_1^2+q_2^2+q_3^2=1$,
 \begin{eqnarray}
 \tensor{R} =
 \begin{pmatrix}
  2(q_0^2+q_1^2)-1 & 2(q_1q_2-q_0q_3) & 2(q_1q_3+q_0q_2) \\
  & & \\
  2(q_1q_2+q_0q_3) & 2(q_0^2+q_2^2)-1 & 2(q_2q_3-q_0q_1) \\
    & & \\
  2(q_1q_3+q_0q_2) & 2(q_2q_3+q_0q_1) & 2(q_0^2+q_3^2)-1
 \end{pmatrix}.
 \end{eqnarray}

The  position $\bm{X}$ and the quaternions $\bm{q}$ of the particle evolve in time according to:
\begin{eqnarray}
\label{eqn_part_pot} 
\frac{d\bm{X}}{dt}=\bm{U},
\end{eqnarray}
and
\begin{eqnarray}
\label{eqn_quaternionsdev}
\dfrac{d\bm{q}}{dt}=\frac{1}{2}\begin{pmatrix}
  0 & -{\bm \varOmega}_{1} & -{\bm \varOmega}_{2} & -{\bm \varOmega}_{3}  \\
  {\bm \varOmega}_{1} & 0  & {\bm \varOmega}_{3}  & -{\bm \varOmega}_{2}  \\
  {\bm \varOmega}_{2} & -{\bm \varOmega}_{3} & 0 & {\bm \varOmega}_{1}  \\
  {\bm \varOmega}_{3} & {\bm \varOmega}_{2} & -{\bm \varOmega}_{1} & 0
 \end{pmatrix} \bm{q}.
 \end{eqnarray}

\noindent The initial conditions of the problem are:
\begin{eqnarray}
 \mb U(t=0)=0, \quad \mb\Omega(t=0)=0,\quad \mb{u}(t=0)=0 \quad\text{in $\Sigma$},
 \label{eqn:boundc}
\end{eqnarray}
and the boundary conditions are given by:
\begin{eqnarray}
 \label{eqn_bcn}
& \mb u = \mb{u_{in}} \quad&\text{on $\Sigma_i$}  \quad \textrm{ (inlet)}, \\
 \label{eqn_bc_out}
& \stress\sigma\cdot\hat{\bm{n}}= 0 \quad&\text{on $\Sigma_o$}  \quad \textrm{ (outlet)}, \\
& \mb u = 0 \quad &\text{on $\Sigma_w$}   \quad \textrm{ (wall \, boundary)}, \\
& \mb u=\mb{U}+\tensor{R}\mb\Omega\times(\mb{x}-\mb{X}) \quad &\text{on $\Gamma_{p}$}  \quad \textrm{ (particle \,surface)}.
\end{eqnarray}
The above formulation is numerically solved and the details are provided in the next subsection.
\subsection{The weak Formulation}

Let $\mathcal{V}$ be the function space given by:
\begin{eqnarray}
 \mathcal{V}=\left\{
 \begin{array} {l}
  \mb V=(\mb U,\mb\Omega,\mb u,p)|(\mb U,\mb\Omega)\in \mathcal{R}^{3},\,\mb u\in \mathcal{H}^1,\,p\in\mathcal{L}^2,\\
  \\
  \mb u = 0 \text{ on }\Sigma_w,\,\mb{u}=\mb{U}+\tensor{R}\mb\Omega\times(\mb x-\mb X) \text{ on }\Gamma_p,\\
  \\
  \mb u = \mb{u}_{in} \text{ on }\Sigma_i,\,
   p = 0\text{ on }\Sigma_o,
 \end{array}\right\}
\end{eqnarray}
where $\mathcal{H}^1$ is the Hilbert space for the fluid velocity field. The test function space $\mathcal{V}_0$ is
the same as $\mathcal{V}$, except that $\mb u=0$ on $\Sigma_i$ and $\Sigma_o$, and hence:
\begin{eqnarray}
 \tilde{\bm{V}}=(\tilde{\mb U},\tilde{\mb\Omega},\tilde{\mb u},\tilde{p})\in\mathcal{V}_0.
\end{eqnarray}
Multiplying equation (\ref{equ_fluid_mome}) by the test function for the fluid velocity $\tilde{\mb u}$, and integrating over the fluid domain at time $t$ yields:
\begin{eqnarray}
 \label{eqn_weak1}
 \int_\Sigma\rho\frac{D\mb u}{Dt}\cdot\tilde{\mb u}\,dv-\int_\Sigma(\nabla\cdot\stress{\sigma})\cdot\tilde{\mb{u}}\,dv=0.
\end{eqnarray}
Upon integration by parts, the second term may be expressed as:
\begin{eqnarray}
\label{eqn_stress_int_part}
 \int_\Sigma(\nabla\cdot\stress{\sigma})\cdot\mb{u}\,dv &=&-\int_\Sigma\stress{\sigma}:\nabla\mb{u}\,dv
 +\int_{\Gamma_p}(\stress{\sigma}\cdot\hat{\mb n})\cdot\tilde{\mb u}\,ds,
\end{eqnarray}
and the last term of equation \eqref{eqn_stress_int_part}  may be rewritten using equations:
\eqref{eqn_part_trans} and \eqref{eqn_part_rot} as
\begin{eqnarray}
\label{eqn_stress_int_part_more}
 \int_{\Gamma_p}(\stress{\sigma}\cdot\hat{\mb n})\cdot\tilde{\mb u}\,ds &=& 
 \int_{\Gamma_p}(\stress{\sigma}\cdot\hat{\mb n})\cdot(\tilde{\mb U}+(\tensor{R}\tilde{\mb\Omega})\times(\mb{x}-\mb{X}))\,ds\nonumber \\
 & & \nonumber \\
 &=& \tilde{\mb U}\cdot\int_{\Gamma_p}\stress{\sigma}\cdot\hat{\mb n}\,ds+(\tensor{R}\tilde{\mb\Omega})\cdot\int_{\Gamma_p}
 (\mb{x}-\mb{X})\times(\stress{\sigma}\cdot\hat{\mb n})\,ds\nonumber\\
  & & \nonumber \\
 &=& -\tilde{\mb U}\cdot\left(m\frac{d\mb{U}}{dt}-\bm G\right)-(\tensor{R}\tilde{\mb\Omega})\cdot\left(\tensor{R}\left[\tensor{I}\frac{d\mb\Omega}{dt} 
 +\mb\Omega\times \tensor{I}\mb\Omega\right]\right)\nonumber\\
  & & \nonumber \\
 &=& -\tilde{\mb U}\cdot\left(m\frac{d\mb{U}}{dt}-\bm G\right)-\tilde{\mb\Omega}\cdot\left(\tensor{I}\frac{d\mb\Omega}{dt} 
 +\mb\Omega\times \tensor{I}\mb\Omega\right).
\end{eqnarray}
From equations \eqref{eqn_stress}, \eqref{eqn_weak1}, \eqref{eqn_stress_int_part}, and \eqref{eqn_stress_int_part_more}, 
we get the weak formulation for the combined fluid-particle momentum equations:
\begin{eqnarray}
\label{eqn_mainweakformula}
 \int_{\Sigma}\rho\frac{D\bm{u}}{Dt}\cdot\tilde{\bm{u}}\,dv -\int_{\Sigma}p\nabla\cdot\tilde{\bm{u}}\,dv
 +\int_{\Sigma}(\mu(\nabla \bm{u}+(\nabla \bm{u})^T)+\tensor{S}):\nabla\tilde{\bm{u}}\,dv
 \nonumber \\
+\tilde{\bm{U}}\cdot\left(m\frac{d\bm{U}}{dt}-\mb G\right)+\tilde{\bm{\Omega}}\cdot
 \left(\tensor{I}\frac{d\bm{\Omega}}{dt}+\bm{\Omega}\times(\tensor{I}\bm{\Omega})\right)=0,
\end{eqnarray}
together with,
\begin{eqnarray}
 \int_{\Sigma}\tilde{p}(\nabla\cdot \mb u)\,dv=0.
\end{eqnarray}
\subsection{Arbitrary Lagrangian-Eulerian (ALE) mesh movement}
An ALE technique is used to handle the movement of the particle in the fluid domain, see \citep{Hu2001a}.
The material derivative of $\mb u(\mb x,t)$ in an ALE formulation is given as:
\begin{eqnarray}
 \frac{D\bm{u}}{Dt}=\frac{\delta \bm u}{\delta t}+[(\bm u-\bm u_m)\cdot\nabla]\bm u,
\end{eqnarray}
where, 
\begin{eqnarray}
\label{eqn_meshvdef}
 \frac{\delta \bm u}{\delta t}=\frac{\partial}{\partial t}\bm{u}(\bm{x}(\phi,t),t)|_{\phi \text{ is fixed}},
 \quad\text{ and }\quad \frac{d}{dt}\bm{x}(\phi,t)=\mb u_m,
\end{eqnarray}
are the time derivatives of the velocity and the mesh velocity, respectively, with the former being defined in a fixed referential frame $\phi$.

The mesh velocity $\bm u_m$ in equation (\ref{eqn_meshvdef}) is set to follow the motion of the particles and the motion of the confined fluid, and is computed using the Laplace's equation in the fluid domain:
\begin{align}\label{eqn_meshvel}
 &\nabla\cdot(\epsilon_e\nabla\bm{u}_m)=0 & \text{ in } &\Sigma,
 \end{align}
subject to boundary conditions:
 \begin{align}
 &\bm u_m = \bm U + \tensor{R}\bm\Omega\times(\bm x-\bm X) &\text{ on } &\Gamma_p, \\
 &\bm u_m = 0&\text{ on } & \Sigma_w+\Sigma_i+\Sigma_o.
\end{align}
Here, $\epsilon_e$ controls the deformation of the mesh and we choose it to be $\epsilon_e=1/V_e$, where $V_e$ is the volume of the tetrahedral element. Similarly, the acceleration $\bm a_m$ of the mesh vertices is chosen to satisfy
\begin{align}\label{eqn_meshacc}
 &\nabla\cdot(\epsilon_e\nabla\bm{a}_m) = 0 &\text{ in }& \Sigma,
 \end{align}
 with boundary conditions:
 \begin{align}
 &\bm a_m = \frac{d\bm U}{dt} + (\mathcal{A}\tensor{R} \bm\Omega+\tensor{R}\frac{d\bm\Omega}{dt})\times(\bm x-\bm X)
     -\tensor{R}\bm\Omega\times\bm U  &\text{ on }&\Gamma_p, \\
 &\bm a_m = 0 &\text{ on }& \Sigma_w+\Sigma_i+\Sigma_o, 
\end{align}
where,
\[\mathcal{A}=\begin{bmatrix}
 0    & -\omega_{z} & \omega_{y} \\
 \omega_{z} &    0 & -\omega_{y} \\
 -\omega_{y} & \omega_{x} & 0    \\ 
\end{bmatrix},\]
and $(\omega_{x},\omega_{y},\omega_{z})^T=\tensor{R}\bm{\Omega}$.

The linear weak formulations for the mesh velocity and acceleration are  solved using the biconjugate gradient stabilized method. The positions of the mesh vertices are updated using the second order forward Euler scheme:
\begin{equation}
\label{eqn_meshpos}
 \bm x^{n+1}_m=\bm x^n_m+\bm u_m^n(\bm x^n)\Delta t+\frac{1}{2}\bm a_m^n(\bm x^n)\Delta t^2.
\end{equation}

\subsection{Temporal and Spatial discretization}
We use an adaptive second-order backward finite difference method to discretize the time derivatives in equation \eqref{eqn_mainweakformula} which are given by:
\begin{eqnarray}
 \frac{D\bm{u}}{Dt} &\approx& C_1\frac{\bm u^{n+1}(\bm x)-\bm u^n(\bm x')}{\Delta t_n}
       +C_2\frac{\delta \bm u^n(\bm x')}{\delta t}\nonumber \\
  &  &\quad + \quad [(\bm u^{n+1}(x)-\bm u_m^{n+1}(x))\cdot\nabla]\bm u^{n+1}(x),
  \label{eqn:discreteDUDt}
  \end{eqnarray}
  
  \begin{eqnarray}
   \frac{d\bm U}{dt} &\approx& C_1\frac{\bm U^{n+1}-\bm U^n}{\Delta t_n}+C_2\frac{\delta\bm U^n}{\delta t}, \\
   & & \nonumber \\
 \frac{d\bm{\Omega}}{dt} &\approx&
  C_1\frac{\bm{\Omega}^{n+1}-\bm{\Omega}^n}{\Delta t_n}
 +C_2\frac{\delta\bm{\Omega}^n}{\delta t}.
 \label{eqn:discreteDwDt}
\end{eqnarray}
where $C_1=\dfrac{\Delta t_n}{2\Delta t_n+\Delta t_{n-1}}$ and $C_2=\dfrac{\Delta t_n+\Delta t_{n-1}}{2\Delta t_n+\Delta t_{n-1}}$, with $\Delta t_n=t_{n+1}-t_n$ being the timestep for integration.

However, we use a second order finite difference scheme to discretize the position and the orientation (represented by quaternions) of the particle as,
\begin{eqnarray}
 \bm X^{n+1}&=&\bm X^n+\Delta t_n\bm U^n+\frac{(\Delta t_n)^2}{2}\frac{d\bm U^n}{dt}, \\
 & & \nonumber \\
 \bm q^{n+1}&=&\bm q^n+\Delta t_n\frac{d\bm q^n}{dt}+\frac{(\Delta t_n)^2}{2}\frac{d^2\bm q^n}{dt^2}.
\end{eqnarray}
The derivatives of $\bm q^{n}$ are computed using equation \eqref{eqn_quaternionsdev}.

Using equations ~\eqref{eqn:discreteDUDt}-\eqref{eqn:discreteDwDt}, the  weak formulation of the governing equations (see equation ~\eqref{eqn_mainweakformula}) may now be expressed as:
\begin{align}
\label{eqn_main_weakfordis}
 &  \int_{\Sigma}\rho\left(\frac{C_1}{\Delta t_n}\bm u^{n+1}(\bm x)+((\bm u^{n+1}(\bm x)-\bm u^{n+1}_m(\bm x))
 \cdot\nabla)\bm u^{n+1}(\bm x)\right)
  \cdot\tilde{\bm{u}}dv -\int_{\Sigma}p^{n+1}(\bm x)\nabla\cdot\tilde{\bm{u}}\,dv \nonumber \\
 & +\int_{\Sigma}\left(\mu(\nabla\bm{u}^{n+1}(\bm x)+
 (\nabla\bm{u}^{n+1}(\bm x))^T)+\mb{S}^{n+1}(\bm x)\right):\nabla\tilde{\bm{u}}\,dv \nonumber \\
 &+\frac{C_1}{\Delta t_n}m\tilde{\bm{U}}\bm{U}^{n+1}
 +\tilde{\bm{\Omega}}\cdot
 \left(\frac{C_1}{\Delta t_n} \tensor{I}\bm{\Omega}^{n+1}+\bm{\Omega}^{n+1}\times( \tensor{I}\bm{\Omega}^{n+1})\right)\nonumber \\
 &=\int_{\Sigma}\rho\left(\frac{C_1}{\Delta t_n}\bm u^{n}(\bm x')-C_2\frac{\delta\bm u^n(\bm x')}{\delta t}\right)
 \cdot\tilde{\bm{u}}\,dv+ \left(\frac{C_1}{\Delta t_n}m\bm{U}^{n} 
 -C_2 m\frac{d\mb U^n}{dt}+\bm{G}\right)\tilde{\bm{U}}^n\nonumber \\
 &+ \tilde{\bm{\Omega}}\cdot
 \tensor{I}\left(\frac{C_1}{\Delta t_n}\bm{\Omega}^{n}-C_2 \frac{d\bm\Omega^n}{dt}\right),
\end{align}
and
\begin{eqnarray}
 \int_{\Sigma}\tilde{p}(\nabla\cdot \mb u^{n+1}(\bm x))\,dv=0.
\end{eqnarray}

The location of the grid in the new domain $\bm x$ and its correspondence to the old domain $\bm x'$ follows eqn.~\eqref{eqn_meshpos}. Since the nodes on the particle surface are also updated by  eqn.~\eqref{eqn_meshpos}, these node positions may move away from the body surface and hence we need to reset the surface nodes at each time step.
\begin{figure} 
\centering
\includegraphics[width=1.0\textwidth]{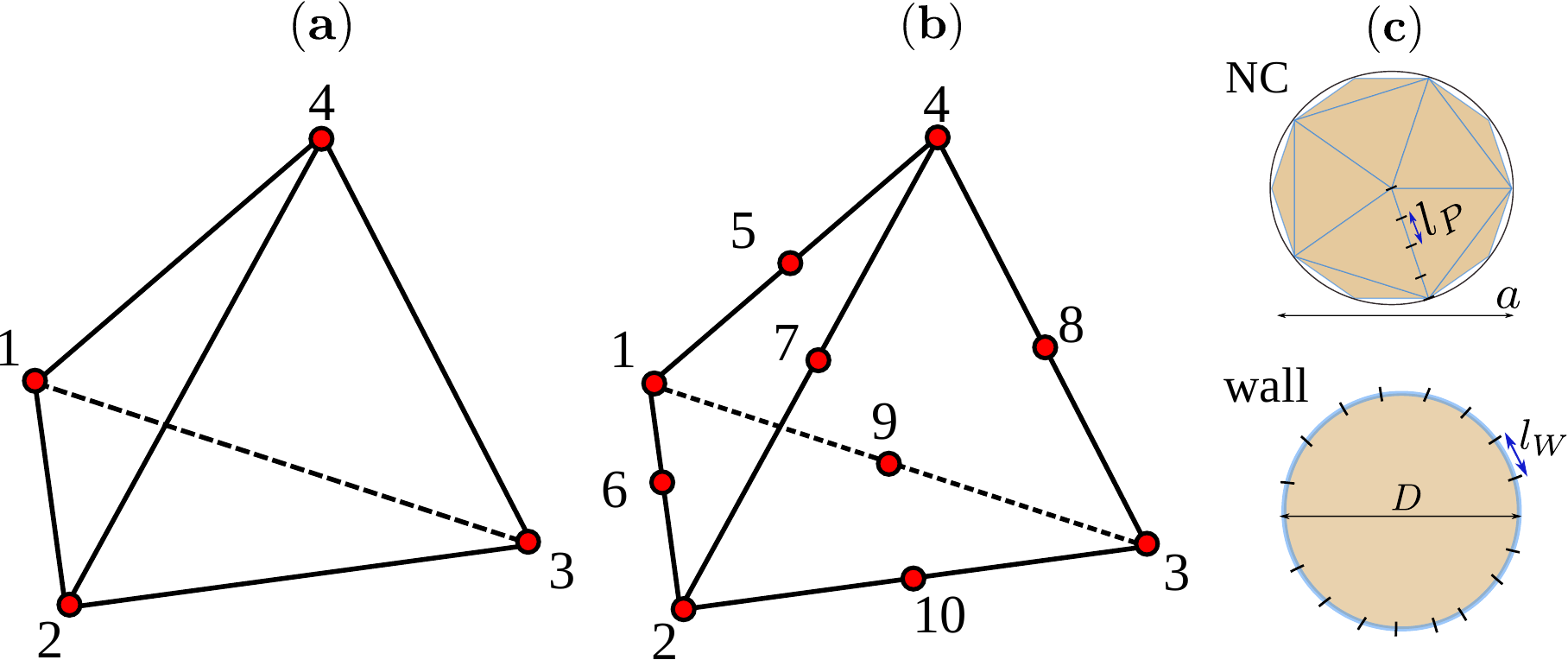}
\caption{The 4-node and 10-node tetrahedrons used in the finite element representation of the computational domain are shown in panels (a) and (b), respectively. The top panel in (c) shows an icosahedron used in the discretization of a spherical particle of diameter $a$ which is later mapped to an ellipsoid --- here $l_P$ denotes the mesh length on the particle surface. The lower panel in (c) shows the cross section of a cylindrical tube of diameter $D$. The mesh size on the particle surface is denoted by $l_W$.}
\label{fig_tetrahedron}
\end{figure}

\subsection{Finite element discretization}
\begin{enumerate}
\item \textit{Surface/boundary mesh:} The boundaries of the computational domain are discretized as described in ~\cite{Hu2001a}. Briefly, as shown in Fig.~\ref{fig_tetrahedron}(c), we start by approximating the surface of a unit sphere by an icosahedron, and further subdivide the faces of the icosahedron into a triangular mesh with a predefined characteristic length $l_P$. The triangular mesh on the icosahedron is stereographically projected to construct the boundary mesh for an ellipsoidal particle with specified values of $a$, $b$, $c$, and $\theta$. Similarly, the cylindrical wall boundary is discretized into a triangular mesh with a characteristic length $l_W$. In the following, we will describe the finite element mesh parameters used in our calculations in terms of $l_P$ and $l_W$.\\
 
\item \textit{Volume mesh:} The fluid domain is discretized by tetrahedral finite elements generated using Delaunay-Voronoi methods. The discrete solution for the fluid velocity is approximated by piecewise quadratic functions and is assumed to be continuous over the domain. We use  10 node tetrahedral elements (Fig.~\ref{fig_tetrahedron}(b)) to locally interpolate the velocity. On the other hand, the pressure and the stress are piecewise linear and continuous, and are interpolated using 4 node tetrahedral elements (Fig.~\ref{fig_tetrahedron}(a)). The 4 node and 10 node elements used to interpolate the stress and the velocity are known to satisfy the Ladyzhenskaya-Babuska-Brezzi conditions for stability~\citep{Hu2001a}. \\
\end{enumerate}

For a given finite element mesh, the combined fluid-solid weak formulation (eqn.~\eqref{eqn_main_weakfordis}) reduces to a nonlinear system of algebraic equations, which is solved by a Newton-Raphson algorithm. Similarly,
the mesh velocity (eqn.~\eqref{eqn_meshvel}) and mesh acceleration (eqn.~\eqref{eqn_meshacc}) can also be reduced to linear 
systems of algebraic equations. These coupled systems are solved by a multigrid preconditioned conjugate gradient method.

\subsection{Random stress tensor for the tetrahedral finite element mesh}
We now describe the procedure to numerically generate the random stresses associated with the unstructured tetrahedral mesh. The random stress at  each node on the computational domain depends on the volumes of the tetrahedrons associated with it.   

The components of the random stress tensor $\tensor{S}^{(i)}$ in the $i$-th tetrahedral element, with volume $V^{(i)}_e$, is approximated from eqn.~\eqref{eqn:fluct-dissip} as

\begin{equation}
 \langle S_{xx}\rangle^{(i)} = \langle S_{yy}\rangle^{(i)} = \langle S_{zz}\rangle^{(i)} = 0, 
\end{equation}

\begin{equation} 
 \langle S_{xy}\rangle ^{(i)} = \langle S_{yz}\rangle^{(i)} = \langle S_{zx}\rangle^{(i)} = 0, 
\end{equation}

\begin{equation} 
\label{eqn:stress_diag}
 \langle S^2_{xx}\rangle^{(i)} = \langle S^2_{yy}\rangle^{(i)} = \langle S^2_{zz}\rangle^{(i)} = \frac{4k_BT\mu}{V^{(i)}_e\Delta t}, 
\end{equation}

\begin{equation} 
\label{eqn:stress_offdiag}
 \langle S^2_{xy}\rangle^{(i)} = \langle S^2_{yz}\rangle^{(i)} = \langle S^2_{zx}\rangle^{(i)} = \frac{2k_BT\mu}{V^{(i)}_e\Delta t},
\end{equation}

\noindent where $\Delta t$ is the time step for the numerical simulation. The total stress on a node is then computed as:
\begin{eqnarray}
 \tensor{S}={\cal C}\sum^{N_e}_{i=1}  \tensor{S}^{(i)},
 \label{eqn:totalstress}
\end{eqnarray}
with ${\cal C}=1$ when the node is inside the computational domain and ${\cal C}=\sqrt{2}$ when the node is on a boundary surface. $N_e$ is the number of tetrahedrons associated with this node. At a boundary node, since we consider the ellipsoidal particles to be solid, the  tetrahedral volume $V_e^{(i)}$  underestimates the total volume defined by the Dirac delta function $\delta(\mb{x}-\mb{x}')$, given in the right hand side of eqn.~\eqref{eqn:fluct-dissip}. Ignoring the effect of the particle curvature on the estimate for $V_e^{(i)}$, we approximate the effective
 volume as $\delta(\mb{x}-\mb{x}')=2V_e^{(i)}$. Using this estimate in eqns.~\eqref{eqn:stress_diag} and \eqref{eqn:stress_offdiag} and summing over all tetrahedral elements linked to a given node leads to the general equation given in eqn.~\eqref{eqn:totalstress}.

\subsection{Deterministic calculations to compute velocity autocorrelations in a quiescent fluid}\label{sec:detmethod}
\par In this study, we also employ a computationally inexpensive calculation to study the velocity autocorrelation function (VACF) and angular velocity autocorrelation function (AVACF)  of a nano-ellipsoid. This method follows from the fluctuation-dissipation relation which states that the temporal correlation in the thermal stresses is equivalent to the correlation in the hydrodynamic memory of a stationary fluid~\citep{Kubo:1966dq}. Earlier works ~\citep{Pagonabarraga:1998ig,Iwashita:2008cj,Yu:2015kh} have shown that the averaged time correlation in the velocity of a Brownian particle, in a stationary medium, is equivalent to that for a driven particle computed in the absence of thermal fluctuations. This technique is called the deterministic method  \citep{Vitoshkin16}.

\par Since the inclusion of the stochastic stresses ($\tensor{S}\neq 0$) in the fluctuating hydrodynamics formulation leads to a large computational overhead, we may use the Deterministic method to investigate the long time behavior of the velocity autocorrelation of the nano-ellipsoid in a quiescent medium. The formulation and numerical techniques for the deterministic method are similar to that for fluctuating hydrodynamics except that:\\

\begin{enumerate}[(1) ]
\item the stochastic stress on each fluid element is taken to be $\tensor{S}=0$, \\

\item the initial value of the particle velocity (eqn.~\eqref{eqn:boundc}) is taken to be $\mb U(0)=\left({\bm U}_{x,0},{\bm U}_{y,0},{\bm U}_{z,0}\right)$ and $\mb\varOmega(0)=\left({\bm \varOmega}_{x,0},{\bm \varOmega}_{y,0},{\bm \varOmega}_{z,0}\right)$ with at least one of the components being non-zero. \\

\end{enumerate}

\noindent It should be noted that, though the deterministic method provides an inexpensive route to compute the long time correlations in the particle velocities, the trajectories obtained in these simulations are not reflective of that for a fluctuating particle.

\subsection{Added masses and moments of inertia as functions of the aspect ratio}
In our computational method, we have employed an incompressible fluid formulation. To account for the effect of incompressibility the effective mass and the moments of inertia must be included in our numerical evaluations, e.g., see \cite{Korotkin:2009ck,Uma2011}. In this section we present the modified expressions for the effective masses and moments of inertia~\citep{Korotkin:2009ck}, for an ellipsoidal particle at a prescribed orientation with respect to the bounding wall.

We define the direction dependent added masses and moments of inertia as:
\begin{eqnarray}
m^*_{\alpha} = (1+k_{\alpha})m,\\
\nonumber \\
I^*_{\alpha\alpha} = (1+K_{\alpha\alpha})I_{\alpha\alpha}.
\label{eqn:effmass}
\end{eqnarray}

Here $k_{\alpha}$ ($\alpha=x$, $y$, $z$) denotes the coefficient of the added mass along the $\alpha$ direction, and $K_{\alpha\alpha}$ ($\alpha=1$, $2$, $3$) denotes the coefficient of the added moment of inertia along the principal direction $\alpha$. The analytical forms of $m^*_{\alpha}$ and $I^{*}_{\alpha\alpha}$, for an ellipsoidal particle fully immersed in a fluid, depend on the aspect ratio of the particle and these expressions are as described below: 

\subsubsection{Oblate spheroids ($a<b=c$), with semi-minor axes along the tube axis}
For an oblate spheroid (with $a<b=c$) and oriented such that the semi minor axes is along the axial direction of the bounding wall, we define coefficients:

\begin{equation}
A_{\rm ob} = \dfrac{2q}{(1-q^2)^{3/2}} \left( \dfrac{\sqrt{1-q^2}}{q}- \sin^{-1}\left(\sqrt{1-q^2}\right) \right),
\label{eqn:Aob}
\end{equation}

\noindent and

\begin{equation}
B_{\rm ob} = C_{\rm ob} = \dfrac{q}{(1-q^2)^{3/2}} \left(\sin^{-1}\left(\sqrt{1-q^2}\right) - q \sqrt{1-q^2} \right),
\label{eqn:Bob}
\end{equation}

\noindent where $p=b/c$ and $q=a/c$. The added mass coefficient may then be expressed in terms of $A_{\rm ob}$, $B_{\rm ob}$ and $C_{\rm ob}$ as:

\begin{equation}
k_x = \dfrac{A_{\rm ob}}{B_{\rm ob}C_{\rm ob}}; \quad k_y = \dfrac{B_{\rm ob}}{A_{\rm ob}+C_{\rm ob}}; \quad k_z = \dfrac{C_{\rm ob}}{A_{\rm ob}+B_{\rm ob}},
\label{eqn:kob}
\end{equation}

\begin{equation}
K_{xx} = \dfrac{(p^2-1)^2}{p^2+1} \quad \dfrac{C_{\rm ob}-B_{\rm ob}}{2(p^2-1)+(B_{\rm ob}-C_{\rm ob})(p^2+1)},
\label{eqn:Kxxob}
\end{equation}

\begin{equation}
K_{yy} = \dfrac{(1-q^2)^2}{q^2+1} \quad \dfrac{A_{\rm ob}-C_{\rm ob}}{2(1-q^2)+(C_{\rm ob}-A_{\rm ob})(q^2+1)},
\label{eqn:Kyyob}
\end{equation}

\begin{equation}
K_{zz} = \dfrac{(q^2-p^2)^2}{q^2+p^2} \quad \dfrac{B_{\rm ob}-A_{\rm ob}}{2(q^2-p^2)+(A_{\rm ob}-B_{\rm ob})(p^2+q^2)}.
\label{eqn:Kzzob}
\end{equation}	

\begin{figure}
	\centering
	\includegraphics[width=10cm]{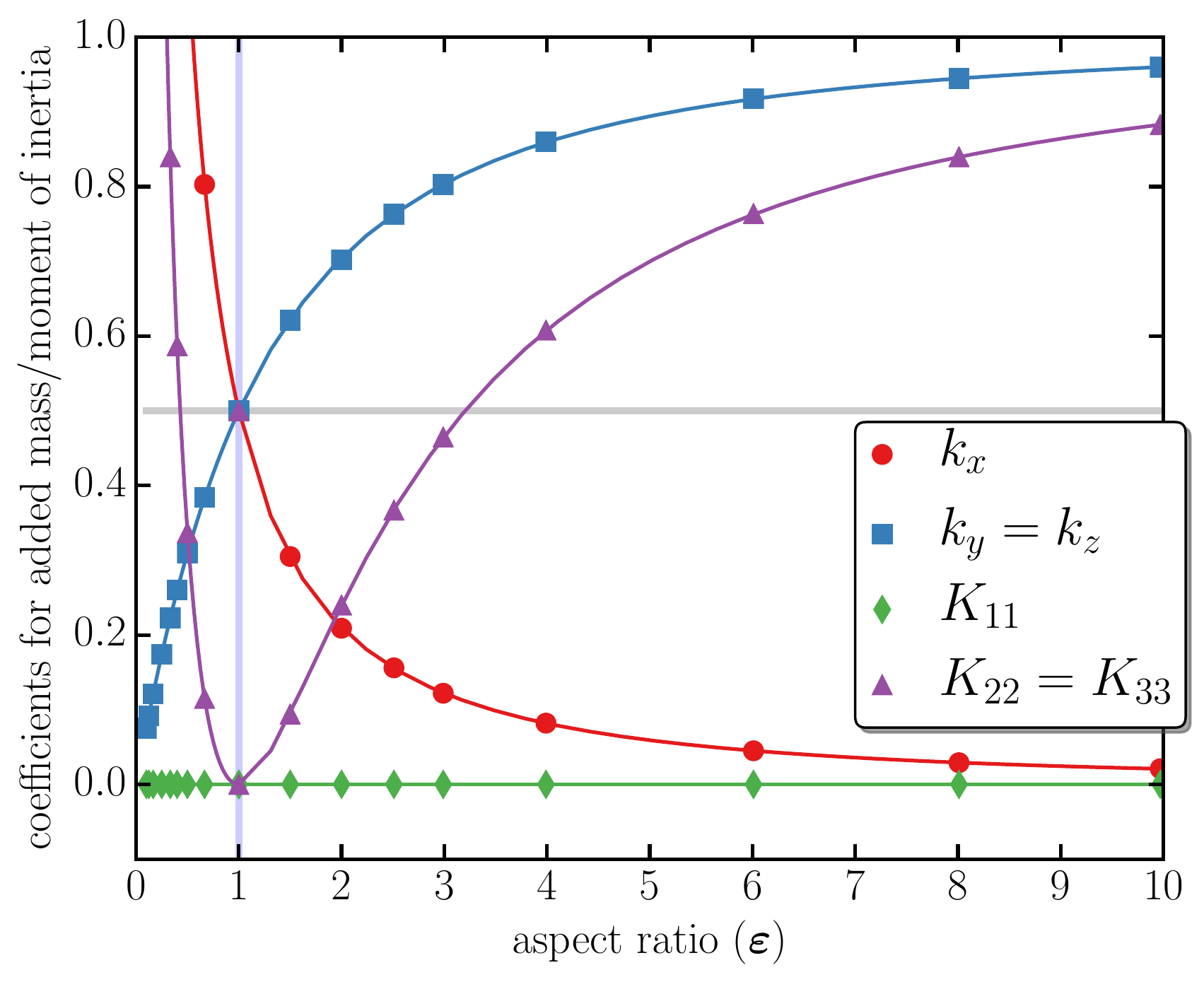}
	\caption{Coefficients of the added masses ($k_x$, $k_y$ and $k_z$) and moments of inertia ($K_{11}$, $K_{22}$ and $K_{33}$) as functions of \asp, the aspect ratio of the particle. The horizontal line represents the added mass component for a spherical particle ($\aspp=1$), for which $k_x=k_y=k_z=0.5$.}

	\label{fig:addedmass}
\end{figure}

\subsubsection{Spherical particles, $a/c=1.0$}
For a spherical particle we use the well known results for the added coefficients 
\begin{equation}
k_x =  k_y =  k_z = \dfrac{1}{2},
\label{eqn:ksp}
\end{equation}

\noindent and
\begin{equation}
K_{xx} =  K_{yy} =  K_{zz} = 0.
\label{eqn:Ksp}
\end{equation}

\subsubsection{Prolate Spheroids ($a>b=c$), with semi-major axes along the tube axis}
For a prolate ellipsoid ($a>b=c$) with semi-major axes along the axial direction of the tube, we define the coefficients:

\begin{equation}
A_{\rm pr} = \dfrac{2v}{(v^2-1)^{3/2}} \left( \log(\sqrt{v^2-1}+v)- \dfrac{\sqrt{v^2-1}}{v}\right),
\label{eqn:Apr}
\end{equation}

\begin{equation}
B_{\rm pr} = C_{\rm pr} = \dfrac{v^2}{v^2-1} \left( 1- \dfrac{\sqrt{v^2-1}}{v} \log(\sqrt{v^2-1}+v) \right),
\label{eqn:Bpr}
\end{equation}

\noindent where $v=a/b$. The coefficients of the added masses and moments of inertia may then be expressed in terms of $A_{\rm pr}$, $B_{\rm pr}$ and $C_{\rm pr}$ as:

\begin{equation}
k_x = \dfrac{A_{\rm pr}}{2-A_{\rm pr}}; \quad k_y = \dfrac{B_{\rm pr}}{2-B_{\rm pr}}; \quad k_z = \dfrac{C_{\rm pr}}{2-C_{\rm pr}},
\label{eqn:kxpr}
\end{equation}

\begin{equation}
K_{xx} = \dfrac{(b^2-c^2)^2}{b^2+c^2} \quad \dfrac{C_{\rm pr}-B_{\rm pr}}{2(b^2-c^2)+(B_{\rm pr}-C_{\rm pr})(b^2+c^2)},
\label{eqn:Kxxpr}
\end{equation}

\begin{equation}
K_{yy} = \dfrac{(a^2-c^2)^2}{a^2+c^2} \quad \dfrac{A_{\rm pr}-C_{\rm pr}}{2(c^2-a^2)+(C_{\rm pr}-A_{\rm pr})(a^2+c^2)},
\label{eqn:Kyypr}
\end{equation}

\begin{equation}
K_{zz} = \dfrac{(a^2-b^2)^2}{a^2+b^2} \quad \dfrac{B_{\rm pr}-A_{\rm pr}}{2(a^2-b^2)+(A_{\rm pr}-B_{\rm pr})(b^2+a^2)}.
\label{eqn:Kzzpr}
\end{equation}

\noindent The aspect ratio and direction dependent added inertial coefficients given by eqns.~\eqref{eqn:Aob}-\eqref{eqn:Kzzpr} are  shown in Fig.~\ref{fig:addedmass}. 

\section{Numerical results and discussion}\label{sec:resuls}
We study the motion of the particle in (i) a quiescent fluid medium, and (ii) in a fully developed Poiseuille flow at the entrance. For particle motion in the presence of a Poiseuille flow, we initially fix the particle at the desired location and subsequently release it only when the flow is fully developed ~\citep{Uma2011}. For the results reported in this study, we take tube diameter $D$ to be in the range  $5-50\mum$ and tube length $L=40\mum$ throughout.  The dynamic viscosity and density of the fluid are taken to be $\mu=10^{-3}$ kg m$^{-1}$s$^{-1}$, and $\rho^{(f)}=10^3$ kg m$^{-3}$, respectively.
%

The presence of stochastic stresses (${\tensor S} \neq 0$) continuously alters the degrees of freedom in the fluid and as a result the ellipsoidal particle is subject to a net force which has contributions from both the hydrodynamic and the stochastic stresses. First, we consider a stationary medium. In the absence of an external flow, the motion of the nanoparticle is solely Brownian.

The fluid temperature is set at $T=310$ K and the thermal energy is given by $\kbt{}$, with the Boltzmann constant $k_{\rm B}=1.3806503\times 10^{-23}$ kg m$^2$/(s$^2$ K). We first consider a neutrally buoyant ellipsoidal NC of aspect ratio \asp=1.5 (with $a=600$ nm, and $b=c=400$ nm and $\theta=0\degree$) initially placed at the center of a fluid filled cylinder with $D=5\mum$ and $L=40\mum$.  The characteristic length of the particle  is take to be the equivalent spherical diameter $d_{\rm eq}=\sqrt[3]{abc}=457.9$ nm and this sets a representative hydrodynamic time scale $t_\nu = (d_{\rm eq}/2)^2/\mu=5.24\times 10^{-8}$ s, and this will be employed for the scaling throughout the treatment. We have also examined the motion of this  particle in the presence of a Poiseuille flow along the $x$ direction. In this case,  flows with maximum inlet velocities in the range ${\bm u}_{\rm max}=10^{-1}$ to $10^{5}$ $\mum$/s, corresponding to particle Reynolds numbers  ($\Rep{}=\rhof{} d_{\rm eq} {\bm u}_{\rm max}/\mu$) in the range  $5\times 10^{-8}$ to $5\times 10^{-2}$, have been investigated.

\subsection{Thermal equilibration of the ellipsoidal NC}
Since this study is a numerical evaluation of a stochastic differential equation formulation, it is very important to set formally correct procedures in place before embarking on the full evaluation. Rigorous requirements in this context consist of guaranteeing thermal equilibration of the NC with the bulk medium and the satisfaction of the Maxwell-Boltzmann distribution for the components of the particle velocity.

As stated earlier, the preset bulk fluid temperature is $T=310$ K. We note that the fluid-particle system is a self thermostat that maintains the equilibrium temperature through the fluctuation-dissipation relation.
Using the equipartition theorem, we numerically estimate the translational and rotational temperatures, denoted by $T^{(t)}$ and $T^{(r)}$ respectively, as:

\begin{eqnarray}\label{eqn_transTemp}
 T^{\rm(t)}=\dfrac{T^{\rm(t)}_x+T^{\rm(t)}_y+T^{\rm(t)}_z}{3} = \frac{1}{3k_{\rm B}}\sum\limits_{\alpha=x,y,z}m^{*}_\alpha \langle \mb U_\alpha^2 \rangle,
\end{eqnarray}
and
\begin{eqnarray}\label{eqn_rotTemp}
 T^{\rm(r)}=\dfrac{T^{\rm(r)}_1+T^{\rm(r)}_2+T^{\rm(r)}_3}{3} = \frac{1}{3k_{\rm B}}\sum\limits_{\alpha=1,2,3}I^{*}_{\alpha\alpha} \langle \bm\Omega_\alpha^2 \rangle.
\end{eqnarray}
In estimating these temperatures, we explicitly account for the effective masses $m^*_{\alpha}$ and moments of inertia $I^*_{\alpha\alpha}$, whose exact forms are given in eqn.~\eqref{eqn:effmass}. 

\begin{figure}
\centering	\includegraphics[width=\textwidth]{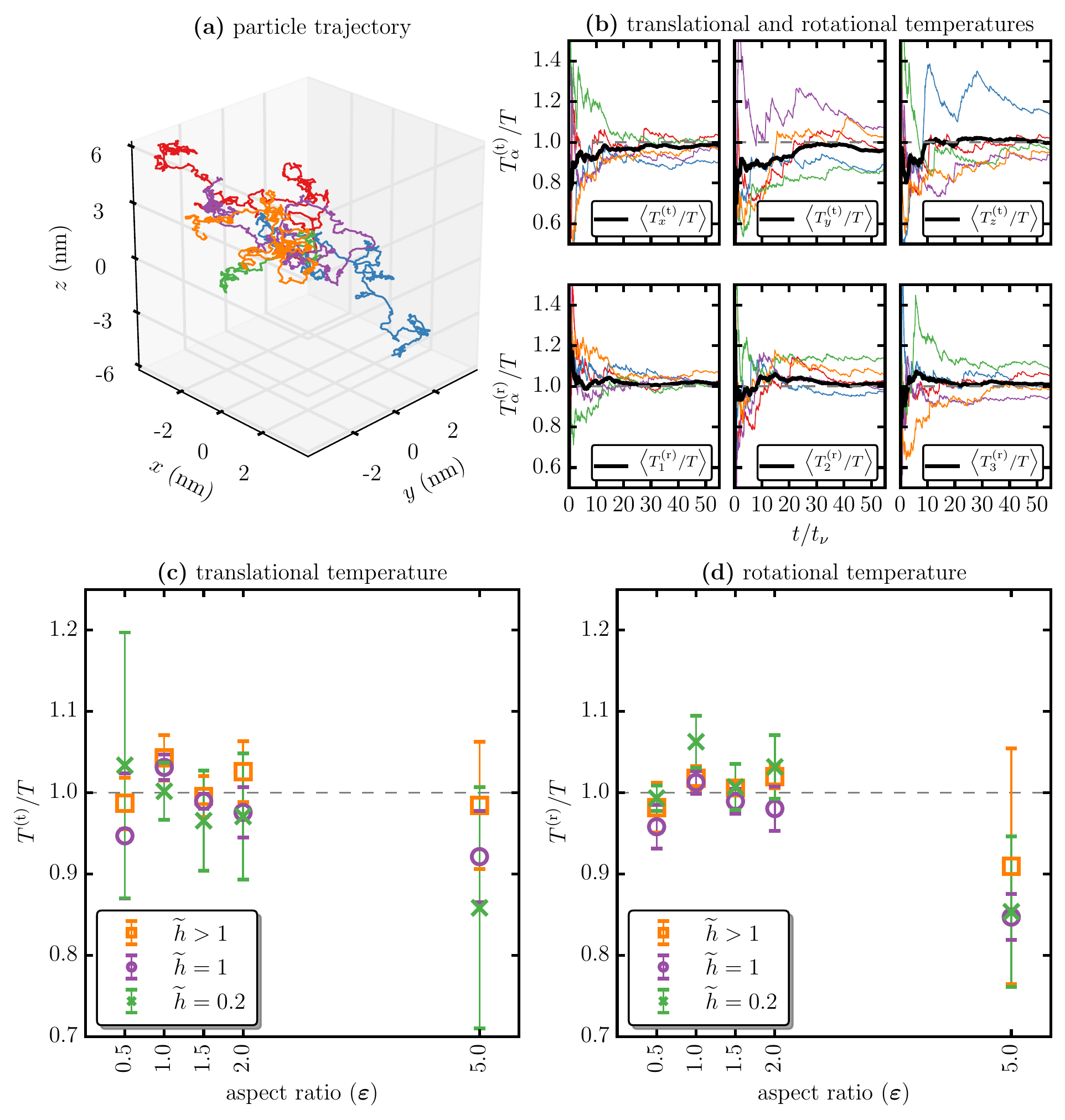}	\caption{(a) Five independent trajectories of an ellipsoidal particle (with $\aspp=1.5$ and $a=600$ nm) immersed in a fluid with fluctuating stresses. (b) Time evolution of the scaled translational and rotational temperatures, along the $x$, $y$ and $z$ directions, for all the five trajectories, along with their ensemble averaged value. (c, d) Translational and rotational temperatures, averaged over all directions, as a function of \asp, the aspect ratio of the ellipsoidal nanocarrier, for three different positions with respect to the bounding wall. Mesh parameters used are $l_P=6$ nm and $l_W=785$ nm. The exact values for $\widetilde{h}>1$ for different $\epsilon$ may be found in Table.~\ref{table:thetazero}.}
	\label{fig:tempevlo}
\end{figure}

Panel (a) in Fig.~\ref{fig:tempevlo} shows five independent trajectories of an NC, initially at the same starting location. The computations were carried out over a period of 3 $\mu$s using a timestep of $\Delta t=10^{-10}$ s. These trajectories demonstrate the Brownian characteristic of the particle. For each of these trajectories, we compute $T^{\rm(t)}_{\alpha}$  and $T^{\rm (r)}_{\alpha}$, the translational and rotational temperatures, respectively, along each principal direction $\alpha$. The time evolution of $T^{\rm(t)}_{\alpha}$  and $T^{\rm (r)}_{\alpha}$ are shown in Fig.~\ref{fig:tempevlo}(b). Both the translational and rotational temperatures of the particle transition to a steady state at very short times ($\sim 500 \Delta t \simeq t_{\nu}$) following their introduction into the fluid. While the temperatures of the individual trajectories fluctuate by as much as $\pm 15\%$, the time averaged temperatures, also shown alongside in each of the panels in Fig.~\ref{fig:tempevlo}(b), show thermal equilibration with the preset bulk temperature.

Next, we investigate the effect of various ellipsoidal NC aspect ratios on thermal equilibration. We consider  five different aspect ratios, \asp=0.5, 1.0, 1.5, 2.0, and 5.0 for which the translational and rotational temperatures are displayed in panels (c) and (d) of Fig.~\ref{fig:tempevlo} --- the complete set of data can be found in Figs. S1.1-S1.15 in the \sis. The NC here all have an equivalent volume of 0.0502 $\mum^3$, which corresponds to an effective spherical diameter of $d_{\rm eq}=457.9$ nm, as before. In this context, we have also studied the effect of confinement on the thermodynamic behavior of the NCs by computing their equilibrations for three wall separation distances, chosen such that the NC is in (i) the bulk regime ($\widetilde{h}>1$), (ii) the near wall regime ($\widetilde{h}=1$), and (iii) the lubrication regime ($\widetilde{h}=0.2$). We note that  $\widetilde{h}$ is a function of $\aspp$ and particle orientation (see Appendix \ref{app:a0derivation}). The computed values of  $T^{\rm (t)}$ and $T^{\rm (r)}$ are averaged over 10 independent 3 $\mu$s trajectories, and these are shown in panels (c) and (d) in Fig.~\ref{fig:tempevlo}, respectively. The evaluated particle temperatures match with that of the bulk fluid within $\pm 15\%$, independent of the  aspect ratios and   confinement effects. The larger deviations seen for $\aspp=5.0$ may further be improved by refining the computational mesh, which we discuss next.

\begin{figure} 
\centering
\includegraphics[width=1.0\textwidth,clip]{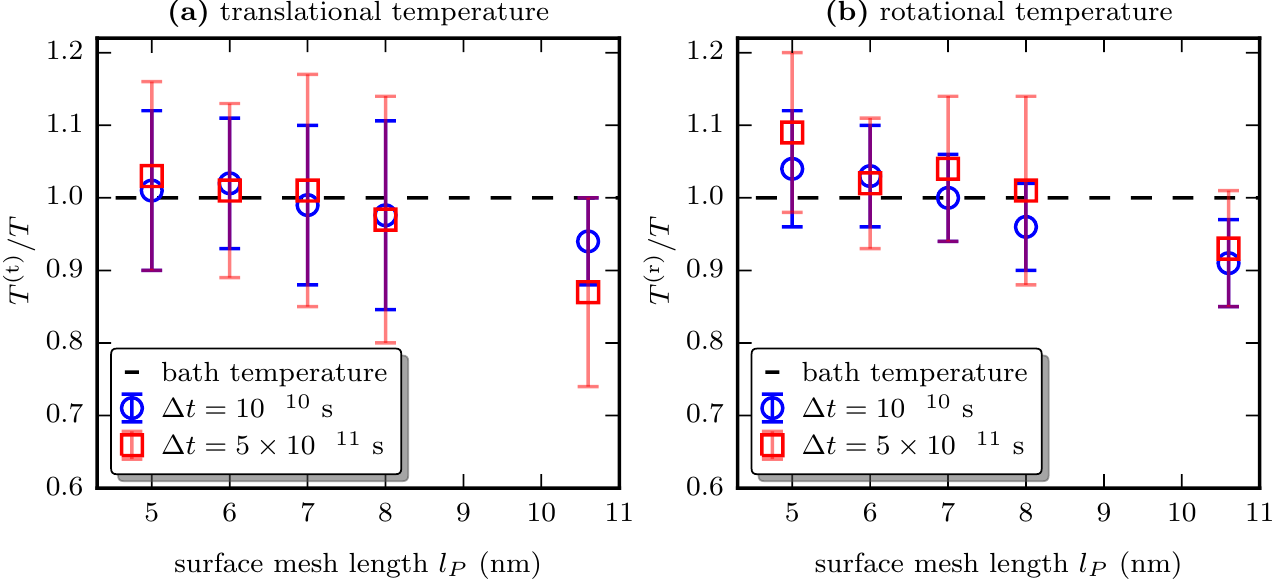}
\caption{Translational and rotational temperatures of the NC as a function of the surface mesh length for two values of the computational timestep: $\Delta t=10^{-10}$ s and $\Delta t=5\times 10{-11}$ s.}
\label{fig_meshtempref}
\end{figure}

The computed values of the equilibrium temperature depend both on the resolution of the computational mesh (i.e., on $l_{P}$ and $l_{W}$ in Fig.~\ref{fig_tetrahedron}(c)) and the timestep $\Delta t$. This is illustrated in Fig.~\ref{fig_meshtempref}, for an NC with $\varepsilon=1.5$, $a=600$ nm, and $\theta=0\degree$, where we show $T^{\rm (t)}$ and $T^{\rm (r)}$ as functions of $l_{P}$ computed for two time steps  $\Delta t=10^{-10}$ s and $\Delta t=5\times 10^{-11}$ s. The error bars correspond to the standard deviation in the temperatures computed from 10 independent ensembles and the maximum error in the predictions is found to be around $15\%$. For the condition of the study, for $l_{P}\sim 8$ nm the  estimates of the equilibrated temperatures  are almost the same as the bath temperature $T$ confirming equilibration. Beyond $l_{P}=8$ nm, equilibrium is not attained because of the inability to determine the stress and the corresponding velocity fields sufficiently accurately. It is important therefore to correctly estimate the mesh length that yields equilibration for the prevailing conditions. These studies establish the criteria for mesh convergence in stochastic hydrodynamic computations.

\begin{figure} 
\centering
\includegraphics[width=1.0\textwidth]{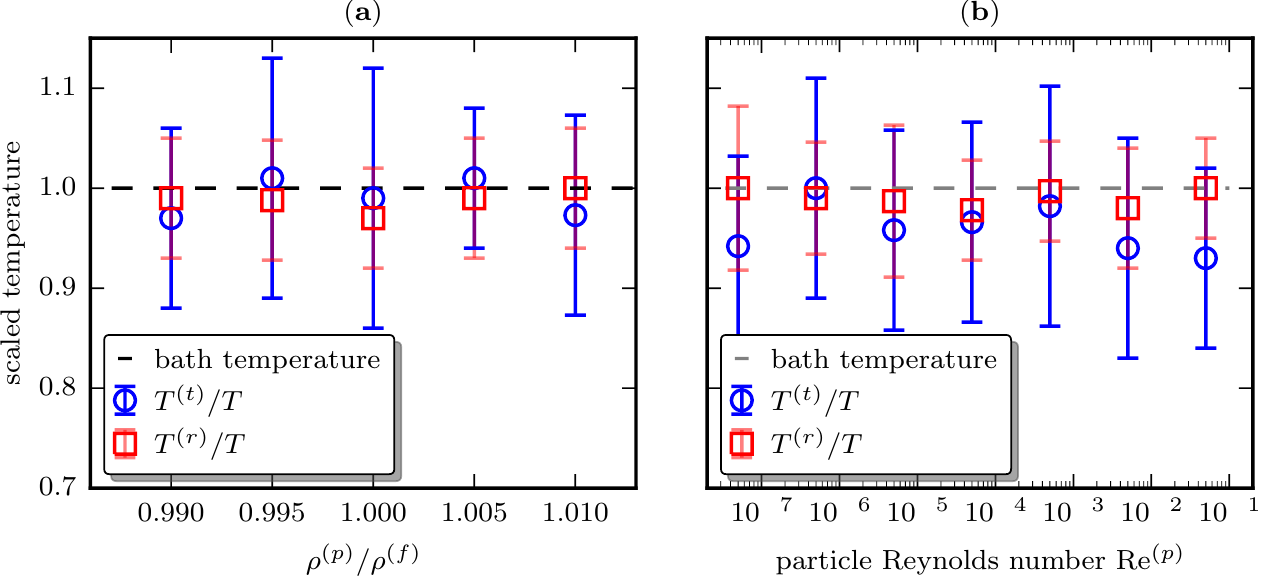}
\caption{Translational and rotational temperatures of the ellipsoidal nanocarrier (a) as a function of the nanocarrier density normalized with fluid density in a stationary fluid medium, and (b) as a function of nanocarrier Reynolds number in Poiseuille flow with $\rho^{(p)}/\rho^{(f)}=1$. Mesh parameters used are $l_P=6$ nm and $l_W=785$ nm. }
\label{fig_tempDenRey}
\end{figure}

Fig.~\ref{fig_tempDenRey}(a) shows $T^{\rm (t)}$ and $T^{\rm (r)}$ for a nearly neutrally buoyant ellipsoidal particle, of similar dimensions as before, for five different particle densities, that are chosen to be in the range $990 \leq \rho^{(p)} \leq 1010$ kg/m$^3$, in thermal equilibrium with a quiescent fluid. We find for the range of densities investigated that thermal equilibration is attained in a manner similar to that described earlier.

We next study how the presence of external flow impacts the stochastic motion of the NC by introducing it at the center of a tube with a well developed incoming Poiseuille flow. We investigate this phenomenon for seven different flow rates with the NC Reynolds numbers in the range $\Rep = 5\times 10^{-8}-5\times 10^{-2}$. Fig.~\ref{fig_tempDenRey}(b) shows the translational and rotational temperatures  as a function of $\Rep{}$ for a neutrally buoyant ellipsoidal NC with a normalized surface mesh length of $l_{P}=5$ nm and $l_W=785$ nm. Our results show that equilibration is attained by the NC in a manner similar to the above even  in the presence of weak Poiseuille flows.

Having shown that the ellipsoidal NC satisfies  the principle of equipartition of translational and rotational energies, we next investigate the behavior of the velocity components. 

To be self consistent, in this section we show that the translational and rotational velocities of the NC in the fluctuating fluid satisfy:

\begin{equation}
P({\bm U}_\alpha)d{\bm U}_\alpha=\dfrac{1}{\sqrt{2\pi}} \exp\left(-\dfrac{{\bm U}^2_\alpha}{2\sigma_{\rm (t),\alpha}^2}\right),
\label{eqn_scaled_tveldist}
\end{equation}
 and 
\begin{equation}
P({\bm{\Omega}}_\alpha)d{\bm{\Omega}}_\alpha=\dfrac{1}{\sqrt{2\pi}} \exp\left(-\dfrac{{\bm{\Omega}}_\alpha^2}{2\sigma_{\rm (r),\alpha}^2}\right),
\label{eqn_scaled_rveldist}
\end{equation}
respectively. Here $\sigma_{\rm{(t)},\alpha}^2=\kbt{}/m^{*}_\alpha$ is the variance in the $\alpha$ component of the particle translational velocity, with $\alpha=x,y,z$ and $\sigma_{\rm{(r)},\alpha}^2=\kbt{}/I^*_{\alpha\alpha}$, with $\alpha=1,2,3$, is the variance in the rotational velocities. 

\begin{figure}
\centering
\includegraphics[width=1.0\textwidth]{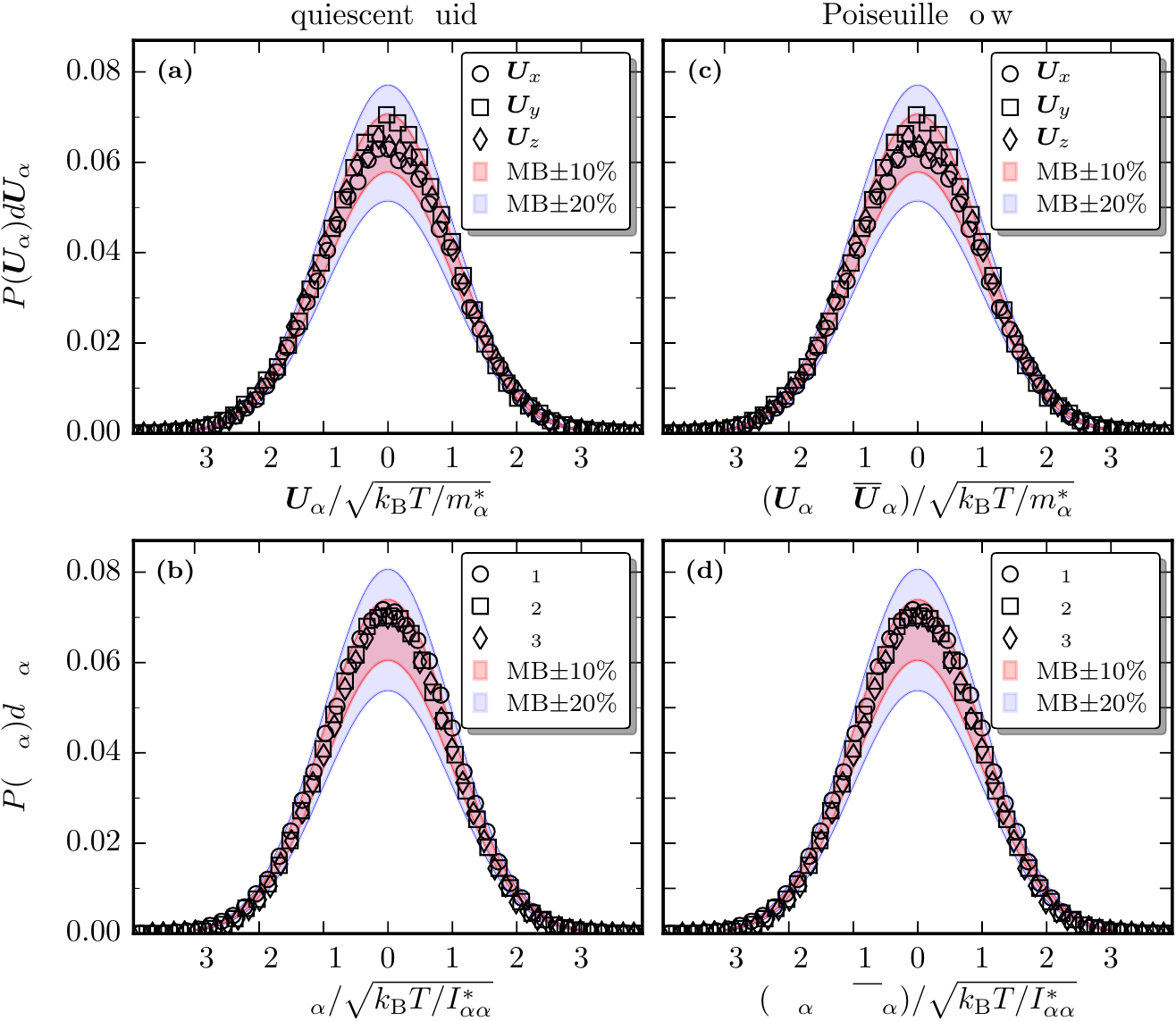}
\caption{Equilibrium probability of the translational and rotational velocities of an ellipsoidal nanoparticle in a quiescent fluid (panels (a) and (b)), and  in a Poiseuille flow (panels (c) and (d)). The two shaded regions in each of the panels represent deviations of $\pm 10 \%$ and $\pm 20 \%$  from the Maxwell-Boltzmann (MB) distribution. Data shown for an NC with $\aspp=1.5$, $a=600$ nm, and $\theta=0\degree$ placed at the center of a cylindrical tube of diameter $D=5\mum$. Mesh parameters used are $l_P=7$ nm and $l_W=785$ nm.}
\label{fig_vHist}
\end{figure}

The corresponding probability distributions for an ellipsoidal particle, with $\aspp=1.5$, $a=600$ nm and $\theta=0\degree$, placed at the center of the tube, is shown in Figs.~\ref{fig_vHist} (a) and (b). $P({\bm U}_\alpha) d{\bm U}_\alpha$ shows a normal distribution, for all values of $\alpha=x$ ,$y$, and $z$. Furthermore, the computed probabilities deviate at most by 10\% from the normal distribution as is shown by  the shaded region that represents a $\pm 10\%$ deviation. We also studied the velocity distribution for an ellipsoidal NC placed at the center of a tube with a steady Poiseuille flow with $\bm{u}_{\rm max}=100$$\mum$/s. $P({\bm U}_\alpha) d{\bm U}_\alpha$ in the presence of flow is shown in Fig.~\ref{fig_vHist} (c) and (d), and both the translational and rotational velocities show a normal distribution consistent with Maxwell-Boltzmann statistics.

\subsection{Translational velocity autocorrelation (VACF) and rotational velocity autocorrelation (AVACF) of the  ellipsoidal NC}
We define the VACF of the NC velocities as
\begin{equation}
\vacf = \frac{\left\langle \bm{U}_{\alpha}(0)\bm{U}_{\alpha}(t)  \right \rangle}{(\kbt/m^{*}_\alpha)} \quad \forall \quad \alpha=x,y,z
\end{equation}
\noindent and the AVACF as
\begin{equation}
\avacf = \frac{\left\langle \bm{\Omega}_{\alpha}(0)\bm{\Omega}_{\alpha}(t)  \right \rangle}{(\kbt/I^{*}_{\alpha\alpha})} \quad \forall \quad \alpha=1,2,3.
\end{equation}

We compute the VACF and AVACF of a Brownian NC using two methods: (i) direct calculations using the Fluctuating Hydrodynamics approach and (ii) from the relaxation of the velocity using the Deterministic method. It has been previously shown that for a sufficiently small initial velocity the decay of the NC velocity is identical to the velocity autocorrelation function, and in scaled units, we express the VACF and AVACF from the deterministic method as:
\begin{equation}
\vacf = \frac{\bm{U}_{\alpha}(t)}{\bm{U}_{\alpha}(0)} \quad \forall \quad \alpha=x,y,z,
\end{equation}
and
\begin{equation}
\avacf = \frac{\bm{\varOmega}_{\alpha}(t)}{\bm{\varOmega}_{\alpha}(0)} \quad \forall \quad \alpha=1,2,3.
\end{equation}

The time correlation in the translational and rotational velocities of a Brownian particle, in the absence of hydrodynamic interactions decays as:
\begin{equation}
\vacf = \exp\left(- \frac{t}{{\cal M}_{\alpha}^{\rm (t)}m_\alpha^{*}}\right),
\label{eqn:tltnu_trans}
\end{equation}
\noindent and
\begin{equation}
\avacf = \exp\left(- \frac{t}{{\cal M}_{\alpha}^{\rm (r)} I_{\alpha\alpha}^{*}} \right),
\label{eqn:tltnu_rot}
\end{equation}
respectively.  Here, we denote the translational and rotational mobilities of the particle, along the $\alpha$ direction, as  ${\cal M}_{\alpha}^{\rm (t)}$ and ${\cal M}_{\alpha}^{\rm (r)}$, respectively. \footnote{The mobilities are  estimated from the towing method described in Sec. S3 in the \sis. In all of our results presented for the VACF we use mobilities computed using this method method to show the duration of the first exponential decay regime.}  However, when the hydrodynamic forces are explicitly taken into account, the exponential decay only holds for times smaller than the viscous relaxation time $t_\nu$ (i.e., for $t \le t_\nu$). The long time behavior ($t>t_\nu$) of the VACF and AVACF for an centrally symmetric ellipsoidal particle, immersed in a bulk fluid, follows an algebraic decay~\citep{Hocquart:1983bh,Cichocki:1995df,Lowe:1995ko,Cichocki:1996ft,Masters:1996dz,Cichocki:1997jk,Masters:1997ic} given by
\begin{equation}
\vacf = \frac{1}{6\sqrt{\pi}}\left(\frac{t}{t_\nu}\right)^{-3/2} \quad \textrm{for translational velocities}\,(\alpha=x,y,z),
\label{eqn:tgtnu-trans}
\end{equation}
\noindent and
\begin{equation}
\avacf =\frac{{\bm \varPsi}^{\rm A}_{\alpha\alpha}}{60\sqrt{\pi}}\left(\frac{t}{t_\nu}\right)^{-5/2} \quad \textrm{for rotational
 velocities}\,(\alpha=1,2,3).
\label{eqn:tgtnu-rot}
\end{equation}

\begin{figure} 
	\centering
	\includegraphics[width=1.0\textwidth]{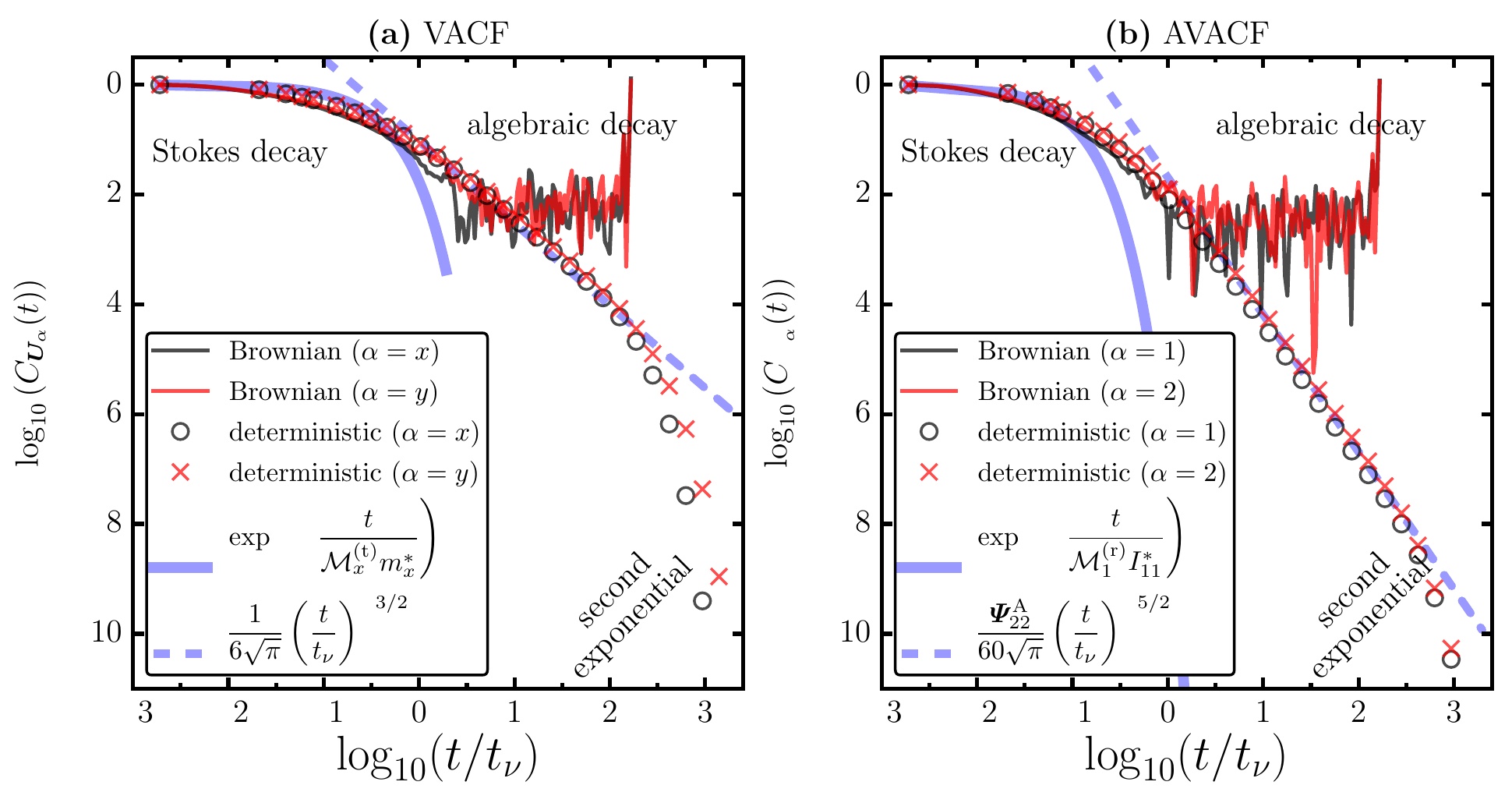}
	\caption{ (a) \vacff, the VACF for $\alpha=x,y$ and (b) \avacff, the AVACF for $\alpha=1,2$, for an ellipsoidal NC ($\varepsilon=1.5$ and $a=600$ nm) placed at the center of a cylindrical tube, with $\theta=0\degree$ and $\widetilde{h}=11.5$. In both the panels, the solid lines correspond to data obtained from stochastic simulations and the symbols denote those obtained using the Deterministic method. The correlations in the particle velocity show a Stokes exponential decay for $t<t_{\nu}$ and an algebraic decay for $t>t_{\nu}$.  }
	\label{fig_vacfcomp}
\end{figure}

\begin{figure}
	\centering
	\includegraphics[width=1.0\textwidth,clip]{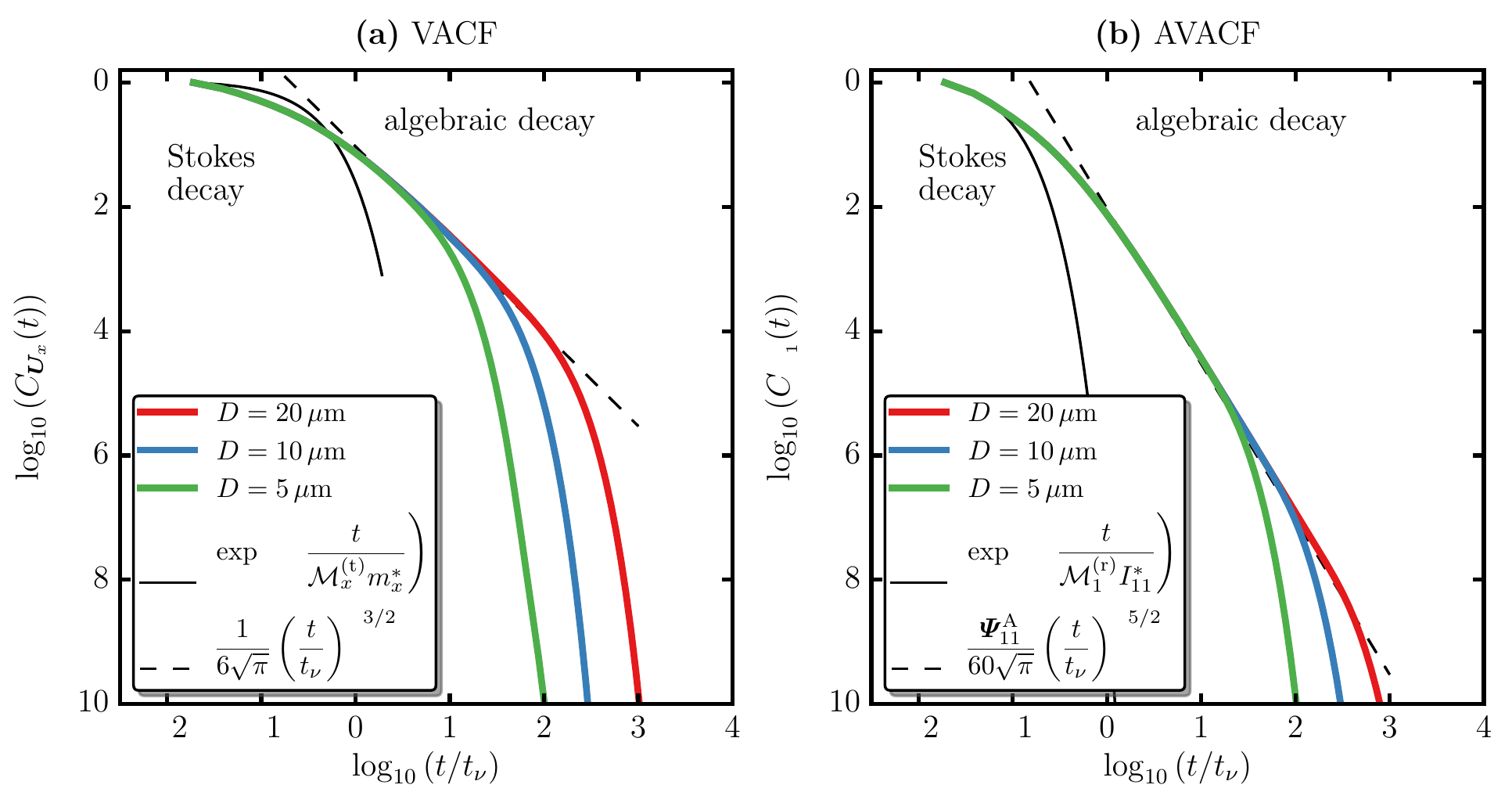}
	\caption{\label{fig:vacf-wall-effect} Effect of wall confinement on the VACF (along the $x$ direction) and AVACF (along the 1 direction) computed using the deterministic method. Data shown for an NC with $\aspp=1.5$, $a=600$ nm and $\theta=0\degree$ placed at the center of a cylindrical tube with $D=5,\,10$, and $20 \mum$. The solid and dotted lines denote the Stokes decay and algebraic decay regimes respectively.
	}
\end{figure}

\begin{figure}
	\centering
	\includegraphics[width=1.0\textwidth,clip]{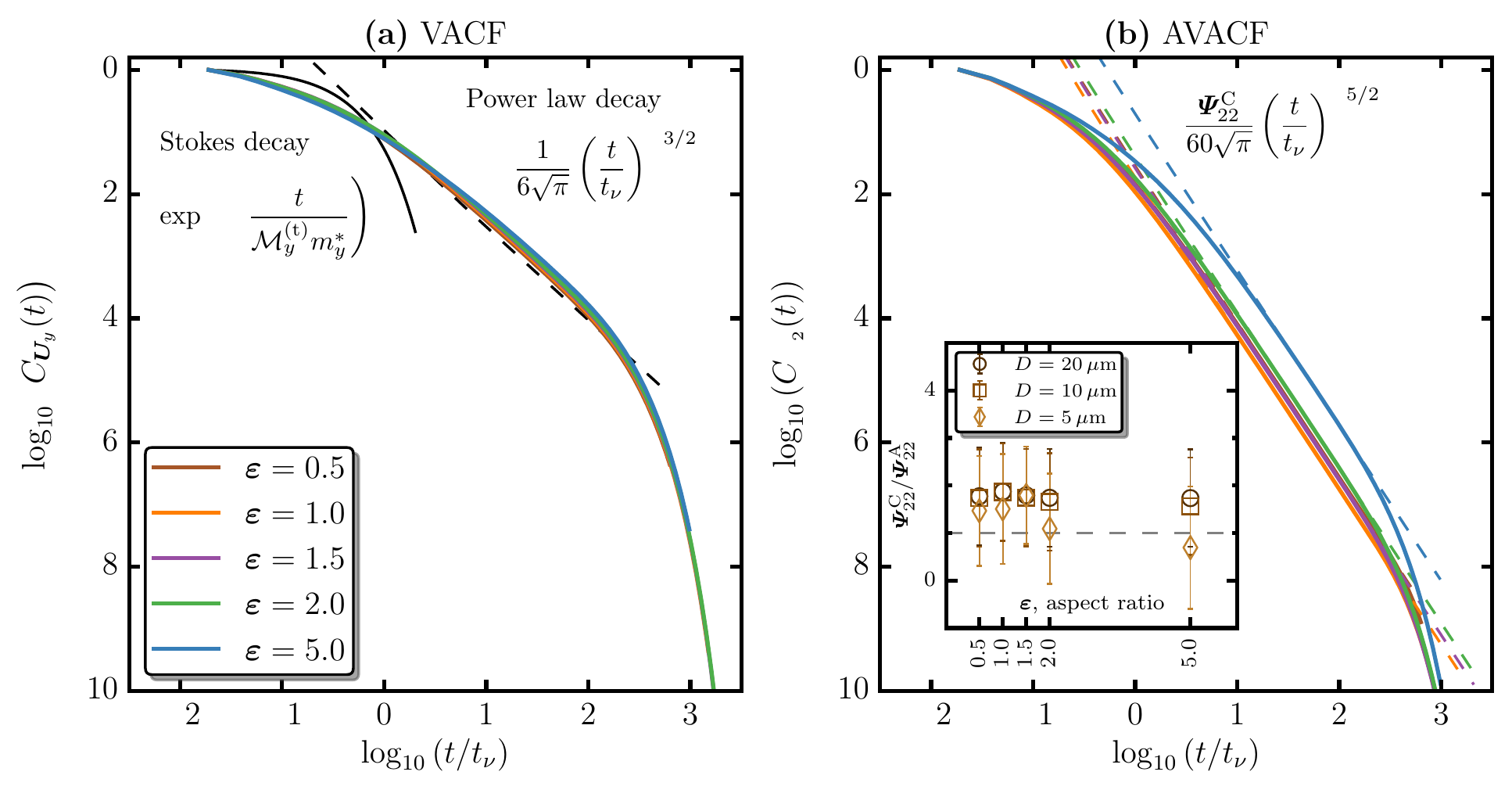}
	\caption{\label{fig:Hinch-res} Effects of aspect ratios on the VACF and AVACF along the $y$ and $2$ directions, respectively. The main plot shows data for NCs confined by a tube with $D=20\mum$. The dotted lines in panel (b) are the best fit curves for the algebraic decay regime which in turn is used to calculate ${\bm \Psi}_{22}^{\rm C}(\aspp)$. The inset to panel (b) shows a comparison of ${\bm \Psi}_{22}^{\rm C}$ to their analytical estimates, as a function of \asp.}
\end{figure}
\noindent Here  
\begin{equation}
{\bm \varPsi}^{\rm A}_{\alpha\alpha} = \dfrac{I_{\alpha\alpha}^*}{I_{\rm sph}} \left(1+\dfrac{3}{5}\left(\dfrac{\epsilon^2_{\beta\gamma}-1}{\epsilon^2_{\beta\gamma}+1}\right)^2 \right).
\label{eqn:Hinch-pfac}
\end{equation}
$I_{\alpha\alpha}^*$ is the added moment of inertia along the $\alpha$ direction and $I_{\rm sph}$ is the moment of inertia of the equivalent sphere. The indices $\alpha,\beta,\gamma$ form a cyclic pair, such that $\beta=2$ and $\gamma=3$ when $\alpha=1$, and $\epsilon_{\beta\gamma}$ is the ratio of the particle dimensions along the $\beta$ and $\gamma$ directions. For example, for a prolate ellipsoid (with \asp=1.5) $\epsilon_{23}=1$, $\epsilon_{31}=2/3$, and $\epsilon_{12}=3/2$. 

In Fig.~\ref{fig_vacfcomp} we show the VACF and AVACF for an ellipsoidal NC, with $\aspp=1.5$, $a=600$ nm and $\theta=0\degree$, placed at the center of the tube with $D=5\mum$ and $L=40\mum$, for which $\widetilde{h}=11.5$. In both panels, the solid lines correspond to data obtained from stochastic simulations and the symbols denote those obtained from the deterministic method. All data for the stochastic simulations have been averaged over 10 independent 3 $\mu$s Brownian trajectories. 

The VACF and AVACF shown in Fig.~\ref{fig_vacfcomp}(a) show an initial Stokes exponential decay for $t<t_{\nu}$ followed by an algebraic decay for $t>t_{\nu}$ culminating in a second exponential decay for large times. For $t<t_{\nu}$, the estimates from both the stochastic and deterministic methods agree very well with each other. The observed exponential decay agrees favourably with that predicted for this regime by eqns. ~\eqref{eqn:tltnu_trans} and \eqref{eqn:tltnu_rot}. When $t>t_\nu$,  the VACF and AVACF from the deterministic method shows a crossover to a power law behavior. The behavior for the VACF scales as $(t/t_\nu)^{-3/2}/(6\sqrt{\pi})$ for all the coordinate directions ($x$, $y$ and $z$). For the AVACF, the scaling law is ${\bm \Psi}_{11}^A (t/t_\nu)^{-5/2}/(60\sqrt{\pi})$, with ${\bm \Psi}_{11}^A=I_{11}^*/I_{\rm sph}$, along the 1 direction and ${\bm \Psi}_{22}^A (t/t_\nu)^{-5/2}/(60\sqrt{\pi})$, with ${\bm \Psi}_{22}^A=1.089(I_{22}^*/I_{\rm sph})$, along the 2 and 3 directions. These scaling laws have been displayed with dotted lines in  Figs.~\ref{fig_vacfcomp}(a,b). Significantly, these predictions compare very favorably with the theoretical estimates of \cite{Hocquart:1983bh}, thus lending credibility for this numerical undertaking. For both VACF and AVACF, at large times a second exponential decay is observed. This behavior is attributable to the presence of the curved boundary of the vessel wall and its interaction with the NC motions. 

It is noteworthy that the predictions of the detailed numerical stochastic calculations and those of the deterministic method compare very well. The detailed stochastic calculations are very time consuming and computationally prohibitively expensive at large times. However, the deterministic calculations that yield essentially the same results have a lower computational overhead. 

In Fig.~\ref{fig:vacf-wall-effect}(a) and (b) we display the translational and rotational VACF, respectively, for a NC with $\aspp=1.5$, $a=600$ nm, and $\theta=0\degree$. The translational quantities are for the $x$ direction while the rotation corresponds to $1$ direction. Three different vessel confinements of diameters of $D=5,\,10,$ and $20 \mum$ with a fixed length of $L=40\mum$  are considered. In all of these cases, the particle is initially located on the central axis of the confining tube, with $\widetilde{h}=11.5$, $24$ and $49$, respectively. As noted in the figures three distinct decay regimes may be identified. An initial exponential decay, followed by an algebraic decay  culminating in a second exponential decay. The first exponential decay regime lasts about nearly the same time for all values of $D$, as would be expected. Larger the $D$ the corresponding algebraic decay regime is of longer duration.  The first exponential decay may be thought as being due to an uncorrelated noise, the algebraic decay is due to the combined effects of the thermal noise and hydrodynamic correlations while the second exponential decay shows the influence of momentum reflections from the confining boundary \citep{Vitoshkin16}. The computed algebraic regimes agree very well with analytical predictions given in eqns.~\eqref{eqn:tgtnu-trans} and ~\eqref{eqn:tgtnu-rot} thus providing an important validation for the comprehensive numerical formulation undertaken in this study. The analytical predictions are displayed by dotted lines and as can be  seen they overpredict both at short and long times. It is worthwhile emphasising that both the short and long time scales have been accessed in these numerical evaluations and this has been possible by the  use of both the fluctuating hydrodynamics and the deterministic methods. The effects of confinements can only be accurately described by employing simulations as carried out in this paper. 

There has been a considerable amount of published literature \citep{Hocquart:1983bh,Cichocki:1995df,Lowe:1995ko,Cichocki:1996ft,Masters:1996dz,Cichocki:1997jk,Masters:1997ic} related to the particle shape dependence on the  VACF and AVACF. Almost all of these studies concern themselves with the long time algebraic decay of AVACF in unconfined systems. Their results also state that the VACF in such circumstances is independent of the NC shape. We  present numerically accurate values for the VACF and AVACF taking into account the effect of confinement.  In this context, the effect of various aspect ratios, $\varepsilon=0.5,\,1.0,\,1.5,\,2.0,$ and $5.0$, on the translational and rotational VACFs along the $y$ and $2$ directions are displayed in Figs.~\ref{fig:Hinch-res}(a) and (b)  for an ellipsoidal NC that is placed at the center of a cylindrical tube with $D=20\mum$ and $L=40\mum$. The VACF (panel (a)) is found to be independent of the aspect ratio of the particle, with the algebraic regime scaling as $(t/t_\nu)^{-3/2}/(6\sqrt{\pi})$, which is shown as dotted lines. The AVACF (panel (b)), on the other hand, shows a strong dependence on the aspect ratio of the particle, particularly in the algebraic decay regime. This dependence has been captured by fitting this regime to ${\bm \varPsi}_{22}^{\rm C} (t/t_\nu)^{-5/2}/(60\sqrt{\pi})$, which are also shown alongside as dotted lines. In the inset, we compare the computed prefactor ${\bm \varPsi}_{22}^{\rm C}$, for five different aspect ratios, to their corresponding analytical estimates given by ${\bm \varPsi}_{22}^{\rm A}$, whose form is given in eqn.~\eqref{eqn:Hinch-pfac}. Again excellent comparison with analytical predictions lend credibility to the numerical study. As discussed earlier, three distinct regimes may be identified: Stokes decay, algebraic decay, followed by a second exponential decay which reflects  the effect of confinement. It is to be noted that the scaling behavior of both the VACF and AVACF may be modified from that given by eqns.~\eqref{eqn:tgtnu-trans} and ~\eqref{eqn:tgtnu-rot} by changes in the NC-wall proximity and wall curvature. These aspects are displayed in section S2 of the \sis{}. \\

\subsection{Diffusion of the ellipsoidal nanoparticle}
\begin{figure} 
	\centering
	\includegraphics[width=1.0\textwidth]{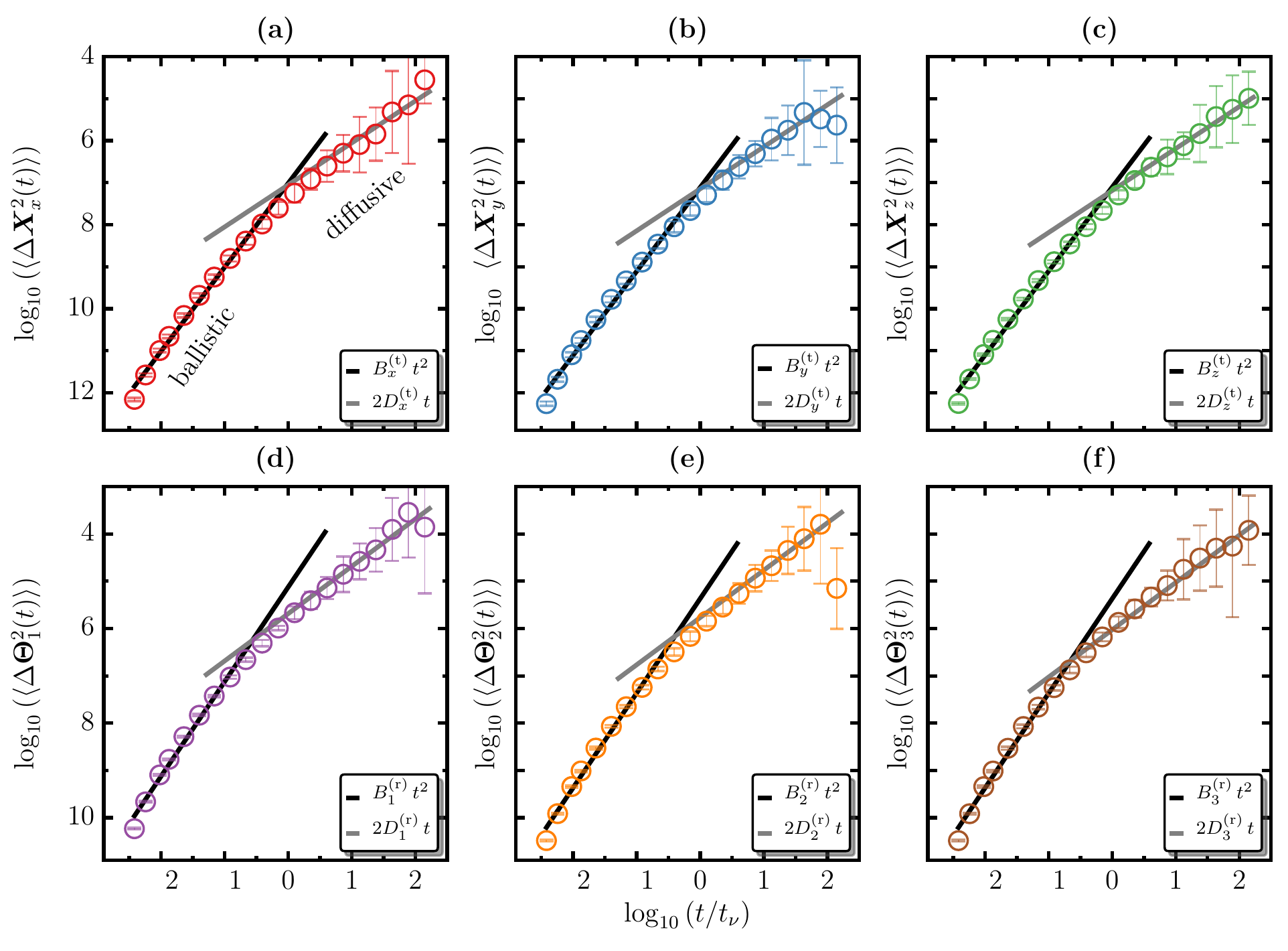}
	\caption{Mean squared displacement (MSD), along various directions, for a neutrally buoyant ellipsoidal NC,  with $\varepsilon=1.5$, $a=600\,{\rm nm}$, and $\theta=0 \degree$, placed at the center of a stationary fluid medium. 	The translational MSD, $\Delta R_{\alpha}^2(t)$ for $\alpha=x,y,z$, and  the rotational MSD, $\Delta {\bm \varTheta}_{\alpha}^2(t)$ for $\alpha=1,2,3$, are shown in panels (a)-(f). The solid lines are the fits to the ballistic and diffusive regimes. Mesh parameters used are $l_P=7$ nm and $l_W=785$ nm.}
	\label{fig_msd}
\end{figure}

\begin{figure}
	\centering
	\includegraphics[width=12.5cm,clip]{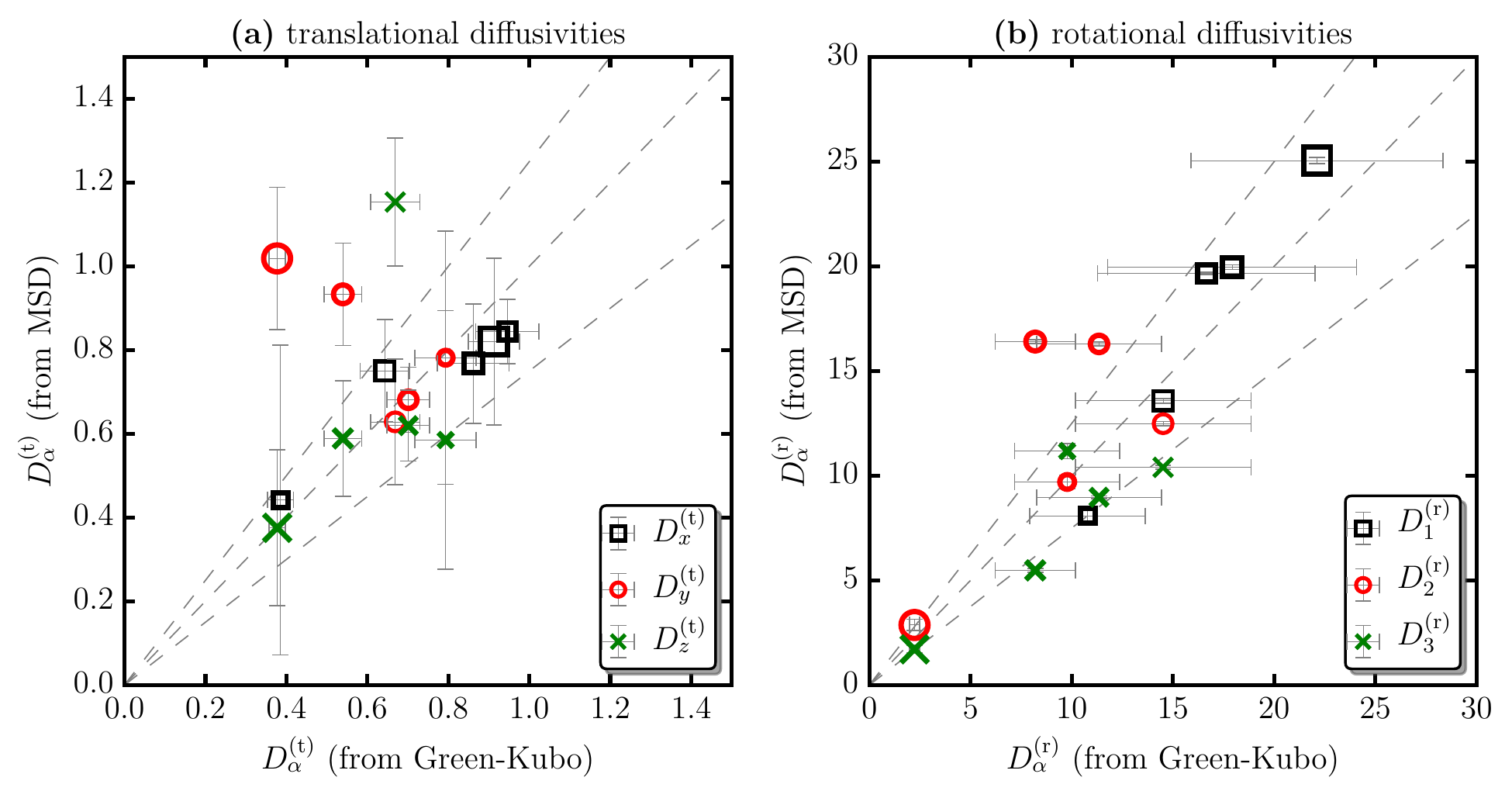}
	\caption{\label{fig:compareD} Comparison of the translational and rotational diffusivities computed from the velocity-autocorrelation, using the Green-Kubo relation, to those estimated from MSDs. Data for shown for NCs with five different aspect ratios and placed at $\widetilde{h}>1$. The central dotted line represents the linear correlation while the rest two represent deviations of $\pm 20\%$. The translational diffusivities (panel (a)) are in units of $\mum^2{\rm s}^{-1}$, and the rotational diffusivities (panel (b)) are in units of ${\rm rad}^2{\rm s}^{-1}$.}
\end{figure}

\begin{figure}
	\centering
	\includegraphics[width=14cm,clip]{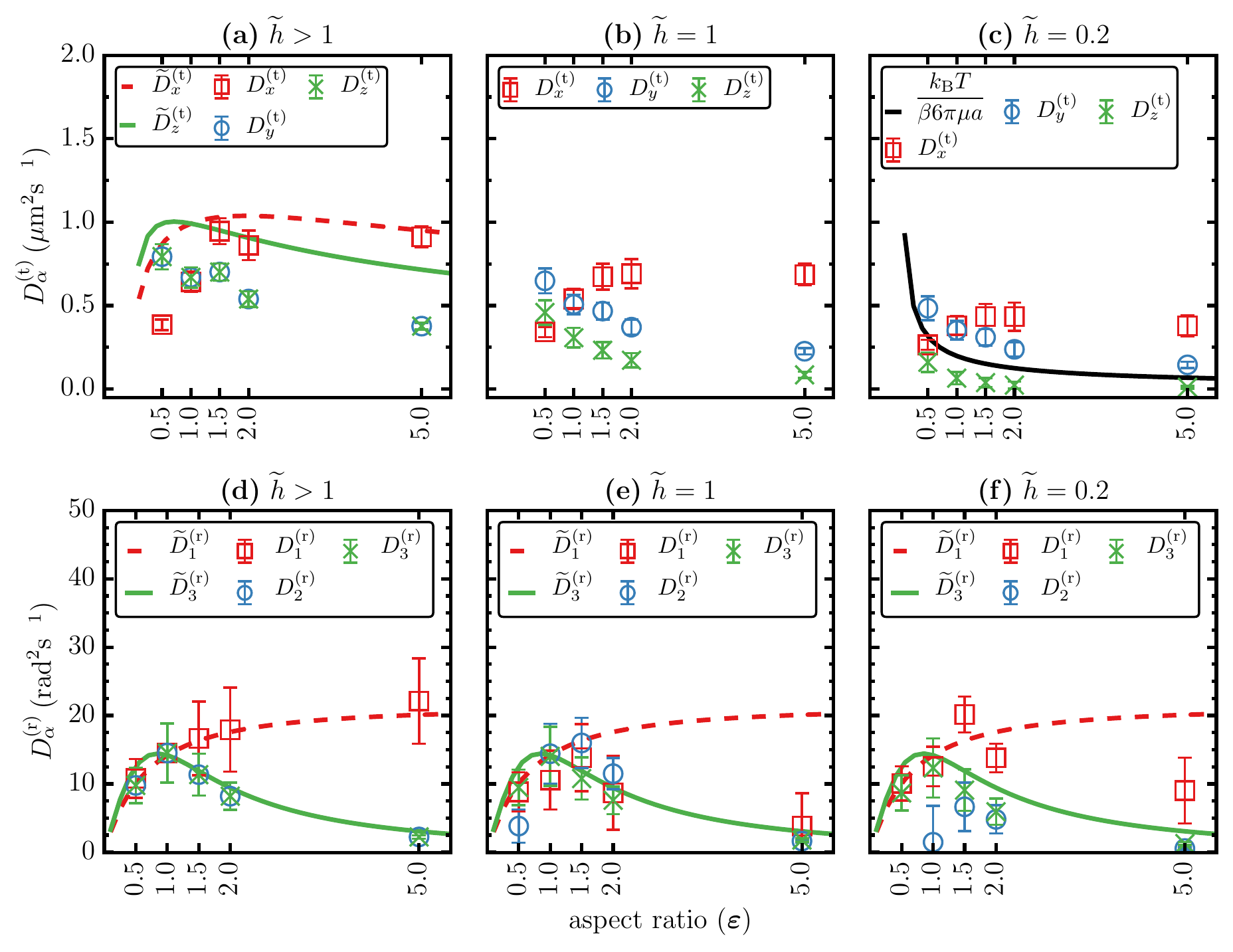}
	\caption{\label{fig:direc-asp-D}Translational diffusivities (panels (a)-(c)) and rotational diffusivities (panels (d)-(f)) as a function of the aspect ratio, for three different NC-wall separations $\widetilde{h}>1$, $\widetilde{h}=1$, and $\widetilde{h}=0.2$. In panel (a), $\widetilde{D}_x^{\rm (t)}$ (dotted line) and $\widetilde{D}_z^{\rm (t)}$ (solid line) are the estimates for the Stokes-Einstein diffusivities along the $x$  and $z$ directions, respectively, which apply for unbounded domains. In panels (d,e,f),  $\widetilde{D}_1^{\rm (r)}$ (dotted line) and $\widetilde{D}_3^{\rm (r)}$ (solid line) are the Stokes-Einstein-Debye diffusivities for an ellipsoidal particle in an unbounded domain.}
\end{figure}

The evaluation of the various velocity autocorrelation functions enable the calculation of the particle diffusivities via the Green-Kubo relation given by ~\citep{Kubo:1966dq}:
\begin{equation}
D_{\alpha} (t) = \int\limits_{0}^{t} \langle {\bm U}_{\alpha}(\tau) {\bm U}_{\alpha}(t+\tau)  \rangle d\tau.
\label{eqn:GK}
\end{equation}

\noindent Alternatively, the diffusive behavior of the NC can also be ascertained from the scaling behavior of the mean squared translational and rotational displacements defined by,
\begin{equation}
\langle \Delta\bm X_\alpha(t)^2\rangle = \langle(\bm X_\alpha(t+\tau)-\bm X_\alpha(\tau))^2\rangle
\end{equation}
for the translational MSDs, with $\alpha=x,y,z$ and
\begin{equation}
\langle (\Delta\bm \varTheta_\alpha(t))^2\rangle =\langle(\bm\varTheta_\alpha(t+\tau)-\bm\varTheta_\alpha(\tau))^2\rangle
\end{equation}
for the rotational MSDs, with $\alpha=1,2,3$. A numerical evaluation of these equations will provide the zero-frequency diffusivity of the particle. We first compute the diffusivity from the MSD calculations and compare them with estimates obtained using the Green-Kubo relation. Here, we numerically evaluate the MSD, along various directions, for a neutrally buoyant ellipsoidal NC,  with $\varepsilon=1.5$, $a=600$ nm, and $\theta=0\degree$, placed at the center of a tube with $D=5\mum$ and $L=40\mum$. The translational MSD, $\Delta {\bm X}_{\alpha}^2(t)$ for $\alpha=x,y,z$, and  the rotational MSD, $\Delta \Theta_{\alpha}^2(t)$ for $\alpha=1,2,3$, are shown in panels (a)-(f) of Fig.~\ref{fig_msd}. We observe that all the mean squared displacements (MSDs) crossover from a ballistic regime ($\sim t^2$) at short times to a diffusive regime ($\sim t$)  at longer times.  For the NC investigated, this crossover for the translational MSD is seen when $0.08 t_\nu <t<5t_\nu$ and for the rotational MSD when $0.05t_\nu<t<t_\nu$. We fit the ballistic regime to the functions $B^{\rm(t)}_{\alpha}t^2$ and $B^{\rm(r)}_{\alpha}t^2$, for the translational and rotational components, respectively.  The corresponding diffusive regimes are fit to $2D^{\rm (t)}_\alpha t$ and $2D^{\rm(r)}_\alpha t$, respectively. The best fits to each of these regimes are shown as solid lines. These enable the direct evaluation of NC diffusivities. Panels (a) through (f) reveal the important feature that the directional diffusivities are different for an ellipsoidal NC and vary along the directions $x$, $y$, $z$ and $1$, $2$, $3$.

In Fig.~\ref{fig:compareD}, we compare the directional diffusivities $D^{\rm (t)}_{\alpha}$ and $D^{\rm (r)}_{\alpha}$ computed from the mean squared displacements with those computed from the VACF and AVACF using eqn.~\eqref{eqn:GK} for NCs with $\widetilde{h}>1$. Results for $\widetilde{h}=1$ and $\widetilde{h}=0.2$ may be found in section. S4 of the \sis.  As may be noted, the predictions from either technique are essentially the same within $\pm 15\%$ (see  dotted lines). This agreement between the predictions of the two techniques are found to be independent of the aspect ratio of the NC \asp{} and the NC location in the fluid medium $\widetilde{h}$. Again this confirms the validity of the numerical scheme.

In Fig.~\ref{fig:direc-asp-D} we show the directional translational and rotational diffusivities as a function of the aspect ratio for various locations of the NC in the vessel. The symbols denote the computed values of $D^{\rm (t)}_{\alpha}$ and $D^{\rm (r)}_{\alpha}$, computed using the Green-Kubo relation from the VACF and AVACF, respectively.  The diffusivities computed from the MSD are shown in Section. S5 in the \sis.

 The Stokes-Einstein long time translational diffusivities for an NC in an unbounded media along the $x$, $y$, and $z$ directions may be computed as $\widetilde{D}_{x}^{\rm (t)}=\kbt/\xi^{\rm (t)}_{x}(\varepsilon)$ and $\widetilde{D}^{\rm (t)}_{y}=\widetilde{D}_{z}^{\rm (t)}=\kbt/\xi^{\rm (t)}_{z}(\varepsilon)$, respectively. Here $\xi^{\rm (t)}_{x}(\varepsilon)$ and $\xi^{\rm (t)}_{z}(\varepsilon)$ are the aspect ratio and direction dependent friction coefficients, computed as in \cite{Clift:1978vm}. These asymptotes are shown in  Fig.~\ref{fig:direc-asp-D}(a) as dotted and solid lines, respectively. For $\varepsilon<1.0$ ($\varepsilon=1$ corresponds to a spherical NC) we consider an oblate spheroidal NC. The $\widetilde{D}_{y,z}^{\rm (t)}$ are both greater than $\widetilde{D}_{x}^{\rm (t)}$. This may be explained as due to the effects of the added mass and added moments of inertia associated with the oblate spheroidal shape. These favor higher diffusivities in the $y$ and $z$ directions. On the other hand when $\varepsilon>1.0$ we have a prolate spheroid diffusing in an unbounded medium. Here, the same physical factors favor the $x$ directional diffusion. Succinctly, for NCs of equal volumes in an unbounded media (and the same equivalent diameter $d_{\rm eq}$), the direction dependent friction coefficients may be shown to scale with the aspect ratio as $\xi^{\rm (t)}_{x}(\aspp) \sim (4+\aspp)/\sqrt[3]{\aspp}$ and $\xi^{\rm (t)}_{z}(\aspp) \sim (3+2\aspp)/\sqrt[3]{\aspp}$. As would be expected, for an NC located at the center line of the vessel, for which $\widetilde{h}>1$, our predictions for $D^{\rm (t)}_{\alpha}$ follows this scaling behavior for all values of $\varepsilon$. At a given $\varepsilon$, the directional diffusivities denoted by the symbols are uniformly lesser compared to the asymptotic values. This is due to the presence of the confining boundary which would serve to retard the diffusion consequent to enhanced viscous effects.
 
 The effects of the presence of the bounding wall and the proximity of the NC to the wall are displayed in Figs.~\ref{fig:direc-asp-D}(b) and (c) for an NC with $\widetilde{h}=1$ and $\widetilde{h}=0.2$, respectively.
In these regimes, we find  the diffusivities along  all the directions  decrease with $\widetilde{h}$ and are smaller compared to the corresponding values for $\widetilde{h}>1$. For $\widetilde{h}>1$ (near wall regime) the effect of the wall is more severely felt for $D^{\rm (t)}_{z}$ and is the least for $D^{\rm (t)}_{x}$.  In fact, $D^{\rm (t)}_{x}$ shows a similar trend to that at $\widetilde{h}>1$. All of these features are due to enhanced viscous effects. These same features with even more reduced  diffusivities due to viscous drag are apparent at $\widetilde{h}=0.2$, and the trends with increasing $\varepsilon$ are similar to those at $\widetilde{h}=1$.  In this panel we also display a solid line that corresponds to $\widetilde{D}_{z}^{\rm (t)}=\kbt/(\beta6\pi\mu a)$ for a particle with  in this regime. The feature that ${D}_{z}^{\rm (t)}$ is severely reduced at $\widetilde{h}=0.2$ and in fact is lower at higher aspect ratios (increasingly prolate shapes) may be attributed to two causes: increased viscous drag and the presence of lift forces in the lubrication regime. ${D}_{y}^{\rm (t)}$ is also affected by increased viscous drag but the effect of lift forces are minimal. It may be recalled that these discussions are applicable only to an NC with an angle of attack $\theta=0$.

The diffusivities along $y$ and $z$, which correspond to the radial directions, are found to depend  on (i) $a$, the NC cross section in the $x$ direction and (ii) the enhanced drag parameter $\beta$ that is a function of the NC-wall separation $\widetilde{h}$. This scaling behavior predicted based on steady lubrication theory \citep{Leal07,Yu:2015kh,Vitoshkin16} is well represented by $\kbt/(\beta6\pi\mu a)$, and this is shown as a solid line in panel (c).

The rotational diffusivities $D^{\rm (r)}_{\alpha}$ for $\widetilde{h}>1$,  display a behavior similar to that described for the translational diffusivities $D^{\rm (t)}_{\alpha}$. In Figs.~\ref{fig:direc-asp-D} (d,e,f)  $\widetilde{D}^{\rm (r)}_1=\kbt/\xi^{\rm (r)}_1$ and $\widetilde{D}^{\rm (r)}_{2,3}=\kbt/\xi^{\rm (r)}_{2,3}$, the rotational diffusivities of an ellipsoidal particle in an unbounded media, are shown as dotted and solid lines, respectively. The rotational friction coefficients  $\xi^{\rm (r)}_1$ and $\xi^{\rm (r)}_{2,3}$ are computed as given by Perrin~\citep{PerrinFrancis:1934jp,PerrinFrancis:1936dy,Koenig:1975tf}. The computed values of the rotational diffusivities are in excellent agreement with the asymptotic values for $\widetilde{h}>1$, where the effect of the bounding walls on the rotational motions is minimal. The effects of added moments of inertia are responsible for the decreased diffusivities in the 2 and 3 directions. Again, with increasing \asp{} the diffusivities decrease in the 2 and 3 directions due to enhanced form drag. For $\widetilde{h}\leq 1$, while $D^{\rm (r)}_{2,3}$ preserve the trend as a function of \asp{}, that is noted for $\widetilde{h}> 1$, the behavior of $D^{\rm (r)}_{1}$ which is significantly different may be explained as a consequence of modifications in the added moment of inertia in the 1 direction caused by confinement.
 
\subsection{Lift force on an NC and its relation to the diffusivity tensor}
Thus far, we have focused on the diagonal components of the mobility/diffusivity tensor as a function of the aspect ratio and confinement. However, in targeted drug delivery applications, the off diagonal elements of the mobility/diffusivity tensor are also of interest in order to quantify the degree of lift/margination of the NC subjected to flow in a confined vessel. While these off-diagonal elements can be estimated using the VACF approach we have utilized thus far in conjunction with the Green-Kubo relationship \citep{Kubo66}, the cross correlation of velocities are not easily computable due to the significant numerical variations that may occur with such calculations, even with the deterministic formulation. An alternative approach which avoids this difficulty is to directly compute the lift and the drag forces on the NC when subject to flow and confinement. This is the approach we will adopt here.

\begin{figure}
\includegraphics[width=12.5cm,clip]{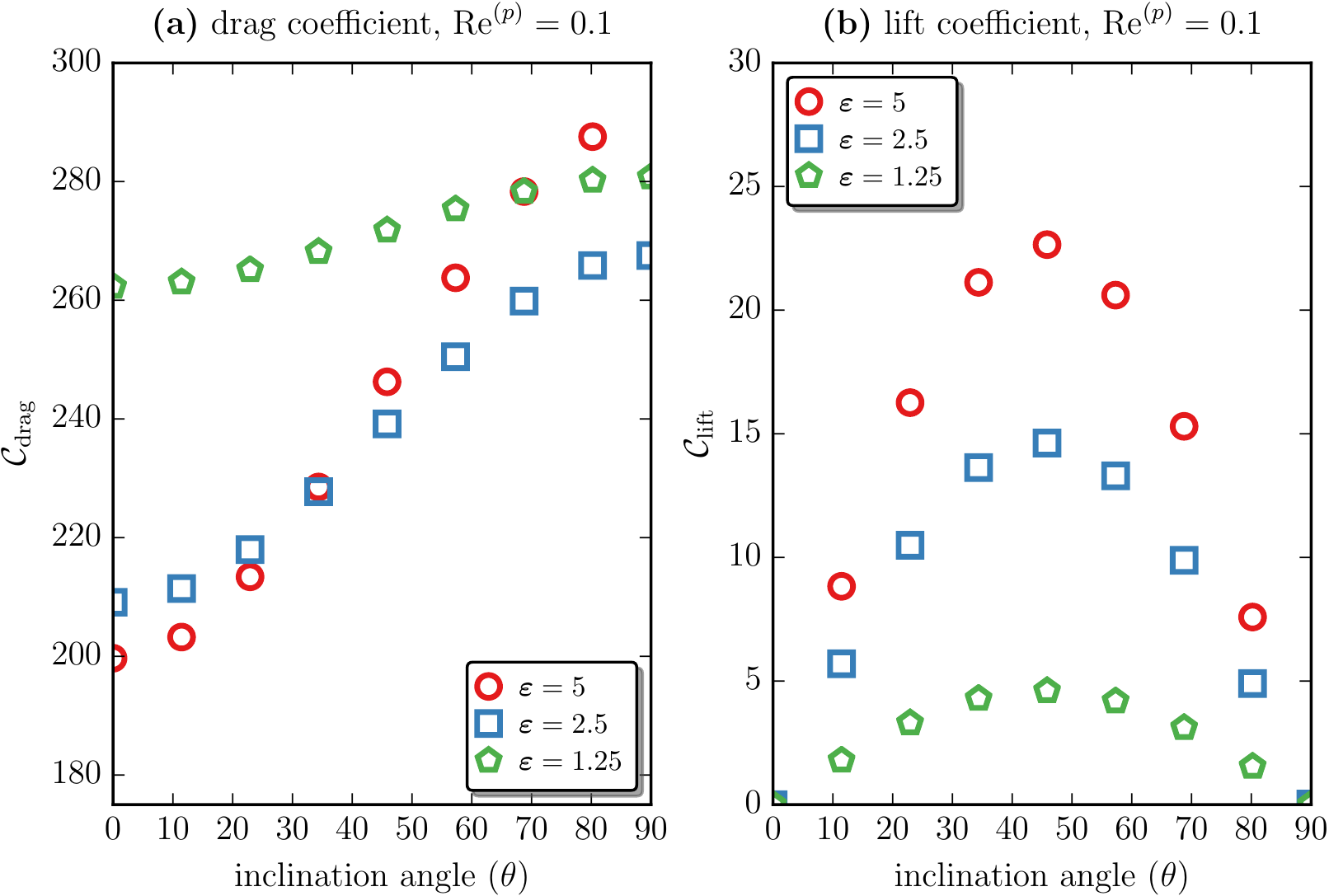}
\caption{\label{fig:compare-drag-lift} A comparison of the drag and lift coefficients  for an ellipsoidal NC (with $a=5\mum$) as a function of the inclination angle and aspect ratio. Data shown for particles, with $\aspp=5$, $\aspp=2.5$, and $\aspp=1.25$, placed at the center of a circular tube of diameter $D=50 \mum$. Flow conditions have been chosen such that particle Reynolds number $\Rep=0.1$. Shown are (a) the drag coefficient ${\cal C}_{\rm drag}$, and (b) the lift coefficient ${\cal C}_{\rm lift}$  as a function of $\theta$. Mesh parameters used in these calculations are  $l_P=75$ nm, $188$ nm, and $473.5$ nm, for aspect ratios $\aspp=5,\,2.5$, and $1.25$, respectively, and $l_W=5.235\mum$.}
\end{figure}

We have explicitly computed ${\bf F}_{\rm drag}$ and ${\bf F}_{\rm lift}$, for a prescribed Poiseuille flow condition using the ALE framework. In these calculations the position ($r$) and orientation ($\theta$) of the nano-ellipsoid are taken to be fixed. We present all our results in terms of the drag and lift  coefficients, ${\cal C}_{\rm drag}(r,\theta)=||{\bf F}_{\rm drag}||(\rho_f |\bm{u}_r|^2 \pi d_{\rm eq}^2/8)^{-1}$  and ${\cal C}_{\rm lift}(r,\theta)=||{\bf F}_{\rm lift}||(\rho_f |\bm{u}_r|^2 \pi d_{\rm eq}^2/8)^{-1}$ respectively. Here $d_{\rm eq}$ is the equivalent sphere diameter and $|\bm{u}_r|$ is the magnitude of the flow velocity at a radial position $r$. At present, results for the drag and lift forces for an ellipsoidal particle bounded by a circular tube are unavailable for comparison. In view of this,  we first validate our calculations with those reported by ~\cite{Ouchene2015}, where the authors study an ellipsoid in  a rectangular tube at particle Reynolds number $\Rep=0.1$. We expect  our estimates for  ${\cal C}_{\rm drag}$ and ${\cal C}_{\rm lift}$  to compare with those results for cases where the particle is located at the center of the vessel and the role of the bounding geometry is a minimum.  In  Fig.~\ref{fig:compare-drag-lift}, we display  ${\cal C}_{\rm drag}$ and  ${\cal C}_{\rm lift}$, as a function of  $\theta$, for the particle bounded by a cylindrical wall. We present results for aspect ratios $\aspp=5$, $2.5$, and $1.25$, and with flow conditions such that $\Rep=0.1$. The parameters used in these calculations are given in appendix~\ref{app:compare-drag}. Excellent comparison is noted.  For all aspect ratios studied, the lift coefficient shows a parabolic profile as a function of $\theta$ (see Fig.~\ref{fig:compare-drag-lift}(b)), with a pronounced peak at $\theta=45\degree$. However, the peak value of ${\cal C}_{\rm lift}$ decreases with a decrease in the aspect ratio (\asp) of the particle. On the other hand, the drag (Fig.~\ref{fig:compare-drag-lift}(a)) shows a monotonic increase as a function of $\theta$. This feature is also seen in ~\cite{Ouchene2015}.

\begin{figure}
\includegraphics[width=12.5cm,clip]{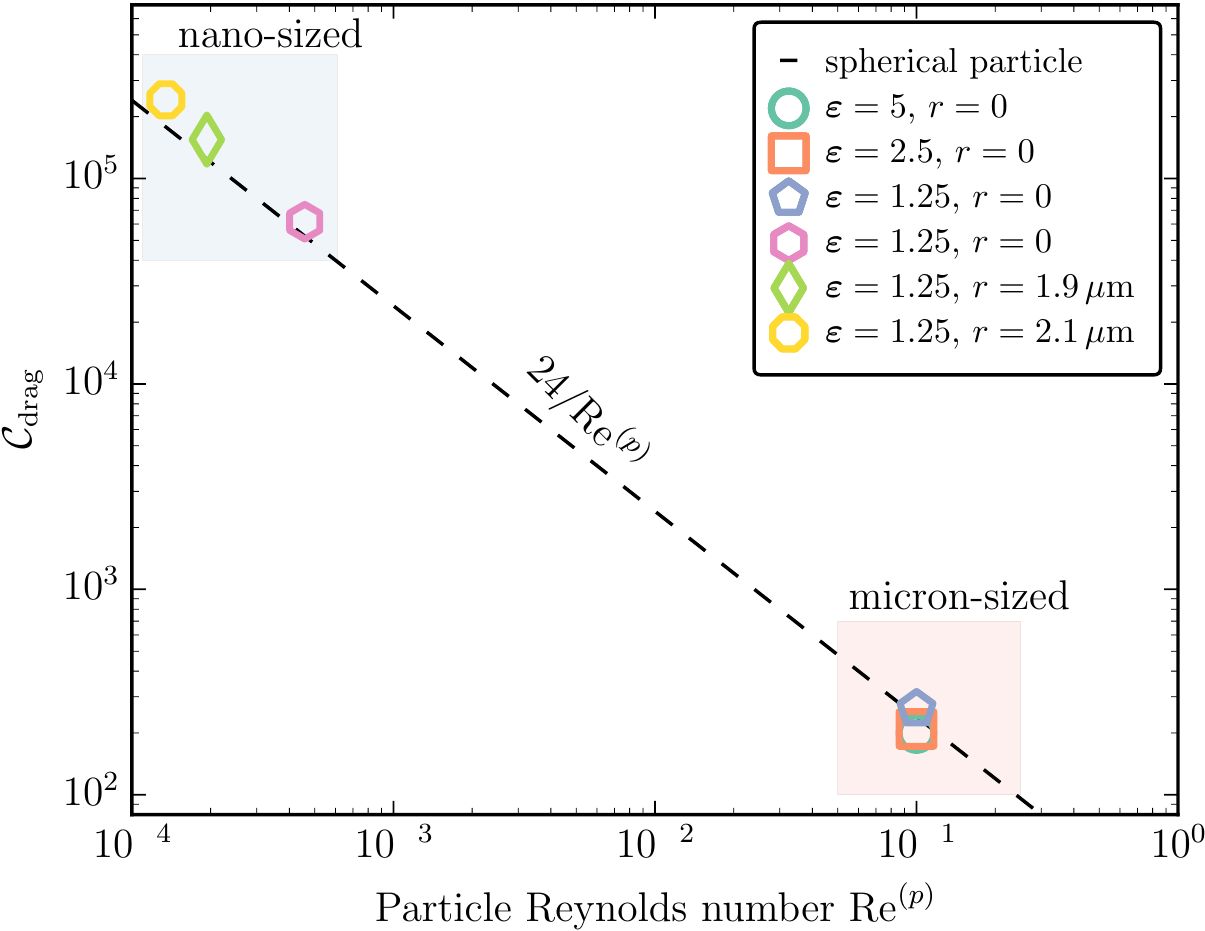}
\caption{\label{fig:drag-analytical} ${\cal C}_{\rm drag}$ as a function of the particle Reynolds number $\Rep$ for micron-sized and nano-sized ellipsoidal particles. Data shown for ellipsoids oriented along the $x$ direction with $\theta=0$. Data for the micron-sized particles are from Fig.~\ref{fig:compare-drag-lift} and for the nano-sized particles are from Fig.~\ref{fig_liftdrag}. The dotted line denotes the predicted scaling relation of $24/\Rep$ for a spherical particle in the bulk.}
\end{figure}

In Fig.~\ref{fig:drag-analytical}, ${\cal C}_{\rm drag}$ is shown as a function of $\Rep$ for $\theta=0\degree$. The dotted line denotes ${\cal C}_{\rm drag}=24/\Rep$ for a spherical particle in an unbounded medium, which in our case to particles located at the center of the pipe ($r=0$).  The figure also displays ${\cal C}_{\rm drag}$ for a micron-sized and a nano-sized particle as a function of aspect ratio and separation distance from the wall. For the micron-sized particle studied, the location is always at the center ($r=0$) while $\aspp=5,\,2.5$, and $1.25$, with $a=5\mum$, $D=50\mum$, and $\Rep$ is taken to be $0.1$ (appendix~\ref{app:compare-drag}). For $\aspp \simeq 1$, ${\cal C}_{\rm drag}$ is noted to approach the value for a sphere. For higher values of \asp, ${\cal C}_{\rm drag}$ is slightly lower than that for a sphere. This is as would be expected since the shape becomes more streamlined with increasing \asp. Next, with nano-sized particle in a vessel of diameter $D=5\mum$,  we fix $\aspp=1.5$ (with a=$600$ nm) and vary the particle location  such that  $r=0.0,\,1.9$, and $2.1\mum$. We study this in the context of a fully developed Poiseuille flow with an inlet velocity $\bm{u}_{\rm max}=0.1$ cm/s, chosen to represent physiological flow rates~\citep{Mazumdar92}. The particle Reynolds numbers at the three radial positions for a fluid with a kinematic viscosity $\nu=10^6 \mum{}^2 s^{-1}$ (which is representative of blood plasma) works out to be around $4\times10^{-4}$, $2\times10^{-4}$, and $1\times10^{-4}$, respectively.   As noted from the figure, with increasing $r$ (closer to the wall) the drag on the nano-ellipsoid increases. Since the shape of the particles are of similar shape, the varying drag values are ascribable to the prevailing velocity profile.

\begin{figure} 
	\centering
	\includegraphics{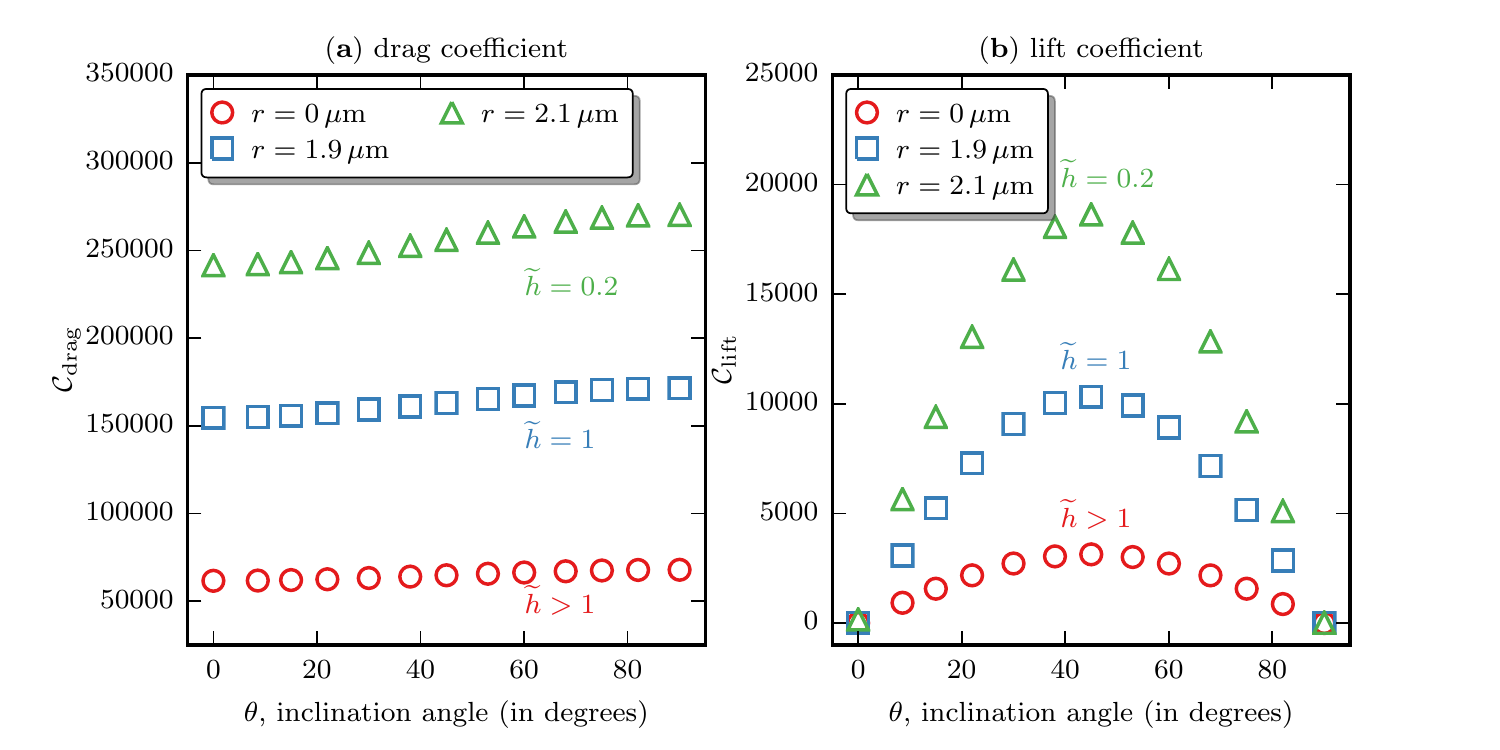}
	\caption{(a) The drag coefficient ${\cal C}_{\rm drag}$ and (b) the lift coefficient ${\cal C}_{\rm lift}$ for a nanoellipsoid, as a function of the inclination angle $\theta$ for three different radial positions: $r=0.0\,\mu{\rm m}$ (bulk), $r=1.9\,\mu{\rm m}$ (near wall), and $r=2.1\,\mu{\rm m}$ (lubrication). Representative values for the mesh parameters are given in Table ~\ref{table:meshlength}. }
	\label{fig_liftdrag}
\end{figure}

\section{Conclusions} \label{sec:conclusions}
We present numerical studies based on the Fluctuating hydrodynamics approach and the Deterministic method to investigate the Brownian motion of ellipsoidal NCs of various aspect ratios, angles of inclination, and proximities to the wall  in a cylindrical fluid filled vessel. The bulk medium may be stationary or may experience a weak Poiseuille flow. The incompressible fluid flow formulation is modified by considerations of added masses and added moments of inertia . A major objective is the evaluation of the effects of the confining boundary. Detailed results  for the VACF and AVACF, mobility, diffusivity, drag and lift forces  as functions of aspect ratio, inclination angle, and proximity to the wall are presented. For the parameters considered, the confining boundary modifies the VACF and AVACF such that three distinct regimes are discernible --- an initial exponential decay, followed by an algebraic decay culminating in a second exponential decay. The effects of shape,  proximity to the wall, and the drag and lift forces  on the  translational and rotational diffusivities of the NC are comprehensively displayed.  The complicated behavior of these quantities are explained in detail. The effects of the off diagonal elements of the mobility/diffusivity tensor that enable the quantification of the degree of lift/margination of the NC have also been evaluated and discussed. Predicted results show excellent comparison with published results for the algebraic regime (the only such results that are available).

~\newline
{\Large \bf Author contributions:} 

\noindent  YW was responsible for implementing the model for ellipsoidal particles and for the static mobility calculations presented in the SI. NR and YW were responsible for implementing the fluctuating hydrodynamics approach. NR was responsible for implementing the deterministic VACF and Green Kubo computations in the main text. DME, PSA, and RR were principally responsible for conceptualizing the method and application along with NR and YW. All authors contributed to critically analyzing the data, interpreting the results, and writing of the paper.

~\newline
\noindent
{\Large \bf Acknowledgments}

The DNS methodology was developed in part by support from National Institute of Health (NIH) grant R01-EB006818. The computation of the equilibrium and transport properties was supported in part by NIH grant U01-EB016027. The authors thank Dr. Hsiu-Yu Yu and Dr. Helena Vitoshkin for useful discussions.

\appendix{
\section{Analytical expression for the radial dimension $\zeta_0$} \label{app:a0derivation}
For an asymmetric particle, the parameter $\zeta_0$ is a measure of the maximum of the projections of the particle dimensions $a$, $b$, and $c$ along a radial direction. Here we give a heuristic expression for $\zeta_0$ that is valid for ellipsoids with one rotational symmetry --- i.e., prolates ($a>b=c$) and oblates ($a<b=c$). In our derivation we consider a prolate ellipsoid (shown in Fig.~\ref{fig_surfacemesh}) whose inclination angle $\theta$ denotes a rotation in the $x-z$ plane. Since the long axis of the cylindrical tube is along the $x$ direction, we take the radial direction for this calculation to be along $z$. Let $(x_0, y_0, z_0)$ denote the position of the center of mass of the particle and $(x(\theta),y(\theta),z(\theta))$ denotes the location of the point on the particle surface with the  minimum value of $z$.  

For a prolate ellipsoid in the $x-z$ plane,  $z(\theta)$ can be computed by computing the projections of the semi-major and semi-minor axis along the radial direction given by $(a\sin \theta)/2$ and $(b\cos \theta)/2$, respectively. Below a critical inclination angle $\theta_0=\tan^{-1}(b/a)$, $z(\theta)$ can be identified with the projection of the semi-minor axis while it is equal to projection of the semi-major axis when $\theta>\theta_0$. This dependence can be expressed as
\begin{equation}
\zeta_0(a,b,\theta) = \dfrac{a\sin\theta}{2} \,\mathcal{H}(\theta-\theta_0) + \dfrac{b\cos\theta}{2} \, \mathcal{H}(\theta_0-\theta),
\label{equ_a0}
\end{equation}
where $\mathcal{H}$ is the Heaviside step function.

\section{Equivalent particle diameters and their radial positions for different aspect ratios} \label{app:equv_dia}
In this section, we present the various parameters for simulations of ellipsoidal particles under three type of confinements, which are classified as:

\begin{enumerate}[(i) ]
\item  bulk: $\tilde{h} = \dfrac{h-\zeta_0}{\zeta_0} >1$, \\
\item  near wall: $\tilde{h}=\dfrac{h-\zeta_0}{\zeta_0} = 1$, \\
\item  lubrication: $\tilde{h}=\dfrac{h-\zeta_0}{\zeta_0} < 1$. \\
\end{enumerate}

The value of $\widetilde{h}$ for the bulk regime depends on $\varepsilon$, $a$, $\theta$, and $D$. For the lubrication regime we take $\widetilde{h}=0.2$ throughout. Tables \ref{table:thetazero}, \ref{table:thetapiby4}, and \ref{table:thetapiby2} show the positions of the NC as a function of its aspect ratio,  for $\theta=0\degree$, $45\degree$, and $90\degree$ respectively. The shown data correspond to confinement by a cylindrical tube with $D=5\mum$ and $L=40\mum$.

\begin{table}
\begin{center} 
	\begin{tabular}{cccccccccccc}
		\multirow{2}{*}{} & & & & \multicolumn{2}{c}{\textbf{bulk}} & & \multicolumn{2}{c}{\textbf{near wall}} & & \multicolumn{2}{c}{\textbf{lubrication}} \\
		& & & & & & & &&  & &  \\
		$\varepsilon$ & $a$ (nm) & $b=c$ (nm) & $\zeta_0=c/2$ (nm) & $h\,(\mum)$  & $\widetilde{h}$ & & $h\,(\mum)$ &   $\widetilde{h}$ & & $h\,(\mum)$ & $\widetilde{h}$ \\ 
		& & & & & & & & & & &  \\
		0.5 & 288.45 & 576.9 & 288.45& 2.5 & 7.667 & & 0.5769& 1.0 && 0.346 & 0.2\\ 
		1 & 457.9 & 457.9 & 228.95& 2.5 & 7.962 & & 0.4579 & 1.0& & 0.274&0.2 \\ 
		1.5 & 600.0  & 400.0 & 200.0 & 2.5 & 11.5 & & 0.4 &1.0 & & 0.24 &0.2 \\ 
		2.0 & 726.8 & 363.4 & 181.7 & 2.5 & 12.75 & & 0.3634 &1.0 & &0.218 &0.2 \\ 
		5.0 & 1338.91 & 267.78 & 133.88 & 2.5 & 17.67 & & 0.26779 &1.0 & &  0.161 &0.2 \\ 
		\end{tabular}
	\end{center}
	\caption{\label{table:thetazero} Simulation parameters for ellipsoidal particles with five different aspect ratios subject to three different confinements in a cylindrical tube with $D=5\mum$ and $L=40\mum$. These parameters correspond to a particle orientation $\theta=0\degree$.} 
\end{table}

\begin{table}
	\begin{center} 
		\begin{tabular}{cccccccccccc}
			\multirow{2}{*}{} & & & & \multicolumn{2}{c}{\textbf{bulk}} & & \multicolumn{2}{c}{\textbf{near wall}} & & \multicolumn{2}{c}{\textbf{lubrication}} \\
			& & & & & & & &&  & &  \\
			$\varepsilon$ & $a$ (nm) & $b=c$ (nm) & $\zeta_0$ (eqn.\eqref{equ_a0}) & $h\,(\mum)$  & $\widetilde{h}$ & & $h\,(\mum)$ &   $\widetilde{h}$ & & $h\,(\mum)$ & $\widetilde{h}$ \\ 
			& & & & & & & & & & &  \\
			0.5 & 288.45 & 576.9 & 203.96 & 2.5 & 11.25 & & 0.408& 1.0 && 0.245 & 0.2\\ 
			1 & 457.9 & 457.9 & 228.95& 2.5 & 9.91 & & 0.4579 & 1.0& & 0.275&0.2 \\ 
			1.5 & 600.0  & 400.0 & 212.13 & 2.5 & 10.78 & & 0.424 &1.0 & & 0.255 &0.2 \\ 
			2.0 & 726.8 & 363.4 & 256.96 & 2.5 & 8.72 & & 0.514 &1.0 & &0.308 &0.2 \\ 
			5.0 & 1338.91 & 267.78 & 473.37 & 2.5 & 4.28 & & 0.947 &1.0 & &  0.568 & 0.2 \\ 
		\end{tabular}
	\end{center}
\caption{\label{table:thetapiby4} Simulation parameters for ellipsoidal particles with five different aspect ratios subject to three different confinements in a cylindrical tube with $D=5\mum$ and $L=40\mum$. These parameters correspond to a particle orientation $\theta=45\degree$. The radial dimension $\zeta_0$ is computed as given in eqn.~\eqref{equ_a0}.}
\end{table}

\begin{table}
	\begin{center} 
		\begin{tabular}{cccccccccccc}
			\multirow{2}{*}{} & & & & \multicolumn{2}{c}{\textbf{bulk}} & & \multicolumn{2}{c}{\textbf{near wall}} & & \multicolumn{2}{c}{\textbf{lubrication}} \\
			& & & & & & & &&  & &  \\
			$\varepsilon$ & $a$ (nm) & $b=c$ (nm) & $\zeta_0=a/2$ (nm) & $h\,(\mum)$  & $\widetilde{h}$ & & $h\,(\mum)$ &   $\widetilde{h}$ & & $h\,(\mum)$ & $\widetilde{h}$ \\ 
			& & & & & & & & & & &  \\
			0.5 & 288.45 & 576.9 & 144.225& 2.5 & 16.334 & & 0.288& 1.0 && 0.173 & 0.2\\ 
			1 & 457.9 & 457.9 & 228.95 & 2.5 & 9.919 & & 0.4579 & 1.0& & 0.274&0.2 \\ 
			1.5 & 600.0  & 400.0 & 300.0 & 2.5 & 7.333 & & 0.6 &1.0 & & 0.36 &0.2 \\ 
			2.0 & 726.8 & 363.4 & 363.4 & 2.5 & 5.879 & & 0.727 &1.0 & &0.436 &0.2 \\ 
			5.0 & 1338.91 & 267.78 & 669.455 & 2.5 & 2.73 & & 1.338 &1.0 & &  0.803 &0.2 \\ 
		\end{tabular}
	\end{center}
\caption{\label{table:thetapiby2} Simulation parameters for ellipsoidal particles with five different aspect ratios subject to three different confinements in a cylindrical tube with $D=5\mum$ and $L=40\mum$. These parameters correspond to a particle orientation $\theta=90\degree$.}
\end{table}

\section{Mesh length on the particle used in the computation}
The values of $l_P$ and $l_W$ (defined in Fig.~\ref{fig_tetrahedron}) used in our calculations are given in Table.~\ref{table:meshlength}.

\begin{table}
\begin{center}
\begin{tabular}{cccccccc}
\multirow{2}{*}{}& & \multicolumn{2}{c}{$\theta=0\degree$}& \multicolumn{2}{c}{$\theta=45\degree$}& \multicolumn{2}{c}{$\theta=90\degree$} \\
& & & & & & & \\
$\varepsilon$& $\widetilde{h}$  & $l_P$ (nm) & $l_W$ (nm) & $l_P$ (nm)& $l_W$ (nm) & $l_P$ (nm) & $l_W$ (nm) \\
& & & & & & & \\
& $7.667$&7  & 628 & 7 & 628 & 7 & 628 \\
0.5  & $1.0$ & 4 & 524 &4  & 524 & 7 & 0.785 \\
  & $0.2$ & 4 & 524 &4  & 628 & 4 & 0.628 \\
& & & & & & &\\

 & $7.962$ & 8 & 785 & - &  - & - &  - \\
1.0  & $1$ & 8 & 785 & - & - & - & - \\
  & $0.2$ &  7 & 628 & - & - & - & - \\
& & & & & & & \\  

     & $11.5$ & 8 & 785 &7  & 628 & 8 & 785 \\
1.5  & $1$ & 8 & 785 & 4 & 524 & 8 & 785 \\
     & $0.2$ & 8 & 785 & 7 & 785 & 8 & 785 \\
& & & & & & & \\  

     & $12.75$ & 7 & 628 & 4 & 524 & 7  & 785 \\
2.0  & $1$ &  7& 628 & 4 & 524 & 7 & 628 \\
     & $0.2$ & 7 & 628 & 7 & 628 & 7 & 628 \\
& & & & & & & \\       
     
     & $17.67$ & 7 & 628 & 4 & 524 & 7  & 785 \\
5.0  & $1$ &  7& 628 & 4 & 524 & 7 & 628 \\
     & $0.2$ & 7 & 628 & 7 & 628 & 7 & 628 \\   
& & & & & & &\\  
\end{tabular}
\end{center}
\caption{\label{table:meshlength} The mesh lengths on the particle and on the tubular wall, $l_P$ and $l_W$, respectively, for five different aspect ratios and a confining wall with $D=5\mum$ and $L=40\mum$.}
\end{table}

\section{Parameters used in the computation of drag and lift forces in the bulk} \label{app:compare-drag}
The target particle Reynolds number is computed as $\textrm{Re}_{p}(r)=|\mb{u}(r)| d_{\rm eq}/\nu$, where $d_{\rm eq}$ is the equivalent sphere diameter and $\nu$ is the kinematic viscosity. We choose $\nu=10^{5}$ and particles with $a=5\mum$ and aspect ratios $\varepsilon=5,2.5$, and $1.25$.  For each of the particles, we compute its equivalent sphere diameter as $d_{\rm eq} = \sqrt[3]{abc}$, see Table \ref{table:Reynum0p1} for details.

\begin{table}
	\begin{center} 
		\begin{tabular}{ccccc}
			$\varepsilon$ & $a\,(\mum)$ & $b=c\,(\mum)$ & $d_{\rm eq}\,(\mum)$ & $\bm{u}_{\rm max}\,(\mum/s)$  \\ 
			& & & & \\
			5 & 5 & 1 & 1.709 & 5851.375 \\
			2.5 & 5 & 2 & 2.71 & 3690.037\\
			1.25 & 5 & 4 & 4.3 & 2325.581\\   
		\end{tabular} 
	\end{center}
	\caption{\label{table:Reynum0p1} Parameter values used in the calculation of drag and lift coefficients for comparison with that from \cite{Ouchene2015}.}
\end{table}
}

\clearpage
\newpage
\begin{titlepage}
\centering
\vspace*{\fill}
\Large{\textbf{SUPPLEMENTARY INFORMATION}}
\vspace*{\fill}
\end{titlepage}

\clearpage
\newpage

\setcounter{section}{0}
\renewcommand{\appendix}{S\arabic{section}}
\renewcommand{\appendixname}{}
\renewcommand{\thesection}{S\arabic{section}}
\renewcommand{\thetable}{S\Roman{table}}
\renewcommand{\thefigure}{S\arabic{figure}}

\section{FHD data for ellipsoidal particles of different aspect ratio}\label{sec:Brownian}
\setcounter{figure}{0}
\renewcommand{\thefigure}{\ref{sec:Brownian}.{\color{blue}\arabic{figure}}}
\setcounter{table}{0}
\renewcommand{\thetable}{\ref{sec:Brownian}.{\color{blue}\arabic{table}}}

In this section, we present data from FHD simulations for ellipsoidal particles, with five different aspect ratios $\varepsilon=0.5,\,1.0,\,1.5,\,2.0,$ and $5.0$, placed at three different locations, inside a cylindrical tube of diameter $D=5\mum$ and length $L=40\mum$. \\

\noindent In each of the panel plots presented here, \textit{columns from left to right} correspond to  \\

\begin{enumerate}[(i) ]
	\item the velocity distribution, $P({\mb U}_{\alpha})\,d{\mb U}_{\alpha}$.
	\item time series of the scaled averaged translational temperature, $T^{\rm (t)}_{\alpha}/T$, and rotational temperature, $T^{\rm (r)}_{\alpha}/T$.
	\item time series of the VACF, $C_{{\mb U}_{\alpha}}(t)$, and AVACF $C_{{\bm \varOmega}_{\alpha}}(t)$.  
	
	\item the mean squared displacement (MSD), $\langle \Delta {\mb X}_{\alpha}^{2}(t) \rangle$. \\
\end{enumerate} 

\noindent \textit{Rows from top to bottom} correspond to the above mentioned measures computed for the  \\

\begin{enumerate}[(i) ]
	\item translational component along the $x$ direction,
	\item translational component along the $y$ direction,
	\item translational component along the $z$ direction,
	\item rotational component along the $1$ direction,
	\item rotational component along the $2$ direction, and
	\item rotational component along the $3$ direction.
\end{enumerate}

\begin{figure}
	\centering
	\includegraphics[width=13cm,clip]{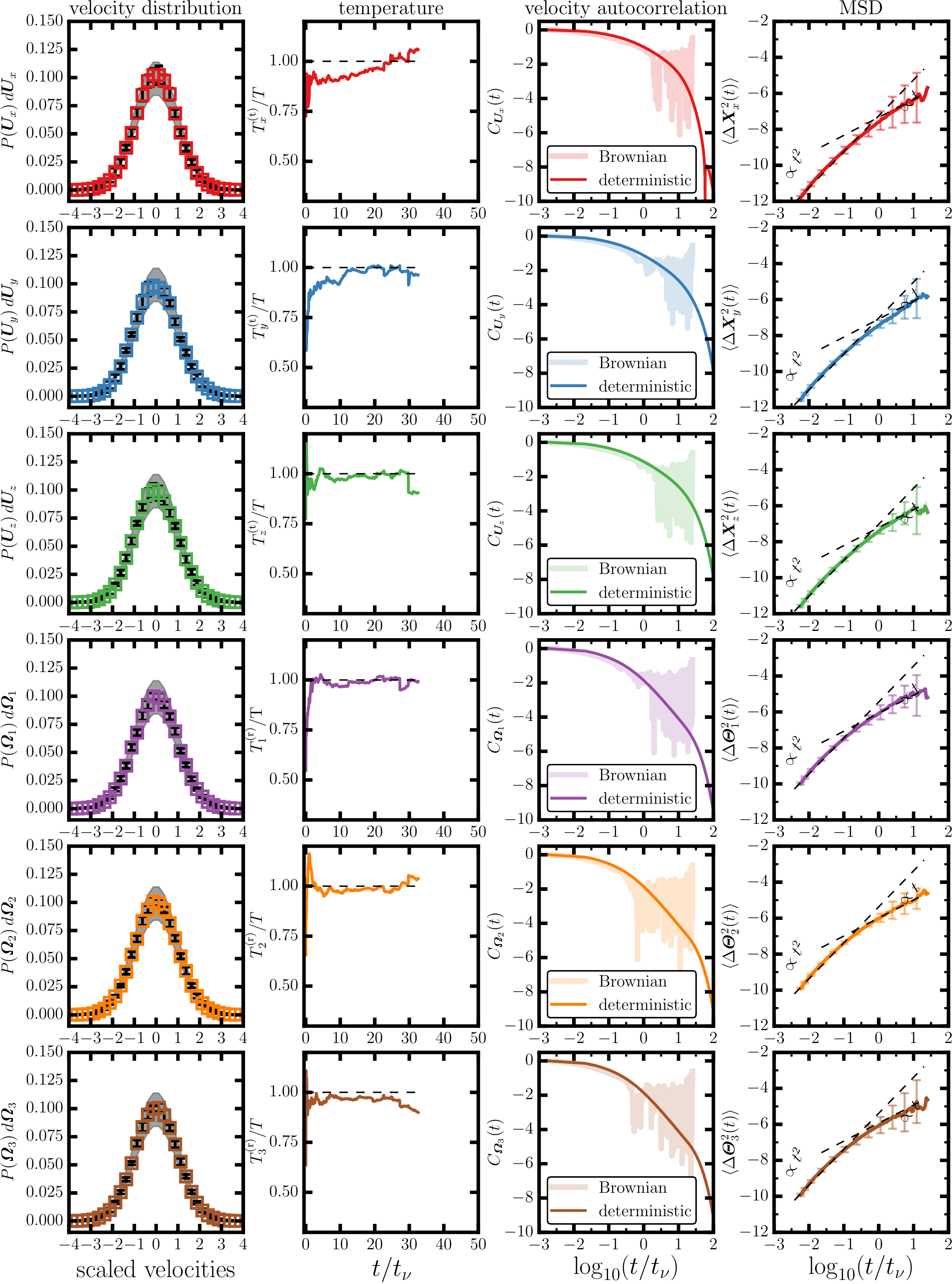}
	\caption{\label{fig:asp0p5-CT} Oblate ellipsoid, with $\varepsilon=0.5$ and $a=288.4$ nm, at the center of a cylindrical tube, with $\widetilde{h}=7.667$.}
\end{figure}

\begin{figure}
	\centering
	\includegraphics[width=14cm,clip]{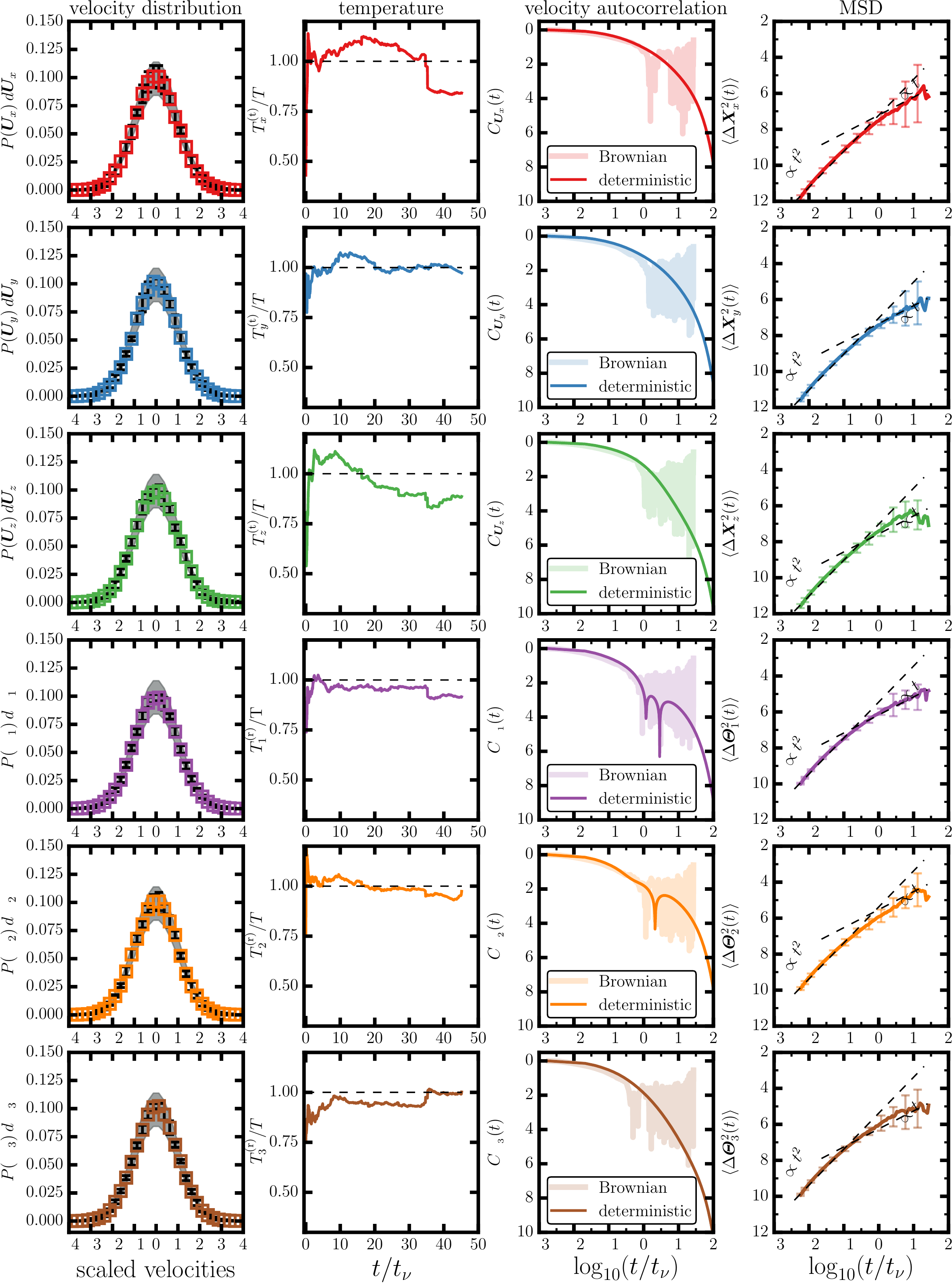}
	\caption{\label{fig:asp0p5-NW}Oblate ellipsoid, with $\varepsilon=0.5$ and $a=288.4$ nm, placed close to the wall of a cylindrical tube, with $\widetilde{h}=1$.}
\end{figure}

\begin{figure}
	\centering
	\includegraphics[width=14cm,clip]{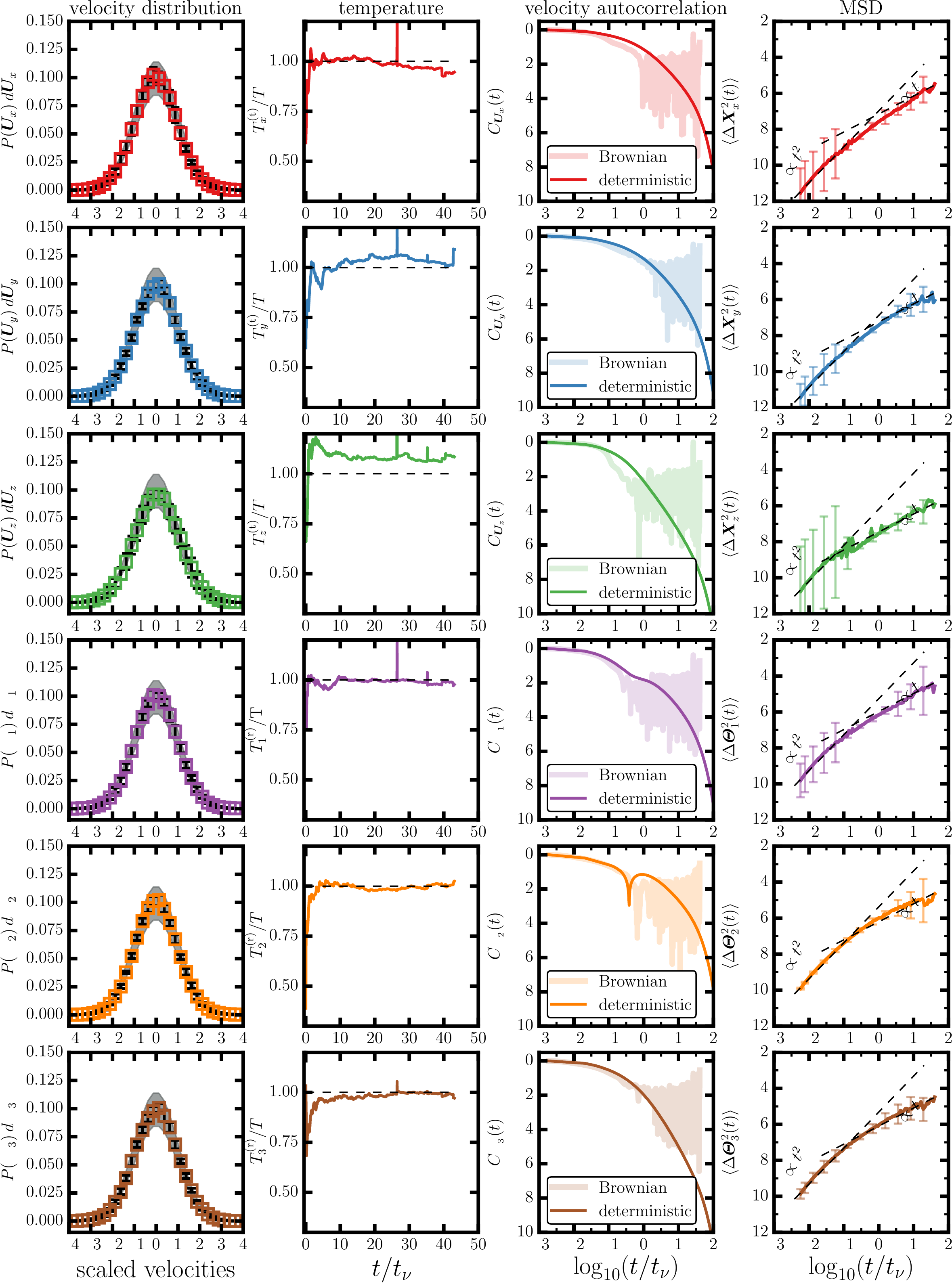}
	\caption{\label{fig:asp0p5-LUB} Oblate ellipsoid, with $\varepsilon=0.5$ and $a=288.4$ nm, placed in the lubrication layer of a cylindrical tube, with $\widetilde{h}=0.2$.}
\end{figure}

\begin{figure}
	\centering
	\includegraphics[width=14cm,clip]{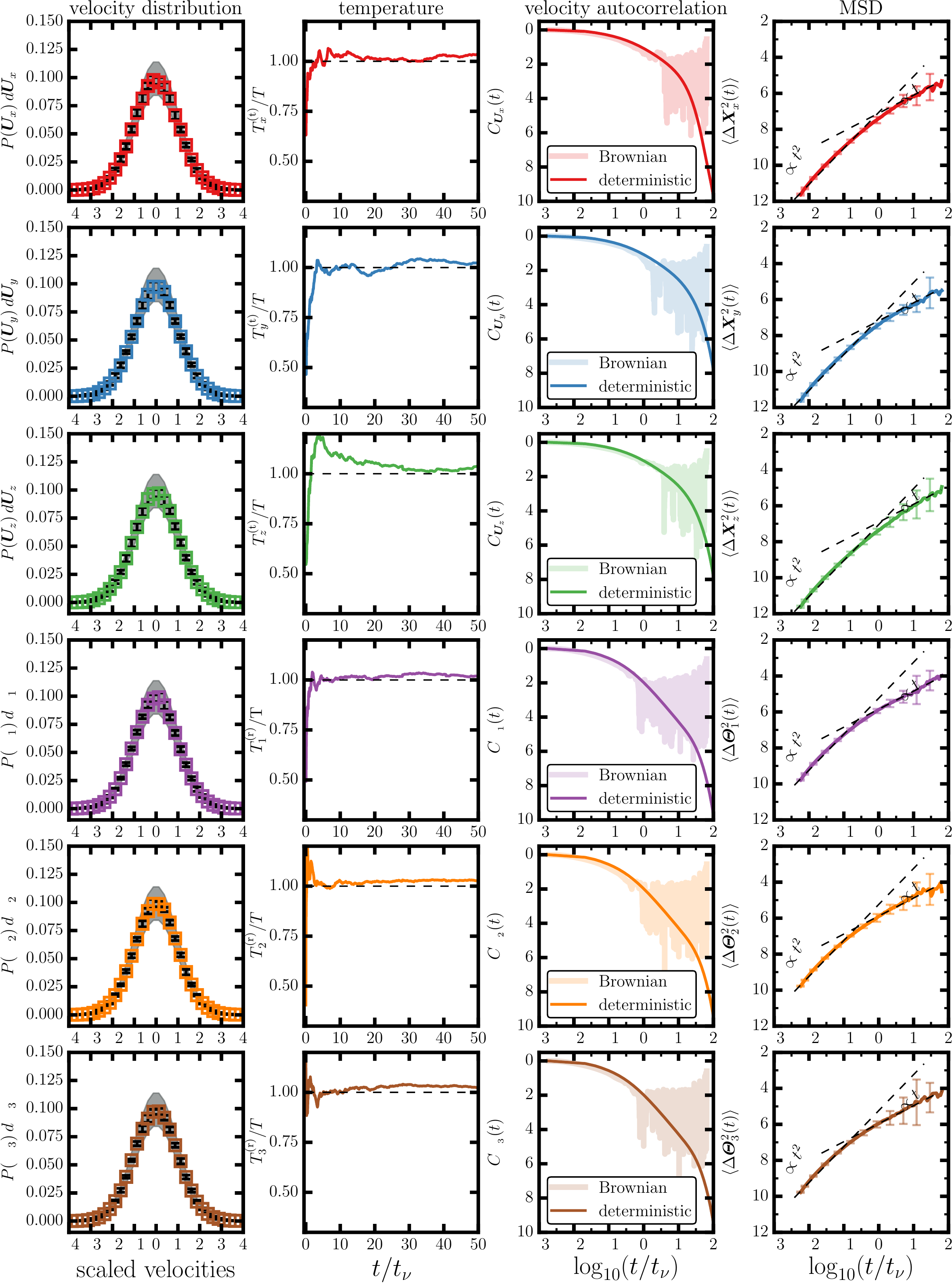}
	\caption{\label{fig:asp1p0-CT} Spherical particle, with $\varepsilon=1.0$ and $a=457.9$ nm, at the center of a cylindrical tube, with $\widetilde{h}=7.962$.}
\end{figure}

\begin{figure}
	\centering
	\includegraphics[width=14cm,clip]{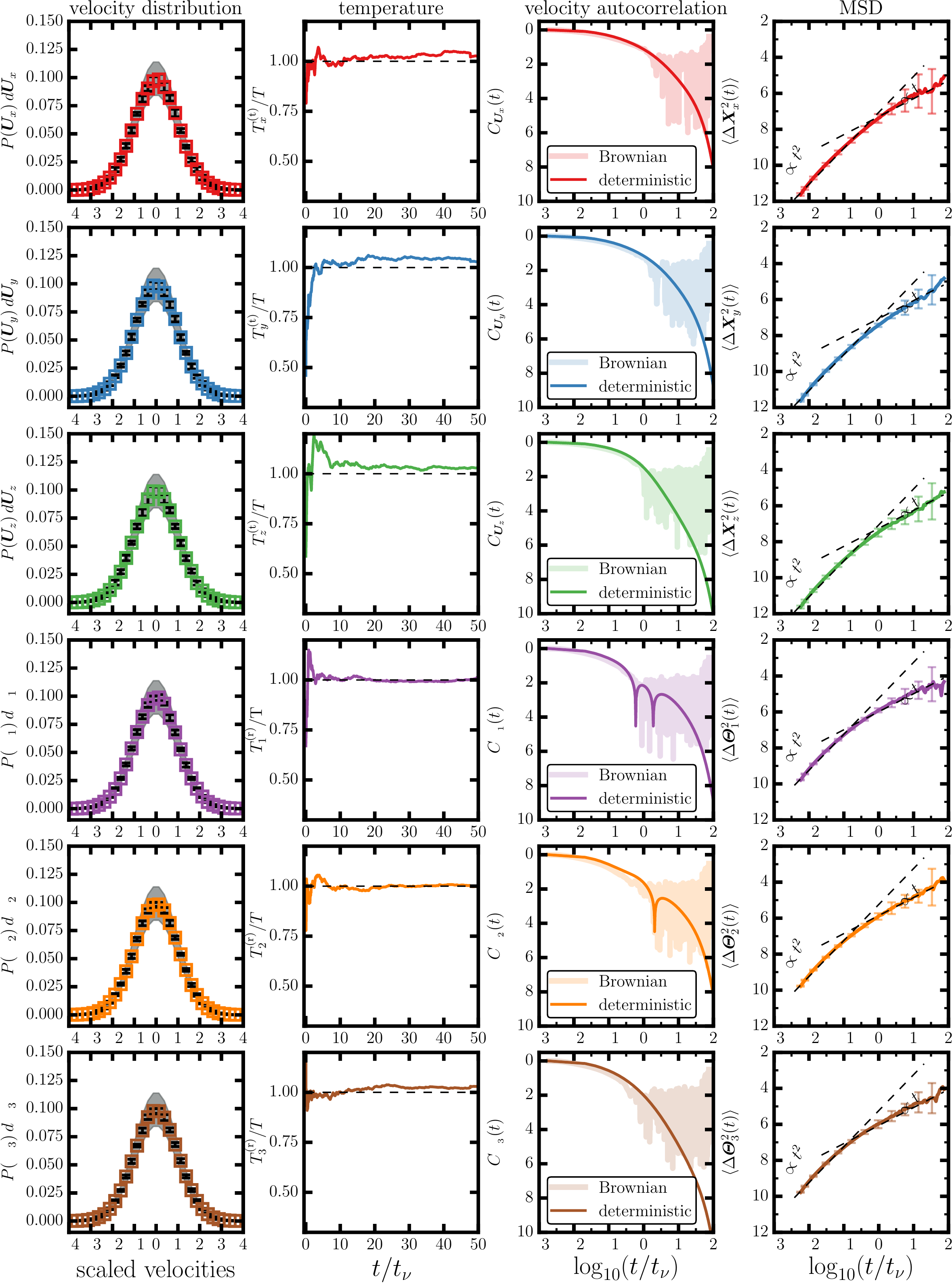}
	\caption{\label{fig:asp1p0-NW} Spherical particle, with $\varepsilon=1.0$ and $a=457.9$ nm, placed close to the bounding wall of a cylindrical tube, with $\widetilde{h}=1$.}
\end{figure}

\begin{figure}
	\centering
	\includegraphics[width=14cm,clip]{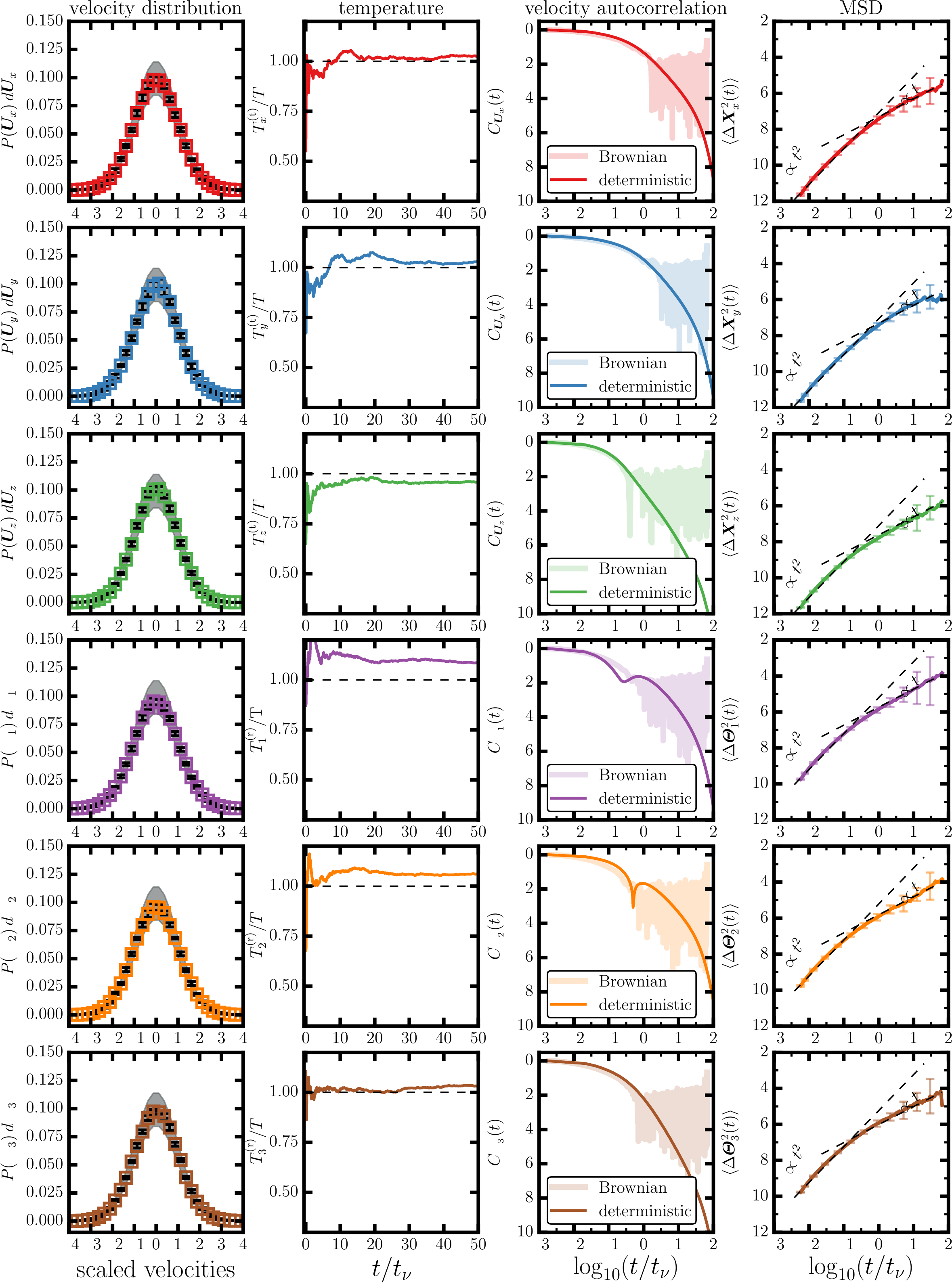}
	\caption{\label{fig:asp1p0-LUB} Spherical particle, with $\varepsilon=1.0$ and $a=457.9$ nm, placed in the lubrication layer of a cylindrical tube, with $\widetilde{h}=0.2$.}
\end{figure}

\begin{figure}
	\centering
	\includegraphics[width=14cm,clip]{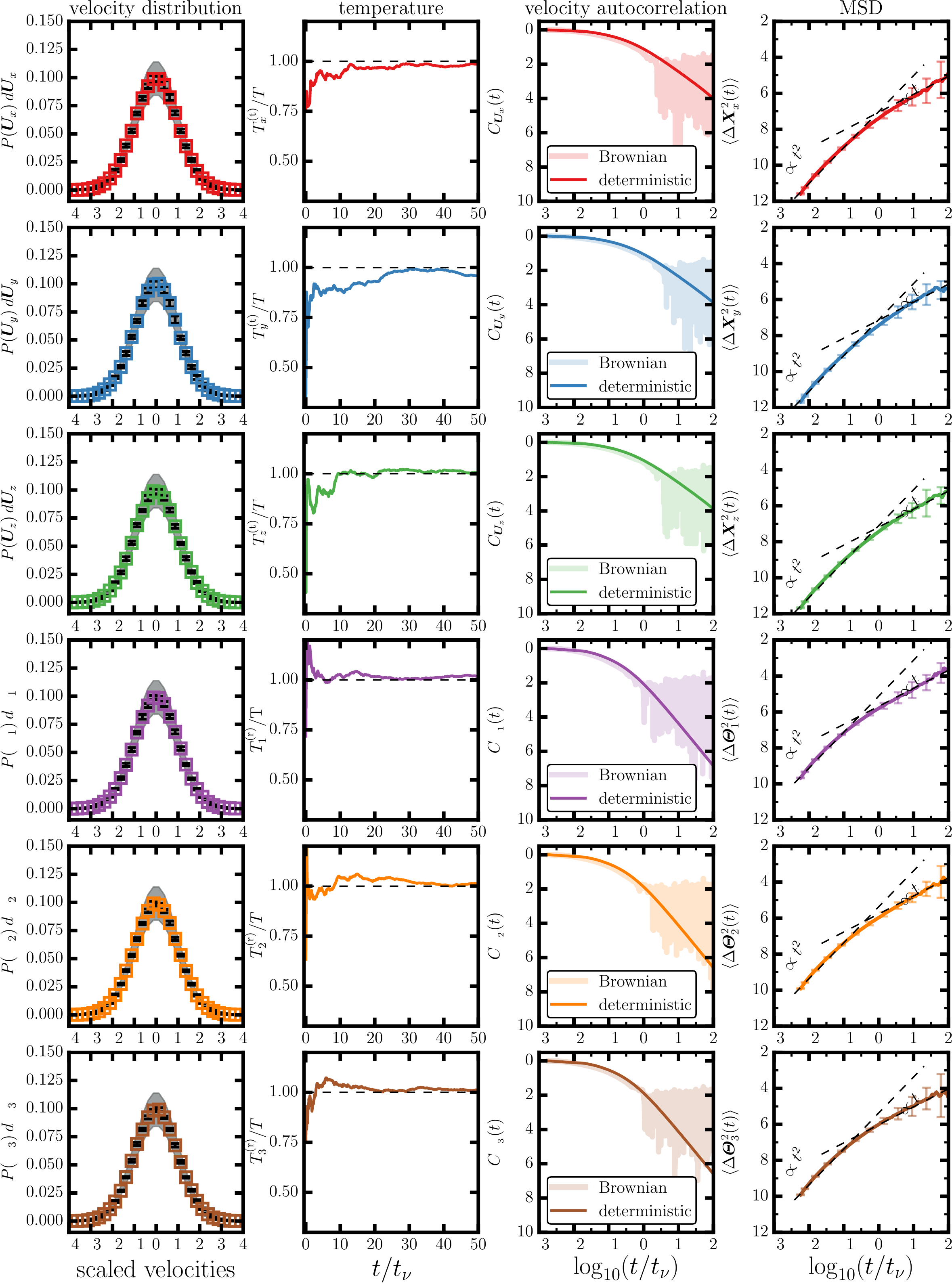}
	\caption{\label{fig:asp1p5-CT} Spherical particle, with $\varepsilon=1.5$ and $a=600$ nm, at the center of a cylindrical tube, with $\widetilde{h}=11.5$.}
\end{figure}

\begin{figure}
	\centering
	\includegraphics[width=14cm,clip]{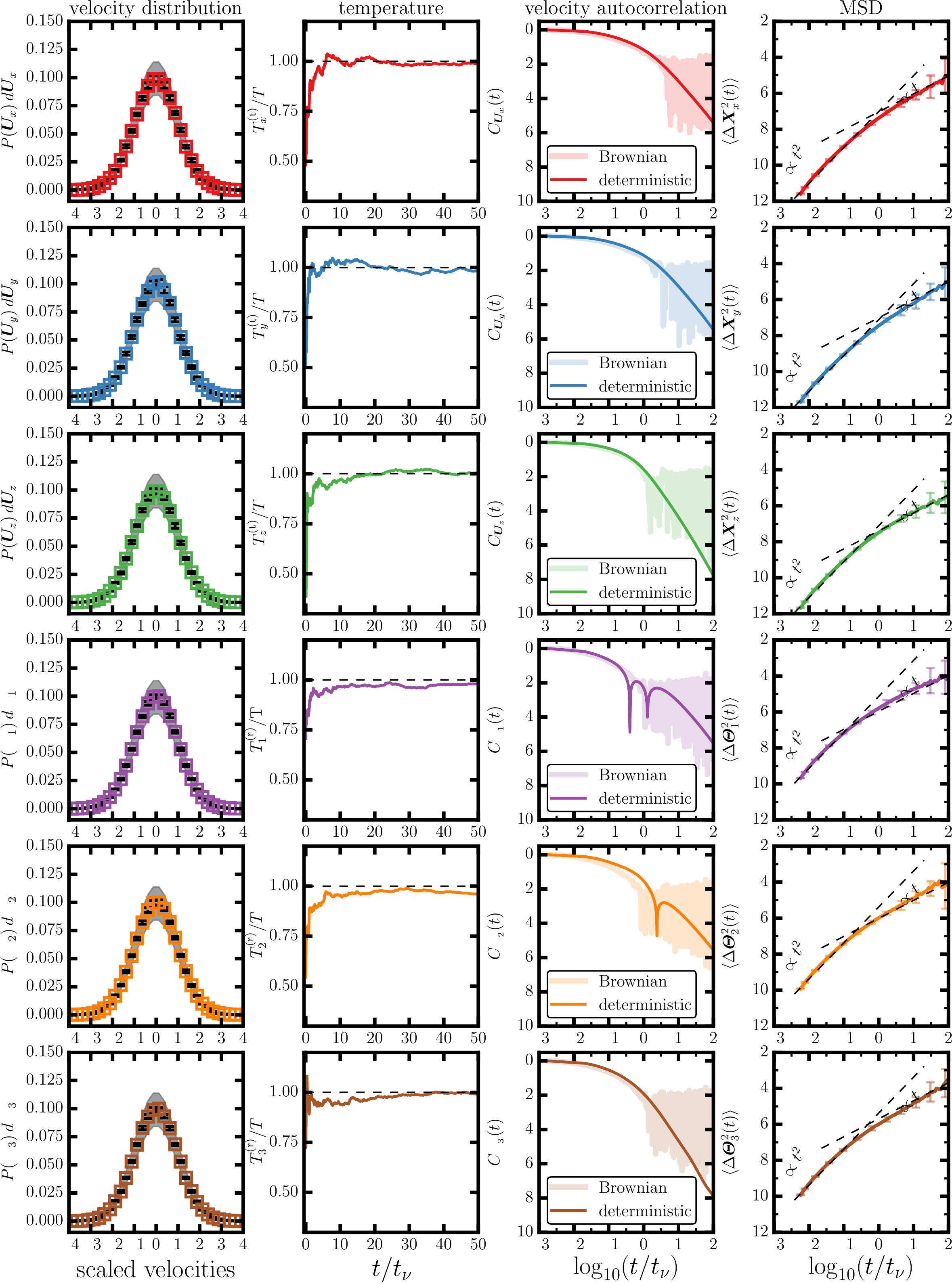}
	\caption{\label{fig:asp1p5-NW} Spherical particle, with $\varepsilon=1.5$ and $a=600$ nm, placed close to the bounding wall of a cylindrical tube, with $\widetilde{h}=1$.}
\end{figure}

\begin{figure}
	\centering
	\includegraphics[width=14cm,clip]{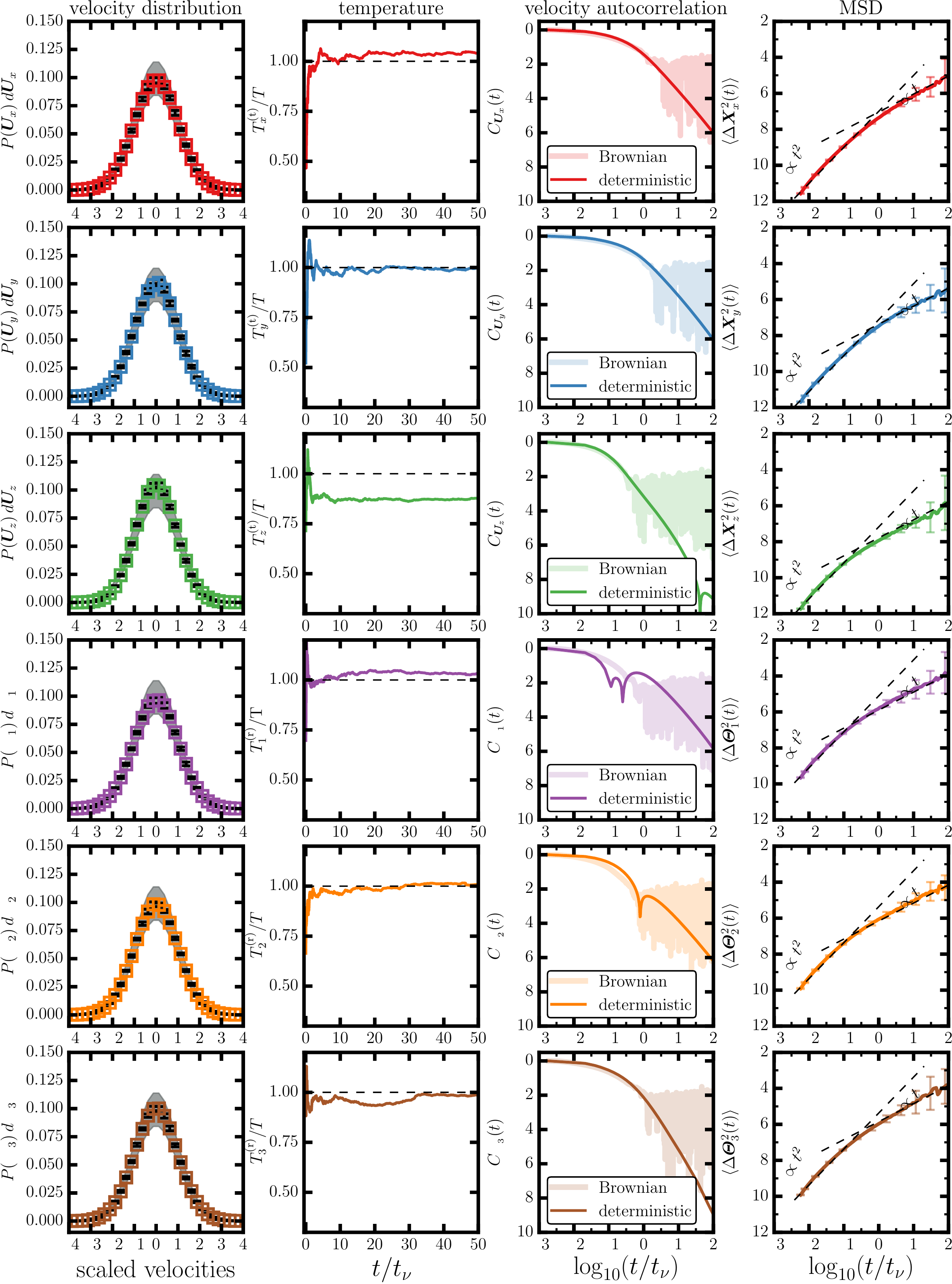}
	\caption{\label{fig:asp1p5-LUB} Spherical particle, with $\varepsilon=1.5$ and $a=600$ nm, placed in the lubrication layer of a cylindrical tube, with $\widetilde{h}=0.2$.}
\end{figure}

\begin{figure}
	\centering
	\includegraphics[width=14cm,clip]{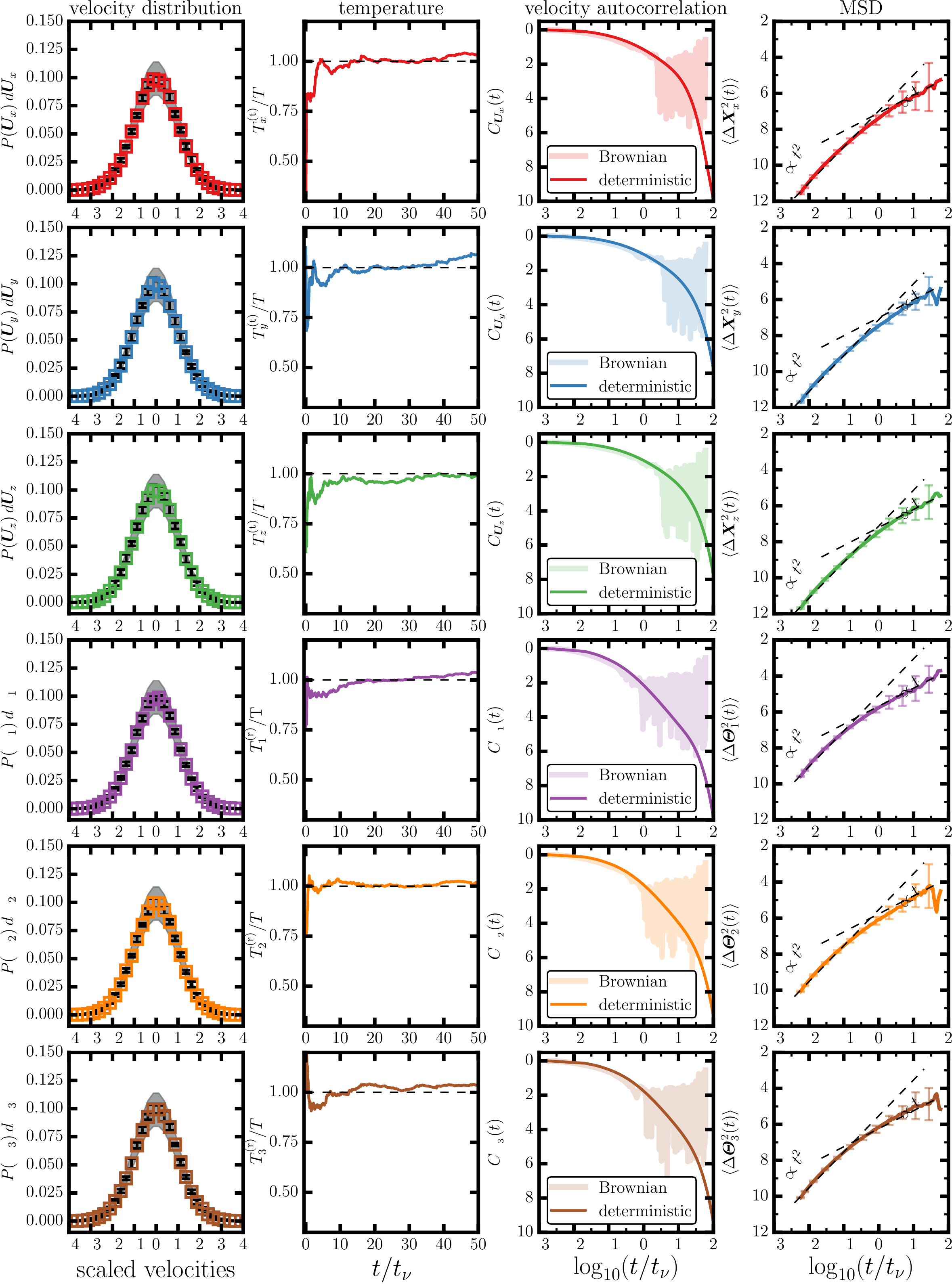}
	\caption{\label{fig:asp2p0-CT} Prolate ellipsoid, with $\varepsilon=2.0$ and $a=726.8$ nm, at the center of a cylindrical tube, with $\widetilde{h}=12.75$.}
\end{figure}

\begin{figure}
	\centering
	\includegraphics[width=14cm,clip]{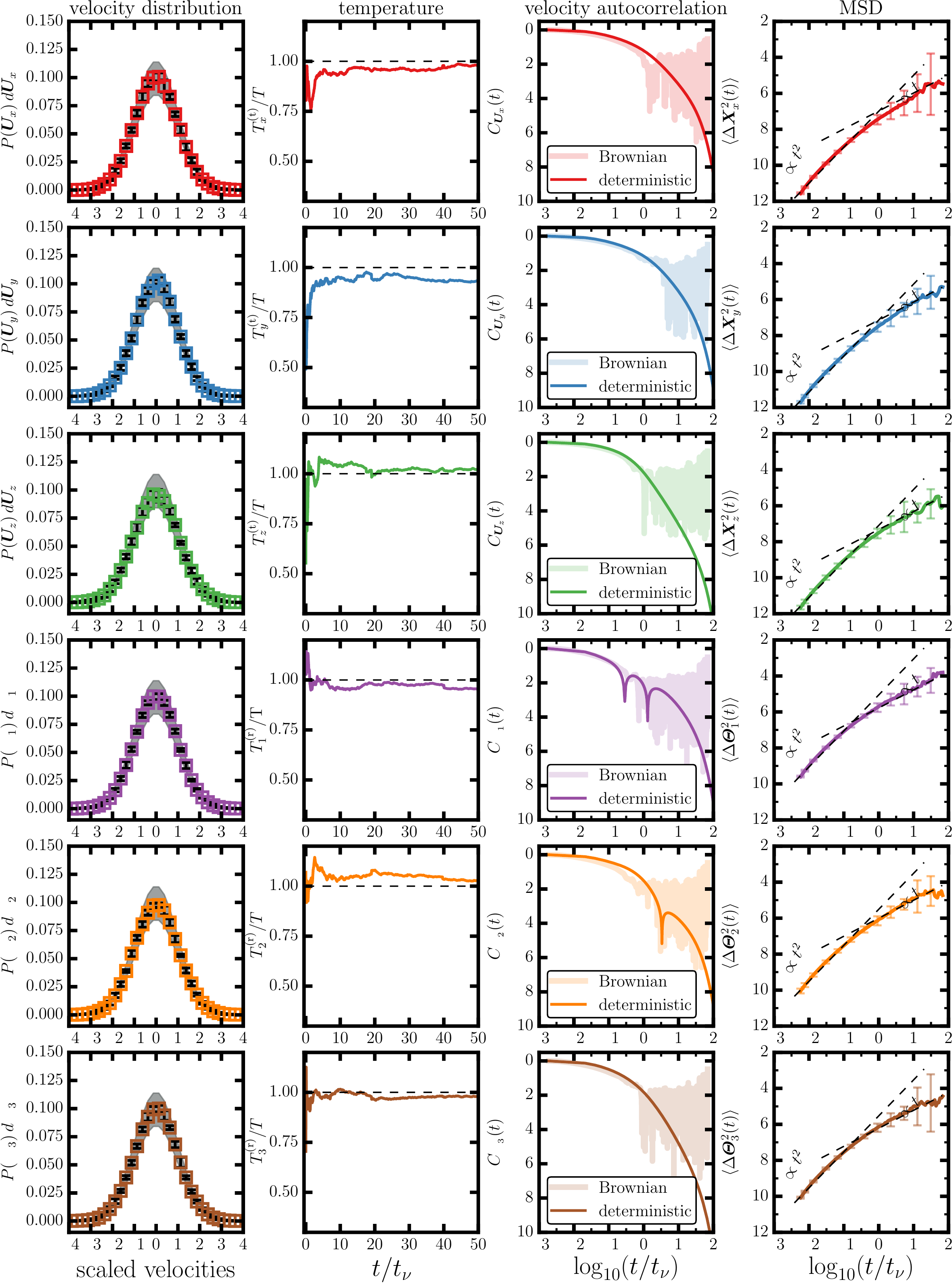}
	\caption{\label{fig:asp2p0-NW} Prolate ellipsoid, with $\varepsilon=2.0$ and $a=726.8$ nm, placed in the near wall region of a cylindrical tube, with $\widetilde{h}=1$.}
\end{figure}

\begin{figure}
	\centering
	\includegraphics[width=14cm,clip]{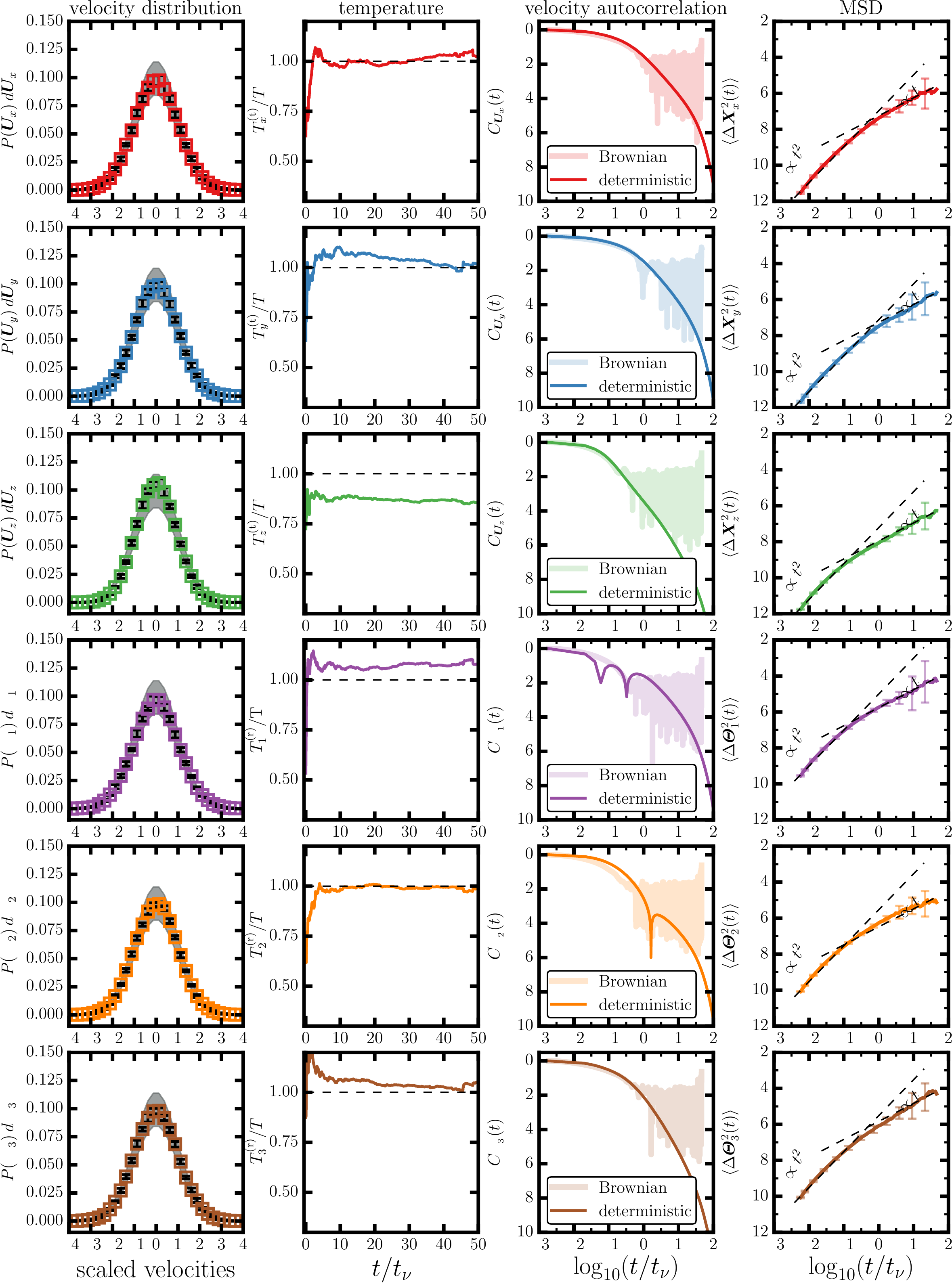}
	\caption{\label{fig:asp2p0-LUB} Prolate ellipsoid, with $\varepsilon=2.0$ and $a=726.8$ nm, placed in the lubrication layer of a cylindrical tube, with $\widetilde{h}=0.2$.}
\end{figure}

\begin{figure}
	\centering
	\includegraphics[width=14cm,clip]{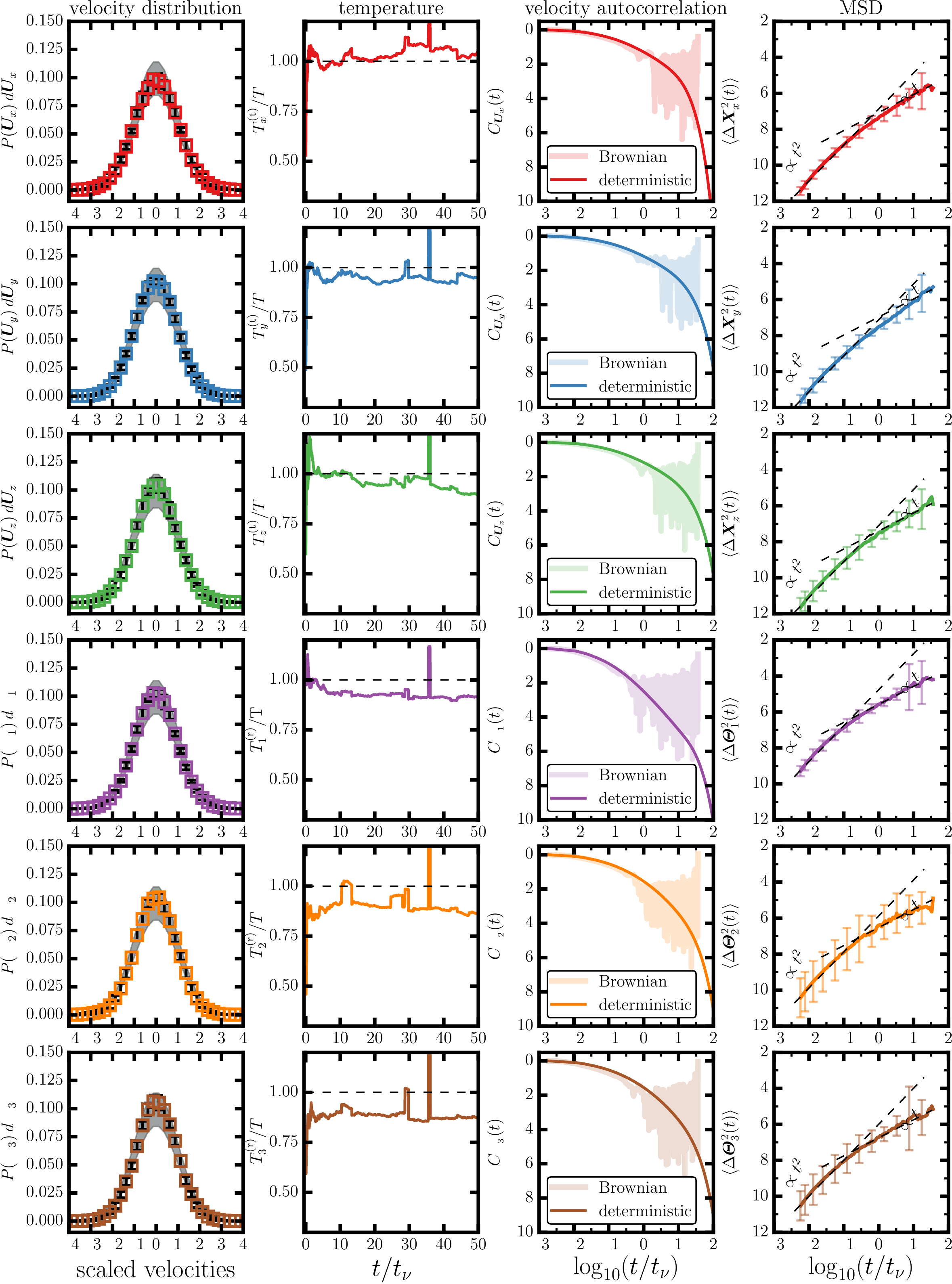}
	\caption{\label{fig:asp5p0-CT} Prolate ellipsoid, with $\varepsilon=5.0$ and $a=1.3389\mum$, at the center of a cylindrical tube, with $\widetilde{h}=17.67$.}
\end{figure}

\begin{figure}
	\centering
	\includegraphics[width=14cm,clip]{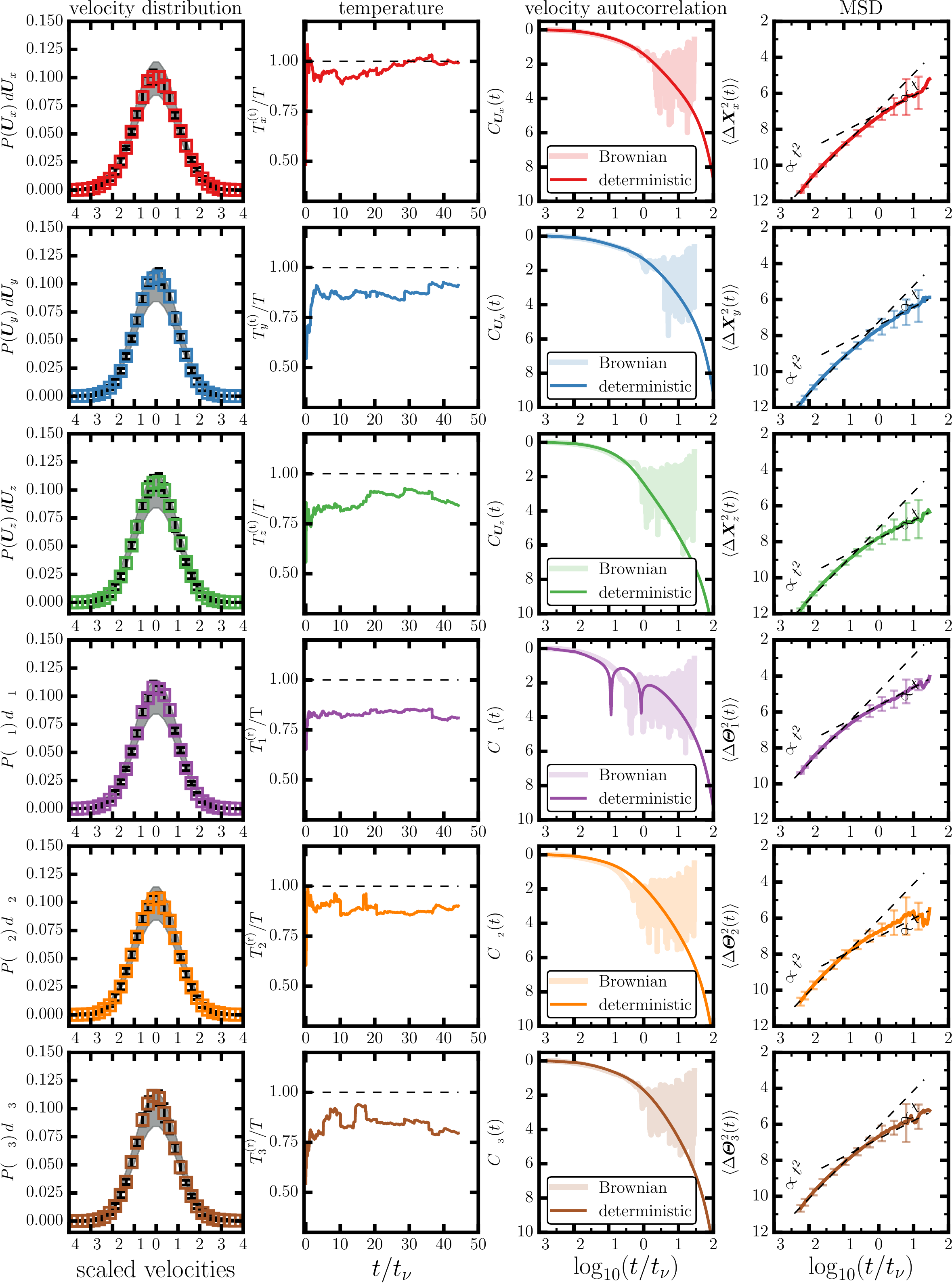}
	\caption{\label{fig:asp5p0-NW} Prolate ellipsoid, with $\varepsilon=5.0$ and $a=1.3389\mum$, in the near wall region of a cylindrical tube, with $\widetilde{h}=1$.}
\end{figure}

\begin{figure}
	\centering
	\includegraphics[width=14cm,clip]{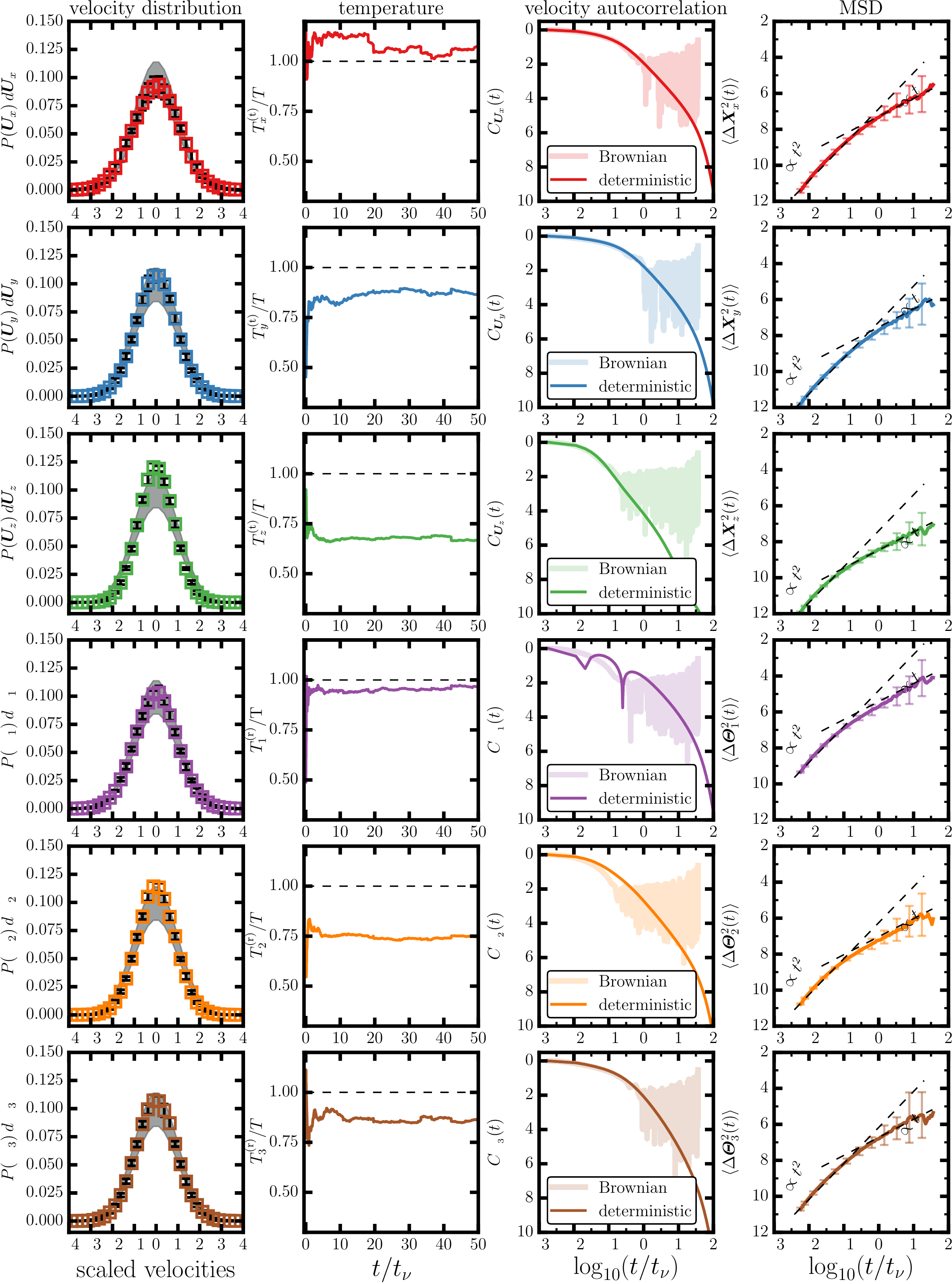}
	\caption{\label{fig:asp5p0-LUB} Prolate ellipsoid, with $\varepsilon=5.0$ and $a=1.3389\mum$, placed in the lubrication layer of a cylindrical tube, with $\widetilde{h}=0.2$.}
\end{figure}

~\\
\noindent The parameters corresponding to Figs.~\ref{fig:asp0p5-CT}-~\ref{fig:asp5p0-LUB} are shown in Table.~\ref{table:param}.

\begin{table}
	\centering
	\begin{tabular}{ccccccc}
		Figure & & $\varepsilon$ & & $a$ (in nm) & & $\widetilde{h}$ \\
		& & & & & &  \\
		Fig.~\ref{fig:asp0p5-CT} & & 0.5 & & 288.4 & & 7.667 \\
		Fig.~\ref{fig:asp0p5-NW} & & 0.5 & & 288.4 & & 1.0 \\
		Fig.~\ref{fig:asp0p5-LUB} & & 0.5 & & 288.4 & & 0.2 \\
		& & & & & &  \\
		Fig.~\ref{fig:asp1p0-CT} & & 1.0 & & 457.9 & & 7.962 \\
		Fig.~\ref{fig:asp1p0-NW} & & 1.0 & & 457.9 & & 1.0 \\
		Fig.~\ref{fig:asp1p0-LUB} & & 1.0 & & 457.9 & & 0.2 \\
		& & & & & &  \\
		Fig.~\ref{fig:asp1p5-CT} & & 1.5 & & 600.0 & & 11.5 \\
		Fig.~\ref{fig:asp1p5-NW} & & 1.5 & & 600.0 & & 1.0 \\
		Fig.~\ref{fig:asp1p5-LUB} & & 1.5 & & 600.0 & & 0.2 \\		
		& & & & & &  \\
		Fig.~\ref{fig:asp2p0-CT} & & 2.0 & & 726.8 & & 12.75 \\
		Fig.~\ref{fig:asp2p0-NW} & & 2.0 & & 726.8 & & 1.0 \\
		Fig.~\ref{fig:asp2p0-LUB} & & 2.0 & & 726.8 & & 0.2 \\
		& & & & & &  \\
		Fig.~\ref{fig:asp5p0-CT} & & 5.0 & & 1338.9 & & 17.67 \\
		Fig.~\ref{fig:asp5p0-NW} & & 5.0 & & 1338.9 & & 1.0 \\
		Fig.~\ref{fig:asp5p0-LUB} & & 5.0 & & 1338.9 & & 0.2 \\				
	\end{tabular}
	\caption{\label{table:param} Parameters for the particle shape and wall proximity used in the fluctuating hydrodynamics calculations for ellipsoidal NCs in a cylindrical tube with $D=5\mum$ and $L=40\mum$.}
\end{table}

\clearpage
\newpage

\section{VACF and AVACF as a function of NC aspect ratio}\label{sec:vacf}
\setcounter{figure}{0}
\renewcommand{\thefigure}{\ref{sec:vacf}.{\color{blue}\arabic{figure}}}

\begin{figure}
	\centering
	\includegraphics[width=14cm,clip]{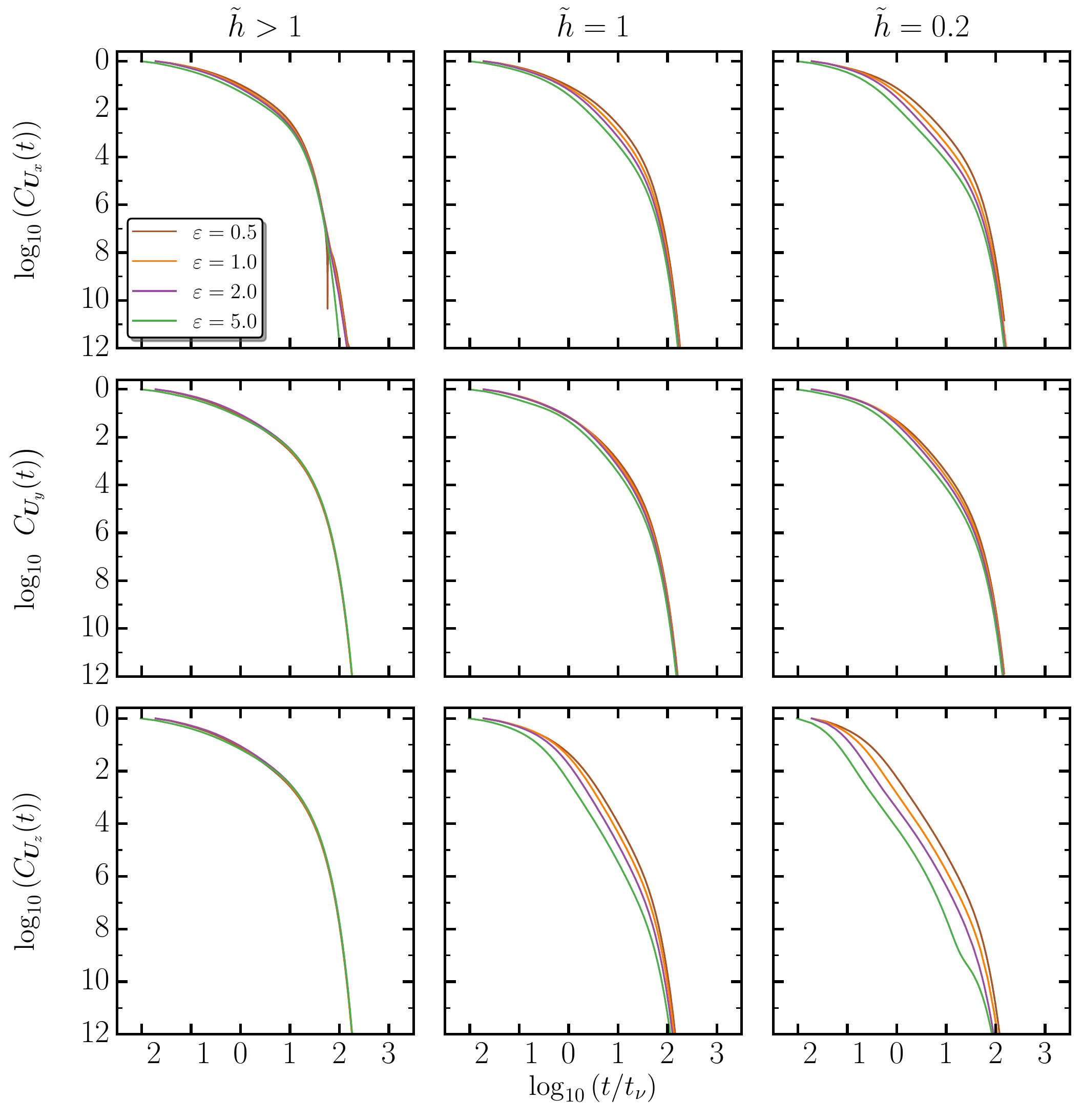}
	\caption{VACF for NCs with $\varepsilon=0.5,1.0,2.0,5.0$ placed at $\widetilde{h}>1$ (see Table 1 in main text for exact values), $\widetilde{h}=1$, and $\widetilde{h}=0.2$, in a tube with $D=5\mum$ and $L=40\mum$.}
	\label{fig:D5vacf}
\end{figure}

\begin{figure}
	\centering
	\includegraphics[width=14cm,clip]{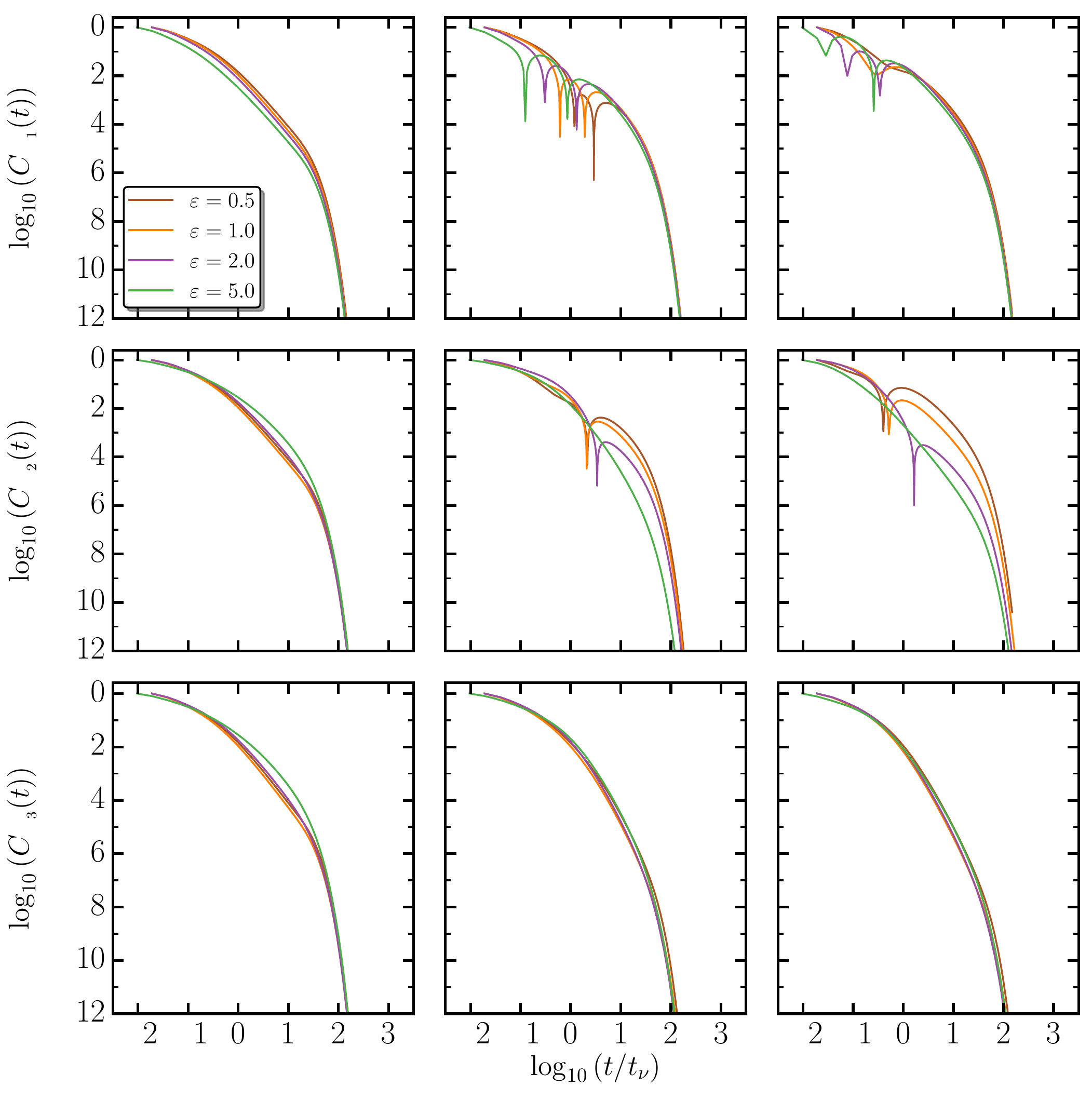}
	\caption{AVACF for NCs with $\varepsilon=0.5,1.0,2.0,5.0$ placed at $\widetilde{h}>1$ (see Table 1 in main text for exact values), $\widetilde{h}=1$, and $\widetilde{h}=0.2$, in a tube with $D=5\mum$ and $L=40\mum$.}
	\label{fig:D5avacf}
\end{figure}

In addition to its shape, the VACF and AVACF for an ellipsoidal NC are also strongly influenced by wall mediated hydrodynamic interactions when the NC is in proximity to the wall. A comparison of the VACF and AVACF are shown in Figs.~\ref{fig:D5vacf} and \ref{fig:D5avacf}, respectively, for NCs with $\varepsilon$=0.5, 1.0, 2.0, and 5.0 placed at $\widetilde{h}>1.0$, $\widetilde{h}=1.0$, and $\widetilde{h}=0.2$ inside a cylindrical tube with $D=5\mum$ and $L=40\mum$. These results have been used to compute the translational and rotational diffusivities presented in Figs.13 and 14 in the main text.

\begin{figure}
	\centering
	\includegraphics[width=14cm,clip]{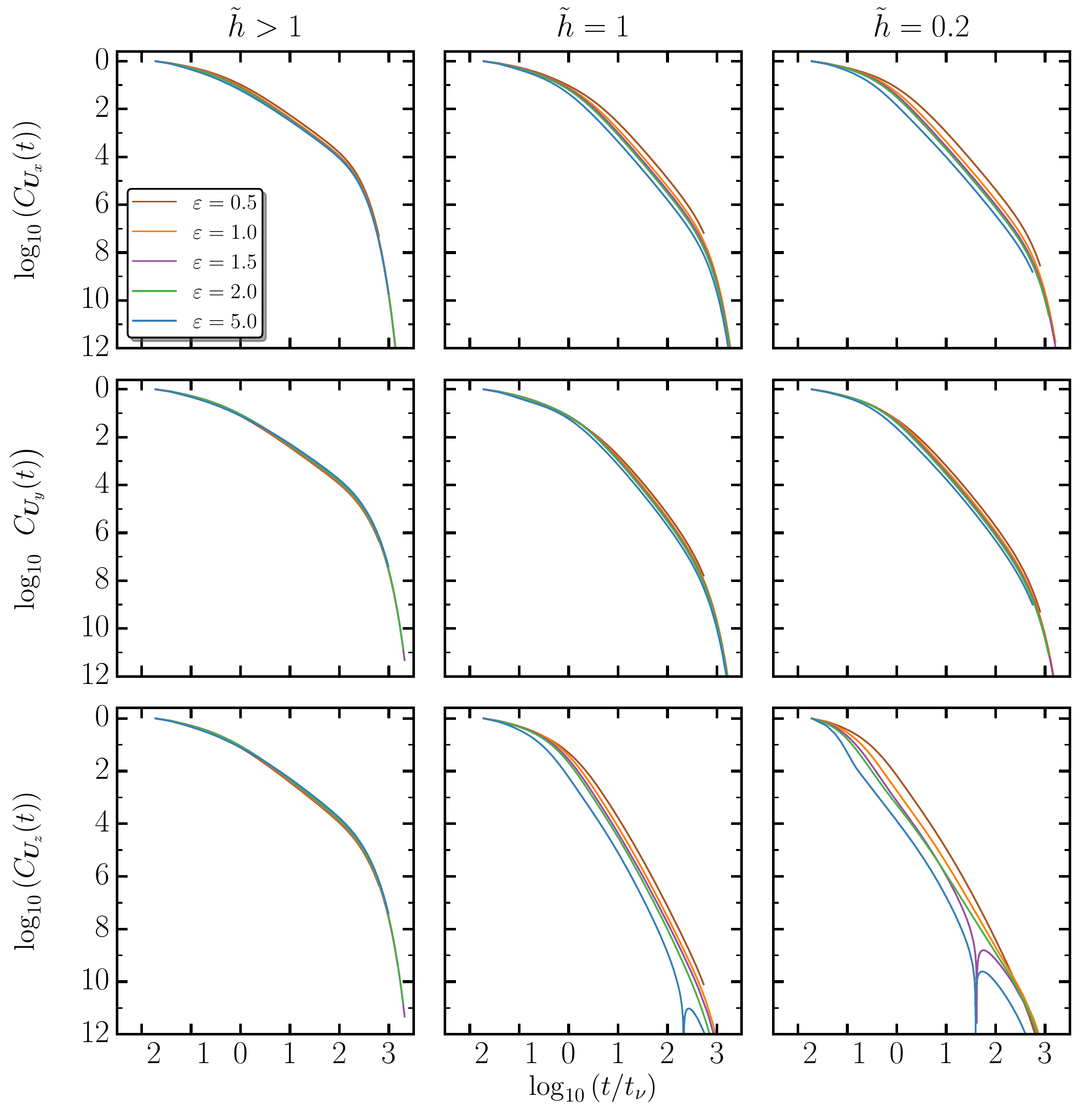}
	\caption{VACF for NCs with $\varepsilon=0.5,1.0,2.0,5.0$ placed at $\widetilde{h}>1$ (see Table 1 in main text for exact values), $\widetilde{h}=1$, and $\widetilde{h}=0.2$, in a tube with $D=20\mum$ and $L=40\mum$.}
	\label{fig:D20vacf}
\end{figure}

\begin{figure}
	\centering
	\includegraphics[width=14cm,clip]{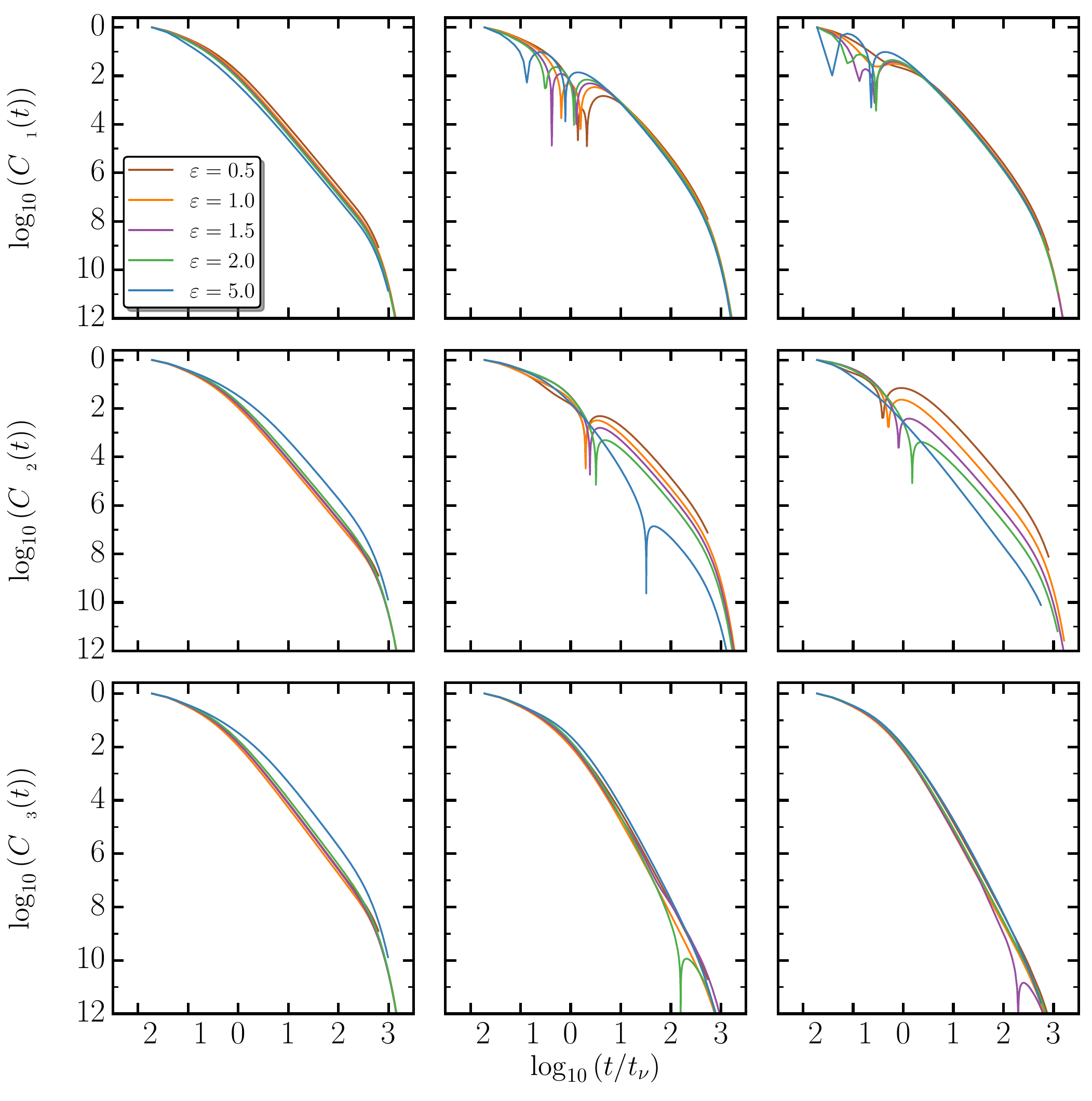}
	\caption{AVACF for NCs with $\varepsilon=0.5,1.0,2.0,5.0$ placed at $\widetilde{h}>1$ (see Table 1 in main text for exact values), $\widetilde{h}=1$, and $\widetilde{h}=0.2$, in a tube with $D=20\mum$ and $L=40\mum$.}		\label{fig:D20avacf}
\end{figure}

Furthermore, the curvature of the bounding wall may also alter the decay of the VACF and AVACF. This is shown in Figs.~\ref{fig:D20vacf} and ~\ref{fig:D20avacf} for for NCs with $\varepsilon$=0.5, 1.0, 2.0 and 5.0 placed at $\widetilde{h}>1.0$, $\widetilde{h}=1.0$, and $\widetilde{h}=0.2$ inside a cylindrical tube with $D=20\mum$ and $L=40\mum$.

\clearpage
\newpage
\section{Computing static mobilities using the Towing method} \label{sec:mobcalc}
\setcounter{figure}{0}
\renewcommand{\thefigure}{\ref{sec:mobcalc}.{\color{blue}\arabic{figure}}}
\begin{figure} 
	\centering
	\includegraphics{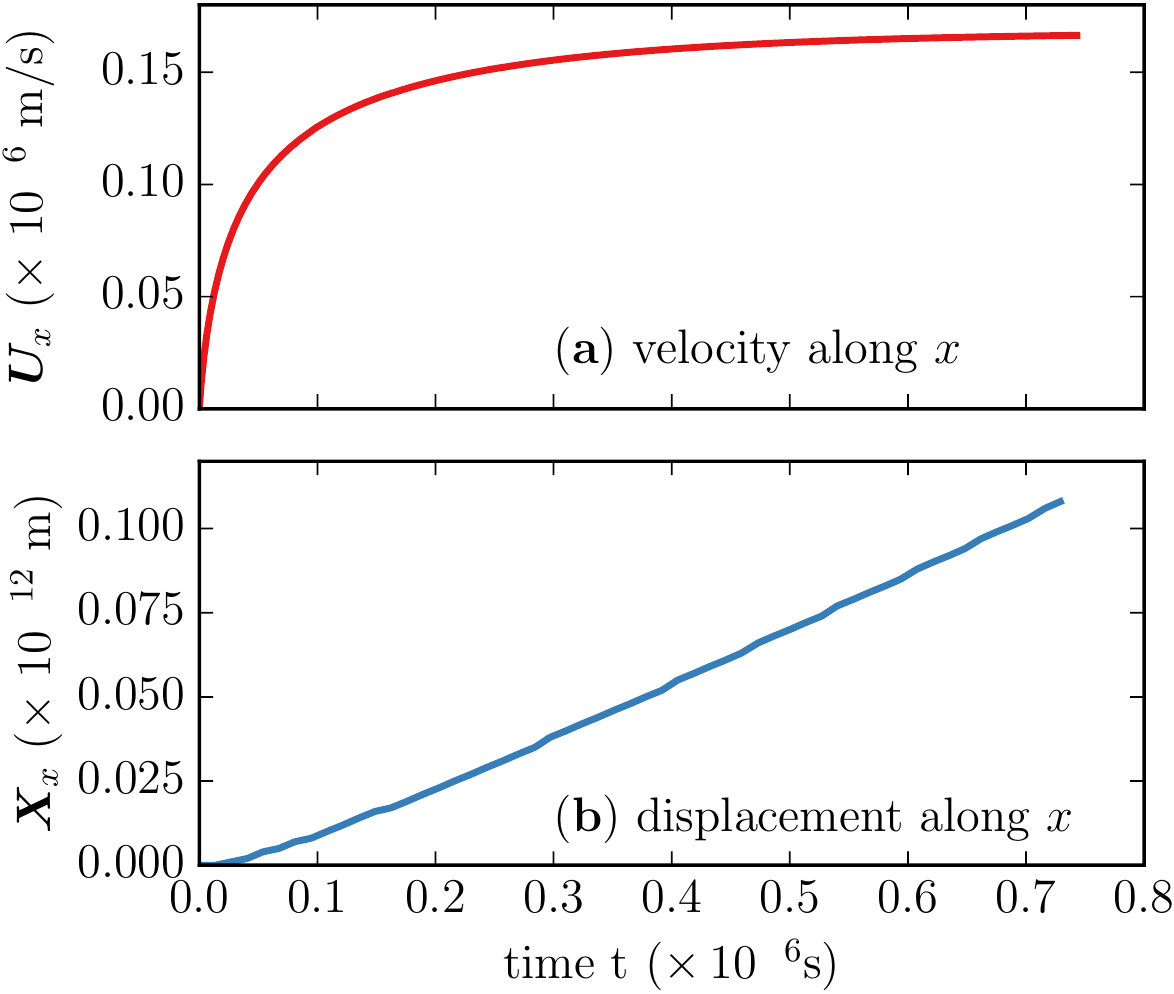}
	\caption{The time evolution of (a) the velocity and  (b) the displacement of an ellipsoid with $\varepsilon=1$ and $a=500$ nm. The particle is placed at the center of a tube of diameter $D=5\mum$ and is dragged along the $x$ direction by applying a constant force ${\bm G}_x=10^{-18}$ N.}
	\label{fig_velposMB}
\end{figure}

Here, we present a computationally inexpensive method to compute mobility of a neutrally buoyant ellipsoidal NC ($\rhop=\rhof$) by assessing the dynamics of the particle in response to a weak applied force. These calculations are performed with the weak formulation by setting the random stress tensor $\tensor{S}$, given in eqn. 3.3 in the main text, to zero. 

The mobility of a nanoparticle $\vindex{\cal M}{\alpha}$, at a radial position $r$ and inclination angle $\theta$ as shown in Fig. 2(c), along any given direction $\alpha$, is computed from its steady state velocity $\bm{U}$ in response to an externally imposed body force ${\bm G}$ (see eqn. 3.6) acting at its center of mass as ${\cal M}_{\alpha}=\vindex{\bm{U}}{\alpha}/\vindex{\bm G}{\alpha}$. Here, $\vindex{\bm U}{\alpha}$ and $\vindex{\bm G}{\alpha}$ are the components of the velocity and force along the $\alpha$ direction. In all the  mobility calculations presented here, the force $\vindex{\bm G}{\alpha}$ is chosen to be $1\, \SI{}{pg}\,\mu\SI{}{m}\,\SI{}{s}^{-2}$. Figs.~\ref{fig_velposMB}(a) and (b) show the time evolution of the $x$ components of the velocity  and displacement, used in computing ${\cal M}_{x}$ for a spherical particle (ellipsoid with $a=b=c$) of diameter $500$ nm, placed at the center of a circular tube, and subjected to a constant force $\vindex{\bm G}{x}=1\, \SI{}{pg}\,\mu\SI{}{m}\,\SI{}{s}^{-2} \equiv 10^{-18}\,\SI{}{N}$. It may be noted from these figures that the velocity reaches a steady state in a very short time ($\approx 10^{-6}\,\rm{ns}$). We ignore the initial transients and use the steady state value of $\vindex{\bm U}{x}$ to compute the mobility. It should also be noted that the net displacement of the particle in the time interval to reach steady state is only about $10^{-11} \mum$ which is negligible compared to the particle diameter ($500$ nm). This method which neglects the transients and only probes the linear response regime of the particle allows us to compute the its zero-freqency mobility at the desired location.

Now we consider comparisons with existing results in two cases: (i) ${\cal M}_x$ as a function of the aspect ratio (\asp) for an ellipsoid placed at the center of the cylindrical tube~\citep{happel1965low}, and (ii) ${\cal M}_y$ for an ellipsoid as a function of its separation ($h$) from the tube wall~\citep{HsuGanatos89}.

For an ellipsoid particle (with $b=c$, and $\theta=0\degree$) at the center of the tube ($r=0$), the analytical form of the translational mobility ${\cal M}_x^{\dagger}$ is given by~\citep{happel1965low}:
\begin{eqnarray}
\label{eqn_happel}
\vindex{{\cal M}}{x}^{\dagger} =\frac{1-\left(\dfrac{3d_{eq}}{8D}\right)\left(5.612-2.0211\left(\dfrac{a}{D}\right)^2-3.5431\left(\dfrac{c}{D}\right)^2 \right)}{3\pi\mu d_{\rm eq}}.
\end{eqnarray}

\begin{figure} 
	\centering
	\includegraphics[width=0.8\textwidth]{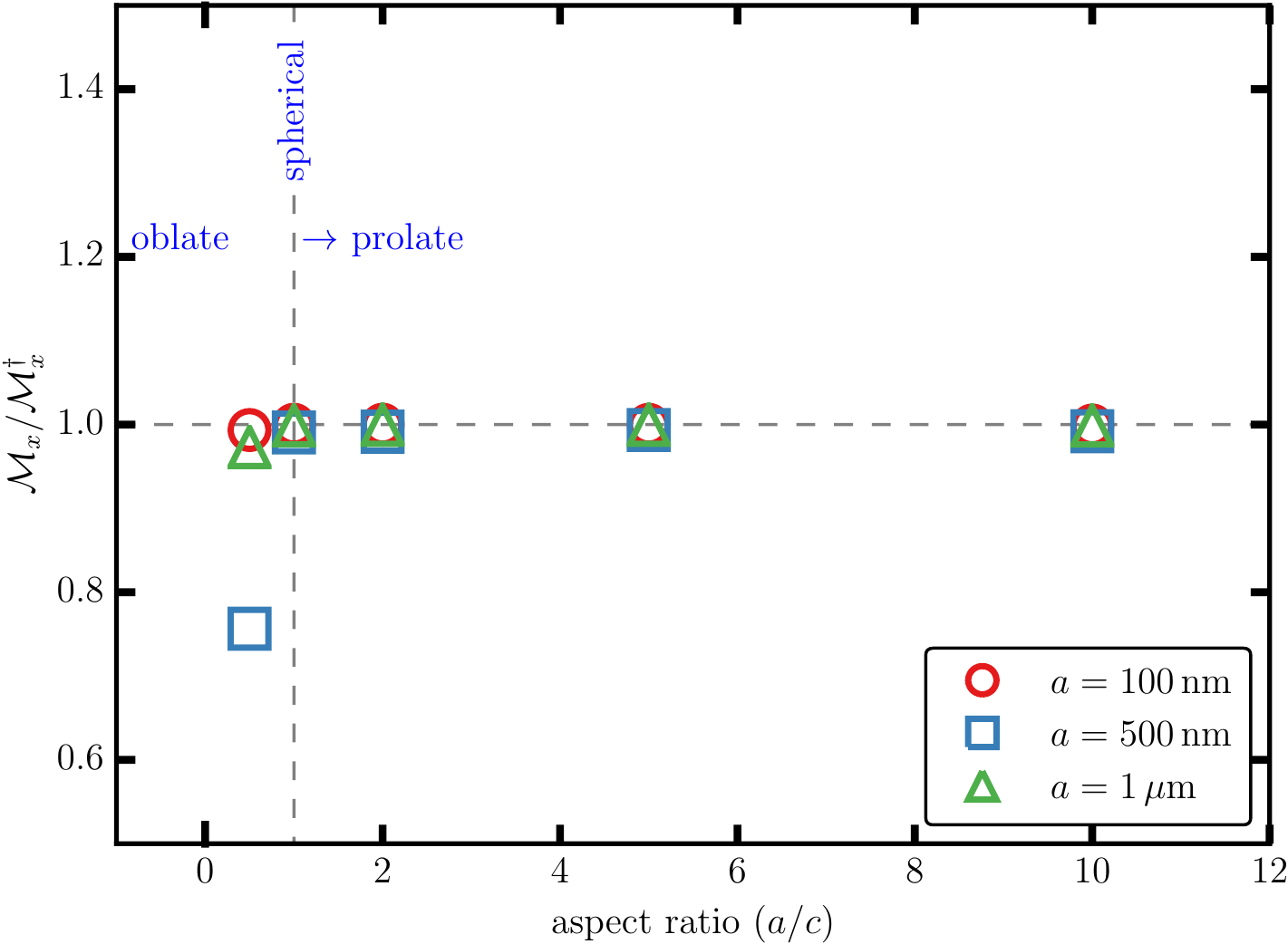}
	\caption{Comparison of the computed values of  $\vindex{{\cal M}}{x}$ to theoretical estimates based on  eqn. (\ref{eqn_happel}). Data shown as a function of the aspect ratio (\asp) for ellipsoidal particles, with $a=1\mum$, $500\,{\rm nm}$, and $100\,{\rm nm}$, placed at the center of a tube with diameter $D=5\mum$. The normalization factor ${\cal M}_x^\dagger$, for each particle size $a$, is chosen to be the analytical estimate (from eqn.~\eqref{eqn_happel})  for  the corresponding spherical particle ($a=b=c$). Mesh parameters used are $l_P=0.026 \mum$ and $l_W=0.524\mum$.}
	\label{fig_tubecenterMobility}
\end{figure}

Here, $d_{\rm eq}$ is the ``equivalent spherical diameter'' for the ellipsoid, whose values have been taken from table 5-11.1 in ~\cite{happel1965low}.  We compare the values of $\vindex{{\cal M}}{x}$ from our simulations to those evaluated from eqn.~\eqref{eqn_happel} for  particles with five different aspect ratios, ranging from $\aspp=0.5$ to $10$. We have chosen particles of three different sizes ($a=1\mum$, $500\,{\rm nm}$, and $100\, {\rm nm}$) and varied their aspect ratios by varying the values of $b$ and $c$. In all these calculations we fix the cylindrical vessel diameter to be $D=5\mum$. Fig.~\ref{fig_tubecenterMobility} shows  the ratio of $\vindex{{\cal M}}{x}$ to  $\vindex{{\cal M}}{x}^{\dagger}$ and it may be seen that the computed values of the mobility are in  excellent agreement with those  given by eqn.~\eqref{eqn_happel} and hence validate the computation for a particle situated away from the wall.


Next, we study the mobility of oblate ellipsoids with a fixed  aspect ratio $\aspp=0.5$, for $a=100$ nm, $500$ nm, and $1\mum$, as a function of the gap length $h$  between the  center of mass of the particle and the tube wall (see Fig.2(c) in the main manuscript). These studies have been performed by varying $h$ along the $y$ direction for an ellipsoidal particle whose axis of symmetry is also oriented along $y$ (i.e., $\theta=90\degree$). This study will validate the numerical scheme where the wall effects are important. Hsu and Ganatos~\citep{HsuGanatos89}  have previously reported a similar study  using the boundary integral approach to compute the mobility of an ellipsoidal particle as  a function of its separation from a plane wall. 

\begin{figure} 
	\centering
	\includegraphics[width=0.7\textwidth]{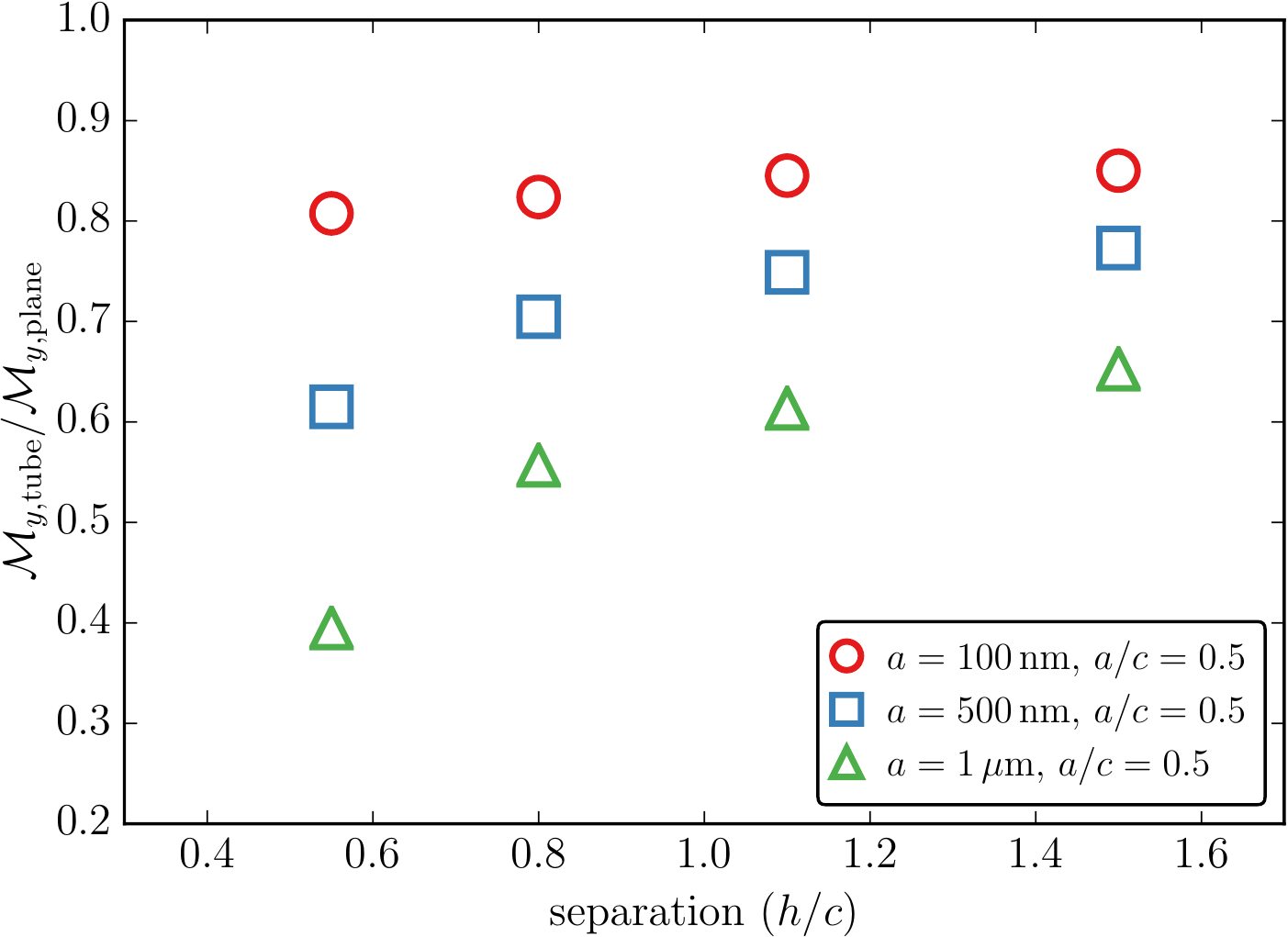}
	\caption{ Ratio of the translational mobilities for an oblate ellipsoid in the vicinity of cylindrical and planar walls  $\vindex{{\cal M}}{y,{\rm tube}}$/$\vindex{{\cal M}}{y,{\rm plane}}$. Data shown as a function of the particle separation from the wall $h$, and in our simulations $h$ is varied by varying the particle position along the $y$ direction. The estimates for $\vindex{{\cal M}}{y,{\rm plane}}$ is from Hsu and Ganatos~\citep{HsuGanatos89}.  We consider three different ellipsoidal particles with a fixed $\aspp=0.5$ and $a=100$ nm, $500$ nm and $1\mum$,  in a cylindrical tube diameter  of $D=5\,\mum$. Mesh parameters for the particle surface are $l_P=6$ nm, $16$ nm, and $64$ nm for the $100$ nm, $500$ nm, and $1\mum$ particles, respectively, and $l_W=524$ nm for the mesh on the cylindrical wall.}
	\label{fig_mobilitywallresult}
\end{figure}

A comparison of results are shown in Fig.~\ref{fig_mobilitywallresult} where we have plotted the ratio $\vindex{{\cal M}}{y,{\rm tube}}/\vindex{{\cal M}}{y,{\rm plane}}$ as a function of $h/c$. For a particle located at the center of the tube, the effect of wall curvature is a minimum and hence, $\vindex{{\cal M}}{y,{\rm tube}}/\vindex{{\cal M}}{y,{\rm plane}}$ $\simeq 1$. Indeed, this is displayed in Fig.~\ref{fig_mobilitywallresult}. For locations closer to the wall boundary, the curvature effects would become increasingly important as we approach the wall. Again, as displayed in Fig.~\ref{fig_mobilitywallresult} the ratio of the mobilities shows significant deviation from unity as we approach the wall and this effect is more pronounced with increasing particle size. This is as would be expected. These complete our validations.

\subsection{Effects of the bounding geometry and particle orientation in the absence of Brownian stresses}
\begin{figure} 
	\centering
	\includegraphics{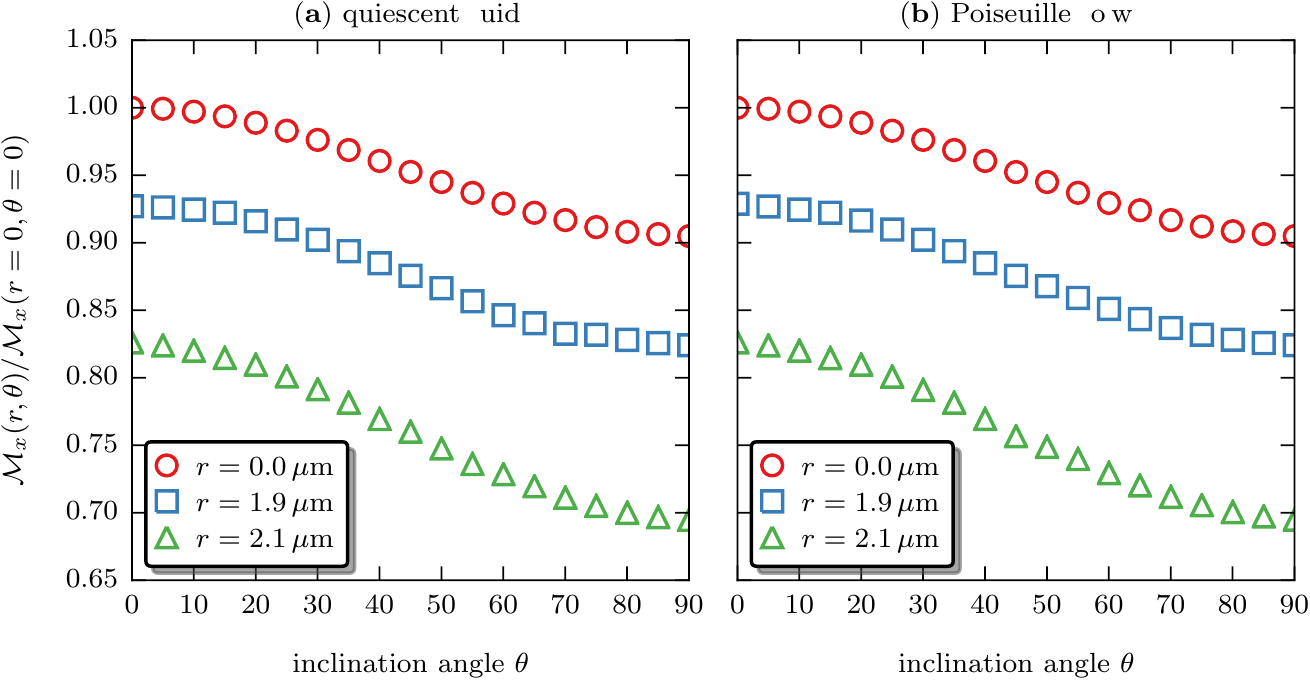}
	\caption{The normalized mobility for an ellipsoidal particle, with $\aspp=1.5$ and $a=500$ nm,  as a function of its inclination angle $\theta$ for three different radial positions -- (i) center ($r=0.0\,\mum$), (ii) near wall ($r=1.9\,\mum$), and (iii) lubrication ( $r=2.1\,\mum$) --  in (a) a quiescent medium and (b) a Poiseuille flow field.  Mesh parameters used are $l_P=5$ nm and $l_W=785$ nm.}
	\label{fig_mobilitywall_pq}
\end{figure}
In order to quantify the effects of the wall curvature and the particle orientation, we have  computed the mobility of an ellipsoidal particle ($a=0.6\,\mu{\rm m}$ and $b=c=0.4\,\mu{\rm m}$) placed at three different positions of the center of mass along the radial direction chosen  as $r=0.0\mum$ , $r=1.9\mum$, and $r=2.1 \mum$, which are representative of a particle in the bulk, near wall and lubrication regimes, for various inclination angles $0\leq \theta \leq 90$.  The mobility calculations are performed as described earlier  for both quiescent and Poiseuille flow conditions. For the Poiseuille flow,  $\bm{u}_{\rm in} (r) = {\bm u}_{\rm max} \left(1-(2r/D)^2 \right)$, where $\bm{u}_{\rm max}$ is the flow velocity at the center of the tube, and $D$ denotes the diameter of the tube. In targeted drug delivery applications since we are interested in capillary flows, we have performed our simulations with $\bm{u}_{\rm max}=0.1\,{\rm cm/s}$, which is representative of flow rates in capillary vessels~\citep{Mazumdar92}.  Figures~\ref{fig_mobilitywall_pq}(a) and (b) show ratio of the mobilities (${\cal M}_x(r,\theta)/{\cal M}_x(r=0,\theta=0)$) of the ellipsoid as a function of $\theta$ and its separation from the wall. We find that the mobility of the particle is strongly dependent both on the orientation $\theta$ and its radial position $r$. The effect of the flow field of the mobility is weak due to the low particle Reynolds number considered here ($\Rep \sim 10^{-4}$). 

\clearpage
\newpage
\section{Comparison of diffusivities for $\widetilde{h}=1$ and $\widetilde{h}=0.2$}\label{sec:diffusivities}
\setcounter{figure}{0}
\renewcommand{\thefigure}{\ref{sec:diffusivities}.{\color{blue}\arabic{figure}}}
Figs.~\ref{fig:comparenw} and ~\ref{fig:comparelub} show a comparison of the translational and rotational diffusivities for ellipsoidal NCs (with $\aspp$=0.5, 1.0, 1.5, 2.0, and 5.0) computed using the MSD approach and from the VACF, using the Green-Kubo relation, for $\widetilde{h}=1.0$ and $\widetilde{h}=0.2$, respectively. Data correspond to ellipsoidal NCs in a tube with $D=5\mum$ and $L=40\mum$.

\begin{figure}
	\centering
	\includegraphics[width=12.5cm,clip]{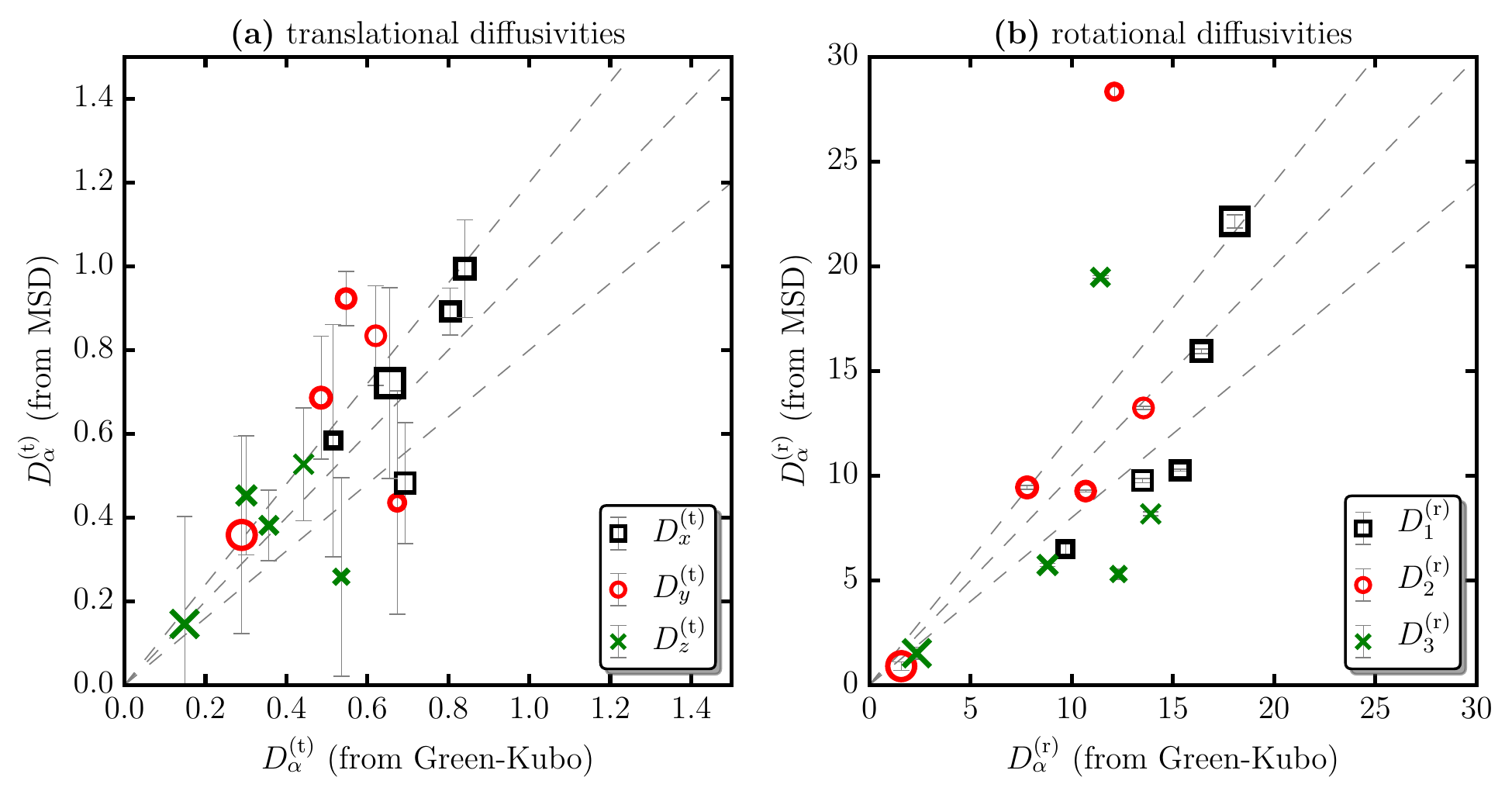}
	\caption{\label{fig:comparenw} Comparison of the translational and rotational diffusivities computed from the velocity-autocorrelation, using the Green-Kubo relation, to those estimated from the MSDs. Data for shown for NCs with five different aspect ratios and placed at $\widetilde{h}=1.0$. The central dotted line represents the linear correlation while the rest two represent deviations of $\pm 20\%$. The translational diffusivities (panel (a)) are in units of $\mum^2{\rm s}^{-1}$, and the rotational diffusivities (panel (b)) are in units of ${\rm rad}^2{\rm s}^{-1}$.}
\end{figure}

\begin{figure}
	\centering
	\includegraphics[width=12.5cm,clip]{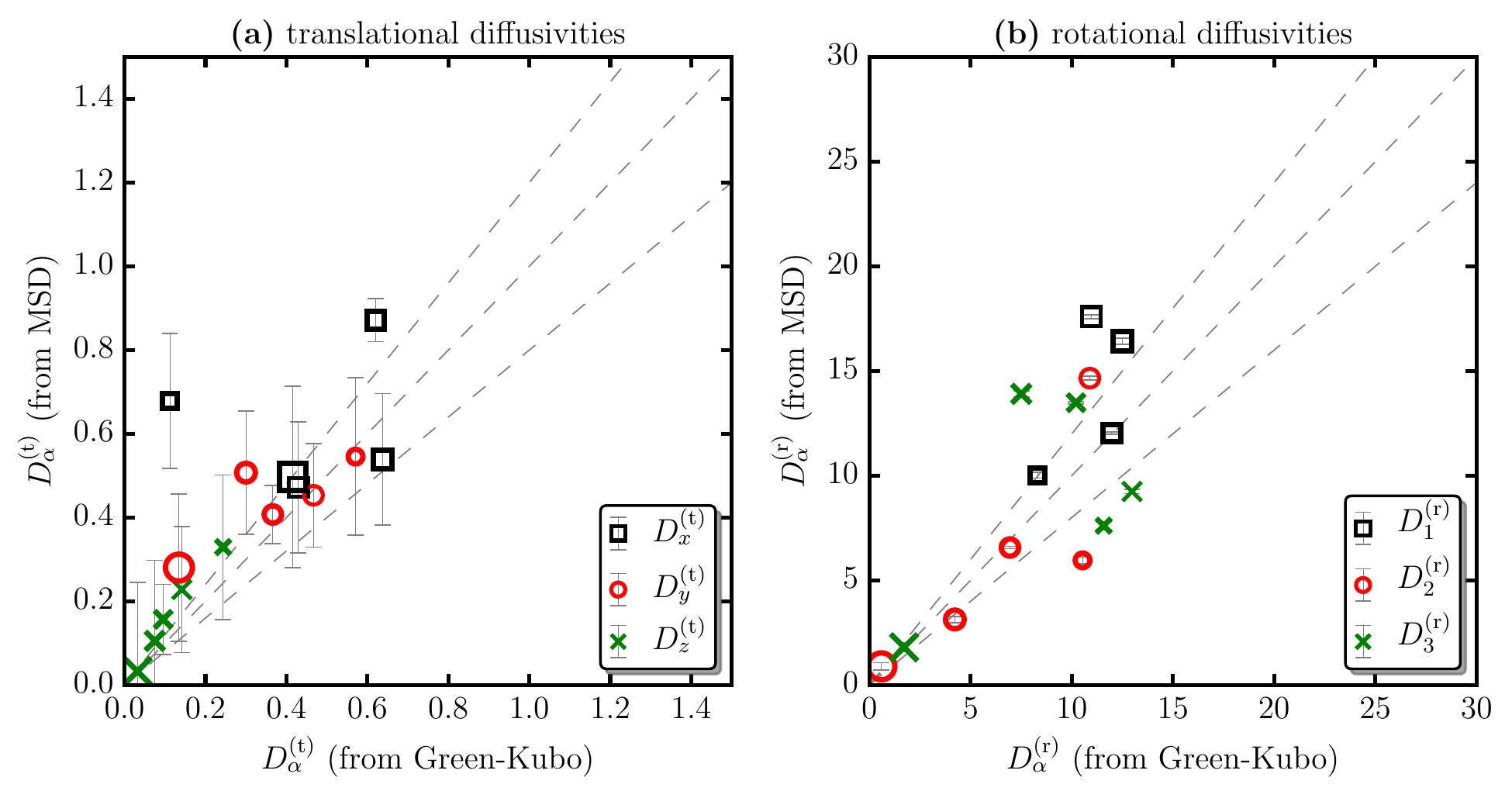}
	\caption{\label{fig:comparelub} Comparison of the translational and rotational diffusivities computed from the velocity-autocorrelation, using the Green-Kubo relation, to those estimated from the MSDs. Data for shown for NCs with five different aspect ratios and placed at $\widetilde{h}=0.2$. The central dotted line represents the linear correlation while the rest two represent deviations of $\pm 20\%$. The translational diffusivities (panel (a)) are in units of $\mum^2{\rm s}^{-1}$, and the rotational diffusivities (panel (b)) are in units of ${\rm rad}^2{\rm s}^{-1}$.}
\end{figure}

\clearpage
\newpage
\section{Comparison of diffusion constants estimated from MSD}\label{sec:diffconst}
\setcounter{figure}{0}
\renewcommand{\thefigure}{\ref{sec:diffconst}.{\color{blue}\arabic{figure}}}
In Fig.~\ref{fig:Dvsh-MSD}, we show the scaling behavior of the translational and rotational diffusivities computed using the MSD approach, as functions of \asp{} and $\widetilde{h}$. For a detailed description of the various scaling behavior see discussions around Fig. 14 in the main text.
\begin{figure}
	\centering
	\includegraphics[width=12.5cm,clip]{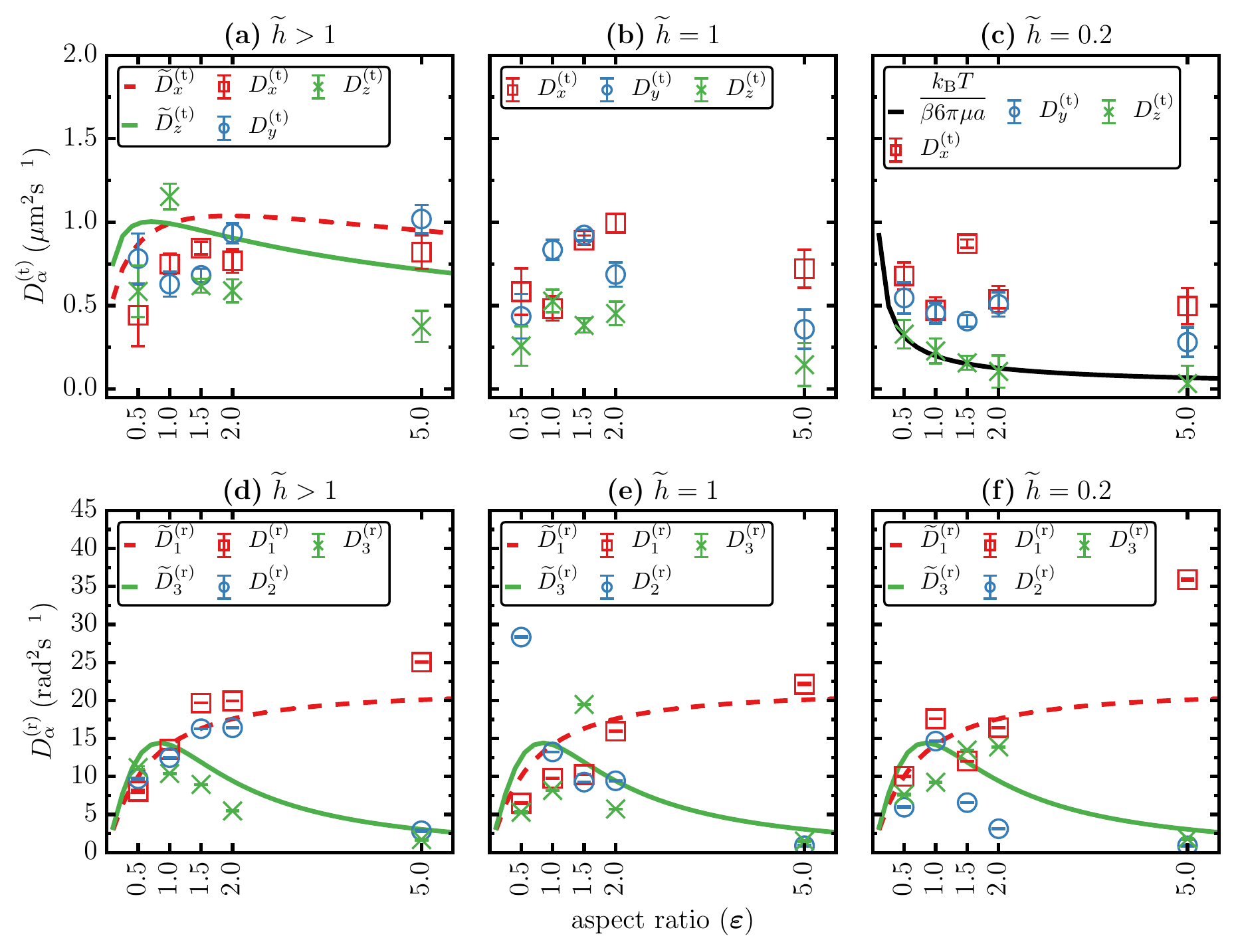}
	\caption{\label{fig:Dvsh-MSD} Translational and rotational MSDs estimated from the MSD, as a function of the aspect ratio and $\widetilde{h}$. The dotted and solid lines in the various panels are as described in Fig.14 in the main text. }
\end{figure}

\bibliographystyle{jfm}


\end{document}